\documentclass[aps,pre,color]{revtex4-2}

\usepackage{natbib,bigints,color,xcolor}
\usepackage{ulem} 
\usepackage{bm} 
\usepackage{amssymb}
\usepackage{amsfonts}
\usepackage{amsmath} 
\usepackage{graphicx}
\usepackage{xcolor} 
\usepackage[pdftex,colorlinks=true,allcolors=blue,breaklinks=true]{hyperref}  

\usepackage[utf8]{inputenc}

\definecolor{g-blue}{rgb}{0.83,0.95,1}
\definecolor{g-yellow}{rgb}{1,1,0.7}
\definecolor{g-green}{rgb}{0.9,1,0.9}
\definecolor{green}{rgb}{0,0.6,0}
\definecolor{cyan}{rgb}{0,0.7,0.7}
\definecolor{black}{rgb}{0,0,0}
\definecolor{grey}{rgb}{0.4,0.4,0.4}
\definecolor{nature-blue}{rgb}{0.0,0.200,0.500}

\def\black#1{\textcolor{black}{#1}}
\def\blue#1{\textcolor{blue}{#1}}
\def\red#1{\textcolor{red}{#1}}
\def\green#1{\textcolor{green}{#1}}
\def\cyan#1{\textcolor{cyan}{#1}}
\def\magenta#1{\textcolor{magenta}{#1}}

\def\white#1{\textcolor{white}{#1}}

\def \ed {\end{document}}

\def\be{\begin{equation}}
\def\ee{\end{equation}}
\def\bea{\begin{eqnarray}}
\def\eea{\end{eqnarray}}
\def\bse{\begin{subequations}}
\def\ese{\end{subequations}}

\let \nn  \nonumber

\let \= \equiv  
\let\*\cdot 
\let\^\widehat 
\let\-\overline

\def\x{\ort x}

\def\1{\bm1}

\def\<{\left\langle}    \def\>{\right\rangle}
\def\({\left(}          \def\){\right)}
\def\[ {\left[}         \def\]{\right]}

\newcommand{\eq}[1]{(\ref{#1})}
\newcommand{\Eq}[1]{Eq.\,(\ref{#1})}
\newcommand{\Eqs}[1]{Eqs.\,(\ref{#1})}
\newcommand{\Fig}[1]{Fig.\,\ref{#1}}
\newcommand{\Figs}[1]{Figs.\,\ref{#1}}
\newcommand{\Sec}[1]{Sec.\,\ref{#1}}
\newcommand{\Secs}[1]{Secs.\,\ref{#1}}

\newcommand{\B}[1]{{\bm{#1}}}
\newcommand{\C}[1]{{\mathcal{#1}}}    

\newcommand{\g}{\gamma}

\def\k{\kappa}

\definecolor{g-blue}{rgb}{0.83,0.95,1}
\definecolor{g-yellow}{rgb}{1,1,0.7}
\definecolor{g-green}{rgb}{0.9,1,0.9}
\definecolor{green}{rgb}{0,0.6,0}
\definecolor{cyan}{rgb}{0,0.7,0.7} 
\definecolor{grey}{rgb}{0.4 ,0.4 ,0.4 }
\definecolor{brown}{rgb}{0.6 ,0  ,0.8 }

\def\brown#1{\textcolor{brown}{#1}}
\def\g-blue#1{\textcolor{g-blue}{#1}}  

\def\black#1{\textcolor{black}{#1}}
\def\blue#1{\textcolor{blue}{#1}}  
\def\red#1{\textcolor{red}{#1}}
\def\green#1{{\textcolor{green}{#1}}}
\def\cyan#1{\textcolor{cyan}{#1}}
\def\magenta#1{\textcolor{magenta}{#1}}

\def\white#1{\textcolor{white}{#1}}
\usepackage{bm}
\usepackage{amssymb}
\usepackage{amsfonts}
\usepackage{amsmath}
 \usepackage{graphicx}
  \usepackage{amsmath,bm,epsfig}
\usepackage[notcite,notref]{showkeys}
\def \ed {\end{document}}
\def \BiG {0.9}
\def \Med {0.4}

\def\Fbox#1{\blue{\vskip1ex\hbox to 8.5cm{\hfil\fboxsep0.3cm\fbox{%
      \parbox{17.30cm}{#1}}\hfil}\vskip1ex\noindent}}  

\newcommand{\jul}

\def\be{\begin{equation}}\def\ee{\end{equation}}
\def\bea{\begin{eqnarray}}\def\eea{\end{eqnarray}}
\def\bse{\begin{subequations}}\def\ese{\end{subequations}}

\let \= \equiv  \let\*\cdot \let\~\widetilde \let\^\widehat \let\-\overline

\def\x{\ort x}  
  \def\1{\bm1} 

\def\<{\left\langle}    \def\>{\right\rangle}
\def\({\left(}          \def\){\right)}
 \def \[ {\left [} \def \] {\right ]}

\renewcommand{\sb}[1]{_{\text {#1}}}  
\renewcommand{\sp}[1]{^{\text {#1}}}  
\newcommand{\Sp}[1]{^{^{\text {#1}}}} 
\def\Sb#1{_{\scriptscriptstyle\rm{#1}}}
\def\He4 {$^4$He~}

\def \k {\bm{k}}

\def \x {\textbf{x}}

\def \x {\B {x}}

\def \q {{\bf q}}

\def \k {\B {k}}

\def \x {\B {x}}

\renewcommand{\sb}[1]{_{\text {#1}}}  
\renewcommand{\sp}[1]{^{\text {#1}}}  
\def\Sb#1{_{\scriptscriptstyle\rm{#1}}}
\begin{document}

\title{
  Energy flux and high-order statistics 
   of   hydrodynamic turbulence
 }

 \author{Yuri V. Lvov$^1$ and Victor S. L'vov$^2$\\
   $^1$ Department of Mathematical Sciences, Rensselaer Polytechnic Institute,
   Troy, NY 12180 \\
$^2$ Dept. of Chemical and Biological Physics, Weizmann Institute of Science, Rehovot 76100, Israel }
\date{\today}
\begin{abstract}  
  We use the Dyson-Wyld diagrammatic technique to analyze the infinite
  series for the correlation functions of the velocity in the
  hydrodynamic turbulence. We {{demonstrate}}
   the fundamental role played by the triple
  correlator of the velocity {{in determining the entire
      statistics of the hydrodynamic turbulence.}}  All
  higher order correlation functions are expressed through
  {{the triple correlator}}.  This  is shown
  through the {{suggested {\it{triangular}}}} resummation of
  the infinite diagrammatic series for multi-point correlation
  functions. {{The triangular}} resummation
  {{ is the next logical step after }}
  the Dyson-Wyld {{\it line}} resummation {{for the
      Green function and the double correlator.}}  {{In particular, it allows us to explain why
    }} the {{inverse cascade of the }}
  two-dimensional hydrodynamic turbulence is close to
  Gaussian.  {{Since the triple correlator dictates the flux
      of energy $\varepsilon$ through the scales, we support the
      Kolmogorov-1941 idea that $\varepsilon$ is one of the main
      characteristics of hydrodynamic turbulence. }}
\end{abstract}
\maketitle

   \section{Introduction}
   Investigation of the statistical properties of the hydrodynamic
   turbulence has a long and exciting history\,\cite{Frisch1995}.  The
   developed hydrodynamic turbulence may be characterized by three
   fundamental quantities: (i) the double correlation of the velocity
    (in the wave-vector, frequency representation  ${\bf q}\=\{\B k,\omega\}$)  $^2\C F(\bf q )$,  characterizing the energy distribution of
   $k$-eddies of scale $\ell\simeq 1/k$; (ii) the characteristic time
   scale $\tau(\B k)$ of the response of the $k$-eddies to the
   external perturbation, given by the Green function $\C G({\bf q})$;
   (iii) The triple correlation $^3\C F({\bf q}_1,{\bf q}_2,{\bf q}_3)$,
   responsible for the energy flux across the scale
   $\ell\simeq k_1^{-1}\simeq k_2^{-1}\simeq k_3^{-1}$.

   A systematic way to analyze these objects was suggested by
   Wyld\,\cite{Wyld1961} who develop a diagrammatic {{method}}
   to treat infinite { perturbation   series for the response
     (Green's) and correlation functions of the velocity field}.
      The essence of a diagrammatic technique is in a graphical
     representation (diagrams) of infinite perturbation series.
         The key advantage of the diagrammatic technique is that it is possible
to draw and analyze diagrams for the higher-order correlation function
without explicitly deriving the corresponding analytical expressions
first.

     {Basic objects in the Wyld technique are {the}
       so-called ``bare" Green's function $\C G_0({\bf q})$ and ``bare
       second-order correlation function $^2\C F_0({\bf q})$.}
     {{These bare objects }} depend on the kinematic viscosity
     $\nu$.  A crucial step forward was the Dyson-Wyld line
     resummation allows one to replace in all remaining diagrams the
     bare kinematic viscosity $\nu$ by what is called ``dressed by
     interaction turbulent viscosity" $\nu\sb{turb}$ that accounts for
     the main mechanism of the eddy damping due to the energy exchange
     between scales.  {{ From physical viewpoint this means
         that besides of accounting for small damping of energy of
         eddies of a given scale due to kinematic viscosity we account
         for much strong effect of their interaction with all the rest
         of the turbulent eddies, in the mean-field approximation,
         known in the physics of turbulence as an approximation of
         turbulent viscosity. Mathematically this is equivalent to
         replacing the initial expansion parameter $ Re\gg 1$, where
         $Re\propto (1/\nu)$ is the Reynolds number, by the parameter
         $ Re \sb{turb}\propto 1/\nu\sb{turb}=O(1)$.  }} As a result,
     the {resummed} diagrams involve only dressed objects:
     Green's function $\C G({\bf q})$, and simultaneous correlators
     $^2\C F({\bf q})$ instead of their bare counterparts
     $\C G_0({\bf q})$, $^2 F_0({\B k})$ involving only
     $\nu\ll \nu\sb{turb}$.  This kind of procedure in diagrammatic
     techniques is called {\it dressing}. {{It is well known
         that the dressing rearranges the terms in the perturbation
         expansion by moving the higher-order terms to lower orders
         and arrange them in the ``dressed'' objects.  Therefore, the
         infinite diagrammatic series becomes better ordered, more
         physically transpent and presumably less divergent. }}
     {{Nevertheless the series for
         $^3\C F({\bf q}_1,{\bf q}_2,{\bf q}_3)$, remains ``undressed"
         in the sense that it can be expressed in terms of the ``bare"
         $^3\C F_0({\bf q}_1,{\bf q}_2,{\bf q}_3)$, proportional to
         the original (``bare") interaction amplitude
         $\B V( \B k_1, \B k_2, \B k_3)$ in the Navier-Stokes
         equations~\cite{Frisch1995,pope2000turbulent}.}}

{ Analyses of the topological
properties of the resulting diagrams allowed us   to suggest in this paper, a natural next logical step after the
Dyson-Wyld line resummation,  the {\it triangle resummation}.  To find the triangle resummation would be impossible, or near-impossible  by studying analytical formulas for the perturbation expansion.  The triangular
  resummation expresses the simultaneous triple correlator
  $^3 F (\B k_1,\B k_2,\B k_3)$ in terms of three {\it dressed}
  objects, $\C G({\bf q})$ and simultaneous correlators $^2 F(\B k)$
  and $^3 F(\B k_1,\B k_2,\B k_3)$ itself. Since this dressing is the
  result of combining higher-order terms into these three dressed
  objects, the resulting infinite diagrammatic series is less likely
  to diverge. }  Moreover, we show that the quadruple and higher-order
correlators $^4 F $, $^5 F $, etc. {are} also proportional to
the powers of $^3 F $.  {Consequently, the fourth quadruple and
  higher order correlators do } vanish if $^3 F =0$. Since in the
thermodynamic equilibrium $^3 F =0$ it means that in the equilibrium
all comulants are zero and statistics of turbulence become Gaussian
order by order.  To reach these goals we carefully revisit the Wyld
diagrammatic approach from the very beginning paying special attention
to the numerical prefactors of the diagrams, crucially important for
their further resummations.

{{The principle  advantage of the proposed triangular resummation is
    that it expresses {\it{all}} simultaneous correlation functions
    through the {\it{dressed}} simultaneous triple correlator. The
    triple orrelator determines  the flux of
    energy {over scales}. Therefore, {\it{all} simultaneous
      correlators depend on the energy flux. This conclusion
      illustrates the unique importance of the energy fluxes through
      the spectral space} and can be considered as a generalization of
    Kolmogorov-1941 dimensional
    reasoning\,\cite{Kolmogorov1941equations,Frisch1995} that related
    the energy distribution over scales (i.e. the second-order
    velocity correlator) with the energy flux.} }
         
Having developed the theory for multiple-point correlators we consider
in more detail the 2D turbulence, which allows the presentation of the
Navier-Stokes equation in a scalar
form\,\cite{Kraichnan1980}. Remarkably, the 2D turbulence serves as an
idealized model for many natural flow phenomena, including geophysical
flows in the atmosphere, oceans, and magnetosphere.  Setups that are
quite close to 2D turbulence were realized
experimentally~\cite{Tabeling2002}.  It is observed in both DNS and
experiments that statistics of 2D turbulence is surprisingly close to
the Gaussian~\cite{Boffeta2000,Boffeta2012}. The natural explanation
of this fact follows { from our results. First, we show that
  $^3 F$ vanishes in the thermodynamic equilibrium. Second, all
  cumulants $^{n\!}F$ are proportional to powers of $^3 F$ and thus
  also vanish in the equilibrium as expected in the Gaussian
  statistics which takes place in the equilibrium (see,
  e.g. \cite{1980Landau}). This exposes the explicit mechanism of how
  Gaussian statistics of turbulence in equilibrium is order-by-order
  consistent with the diagrammatic expansion. Finally, because in
  {fractional} dimension $d=4/3$ the scaling index of the
  inverse energy cascade $^{2\!}F(k)\propto k^2$ coincides with that
  in the thermodynamic equilibrium (with the enstrophy equipartition
  between scales) we demonstrate Gaussianity of the inverse energy
  cascade in $d=4/3$.  We show also that near $d=4/3$ the triple
  correlator $^3 F\propto (d-4/3)$ and thus all cumulants $^{n\!}F$
  are small near $d=4/3$, being proportional to the powers of
  $(d- 4/3)$. This explains the closeness of the inverse cascade of
  the 2D turbulence in close to the Gaussianity also in the physical
  case $d=2$, as noticed in \cite{Lvov2002b}.}
  
The paper is organized as follows: in \Sec{ss:ME} we set up the
   stage by introducing a scalar equation for the 2D  and 3D turbulence. In
   \Sec{ss:field} we discuss the perturbation expansion for the field
   amplitudes showing that the prefactors in resulting tree diagrams
   are equal to $\dfrac 1N $, where $N $ is the number of elements in
   the symmetry group of each particular tree diagram. {{
   Many diagrams do not  have any symmetries apart from the identity
       transformation, so the $N=1$. If the diagram is symmetric with respect
       to a certain line, there will be two symmetry elemets: reflection and identity,
       so that $N=2$ and so on. This factor will be considered in details in the
       body of the paper. }}  We refer to
   this fact as the "$\frac{1}{N}$-symmetry" rule.  We show that the
   $\dfrac 1N$-symmetry rule is valid for all types of diagrams and for
   any of its fragments.
   
  The next step, presented in \Sec{ss:DT-CF}, is the procedure of
  ``gluing" of the $n$ tree diagram that results in diagrams for the
  $n$-point, different-time correlation functions $^n \C F$ for which
  the symmetry rule for the prefactors is also applicable.
  
  Analysis of the resulting diagrams leads to formulations in
  \Sec{ss:hight} of diagrammatic rules for $^n \C F$ that allow one to
  find them in arbitrary order without sequential analysis of all
  previous orders in the expansion. In principle this allows one to
  skip reading \Secs{ss:field}, \ref{ss:DT-CF} and \ref{ss:hight} and
  to look only at the final diagrams for the correlation functions.

  In \Sec{s:one-time} we reduce the resulting diagrams for the
  different-time correlations $^n F$ (in the
  ${\bf q}\= (\B k,\omega)$-representation) to the single-time domain,
  denoted as $^n F$. For this goal, we used the relation
  $^2 \C F(\B k, \omega)\propto \mbox{Re} \{\C G(\B k, \omega)\}$
  {{where Re denotes the real part of a complex quantity. This
      expression}} follows from Wyld resummation. The resulting
  ``extended" set of diagrams for simultaneous correlators $^n F$
  involve simultaneous $^2 F(\B k)$ and {the} Green's functions
  $\C G(\B k, \omega)$. Once again, the prefactors are given by the
  $\dfrac{1}{N}$ symmetry rule.
  
  The numerical value of the prefactors in the extended set of
  diagrams for $^n F$, given by the $\dfrac1N$-symmetry rule, allows us
  to group them into groups of three (triads) such that each group
  appears as a diagram for $^3 F$. {{ Interestingly, some
      diagrams participate in more than one triad. Consequently,
      grouping diagrams into triads to form a triple correlator is a
      nontrivial task.  Finally, we discovered how to find a set of
      triads that can be summed up to the full series for $^3 F$.}}

  Note that the topological structure of the diagrammatic series is
  defined by the quadratic nonlinearity of the Navier-Stockes equation
  with the interaction {{vertex}} satisfying {{Jacobi}}
  identity.  The Jacobi identity is a mathematical manifestation of
  energy conservation in hydrodynamic turbulence. The 2D turbulence
  has an additional {{Jacobi}} identity manifesting the
  enstrophy conservation.  Therefore our conclusions are applicable to
  both three- and two-dimensional turbulence.

\section{\label{s:DT} Diagrammatic technique for strongly interacting fields}

\subsection{\label{ss:ME} Basic equation of motion {for 3D and 2D} hydrodynamics}
This paper is based on the Wyld diagrammatic technique for
hydrodynamic turbulence\,\cite{Wyld1961} generalized by Martin,
Siggia, Rose\cite{Martin1973} and by Zakharov,
L'vov\,\cite{Zakharov1975b}. Its detailed review is available in
\cite{Lvov1995d}.  {Generally speaking, the proposed technique
  can be applied straightforwardly to any integer dimensions,
  including either two-dimensional or three-dimensional turbulence
  that differs in the analytical form of the Navier-Stokes equations,
  as well as to other problems, for example, passive scalar. Its
  application for non-integer dimensions is more tricky and requires
  understanding how to perform integrations in non-integer dimensions,
  see e.g. \cite{Lvov2002b}.}

{In the three dimensional case (3D) Euler equations for
  velocity $\B v(\B r,t)$ of an incompressible fluid with the density
  $\rho=1$ has well known form~\cite{landau2013Fluid}:
\begin{subequations}\label{NSE}
\begin{equation}\label{NSEa}
\frac{\partial \B v(\B r,t)}{\partial t} + (\B v \cdot \B \nabla)\B v + \B \nabla  p= 0\,, \quad \B \nabla \cdot \B v= 0\ .
\end{equation}
In the $(\B k, t)$ representation for the vector components $u^\alpha (\B k, t)$   \Eq{NSEa} can be rewritten as follows:
\begin{align}\begin{split}\label{NSEb}
\frac{\partial v^\alpha (\B k, t)}{\partial t}=   &= \frac 12 \int
 \frac {d^3 k_1 d^3 k_2}{ (2\pi)^2} \delta (\k+\k_1+\k_2)\Gamma_{k12}^{\alpha\beta\gamma}~
   u^{* \,\beta  }_{\k_1}u^ {*\,\gamma}_{\k_2}\,, 
\end{split}\end{align}
see e.g.~\cite{Lvov1995d}.
Here $\Gamma_{k12}^{\alpha\beta\gamma}$  is the interaction amplitude
\begin{equation}\label{NSEb}
 \Gamma_{k12}^{\alpha\beta\gamma}=i\sum_{\alpha'} \Big( \delta_{\alpha \alpha'} -\frac {k^\alpha k^{\alpha'}}{k^2}  \Big)( k^{\beta}\delta_{\alpha'\gamma}+ k^\gamma \delta_{\alpha'\beta}) 
 \end{equation}
 and $\delta_{\alpha\beta}=1$ if $\alpha=\beta$ and vanishes
 otherwise.  Euler equation \eqref{NSEa} preserves the total energy of
 the flow
 \begin{equation}
 \C E = \int |\B v(\B r,t)|^2 d^3 r= \int |\B u(\B k,t)|^2 \frac{d^3 k}{(2\pi)^3}\ . 
 \end{equation}
 Therefore  $\Gamma_{k12}^{\alpha\beta\gamma}$ satisfies   Jacobi identity 
 \begin{equation}\label{1e}
 \Gamma_{k12}^{\alpha\beta\gamma}+\Gamma_{2k1}^{\gamma\alpha\beta}+
 \Gamma_{12k} ^{\beta\gamma\alpha}=0 
 \end{equation}
 on the surface $\B k+ \B k_1+ \B k_2=0$. 
\end{subequations}
}

{The basic equations of motion for two-dimensional (2D)
  turbulence has a structure, similar to 3D case \Eqs{NSE}.} The 2D
turbulence may be represented as a scalar equation for the vorticity,
which simplifies analytical
expressions. Therefore for the transparency of the presentation we
illustrate our formalism {{for}}  the 2D
turbulence.  In the present work we, following\,\cite{Lvov2002b},
consider the Euler equation for the vorticity equation in 2D:
\begin{equation}
\label{NS3}
\partial {\omega}/ \partial t+ ({\B u}\cdot {\bf \nabla})\, \omega=0\ .
\end{equation}
The velocity and vorticity of a two-dimensional (2D) flow may be
derived from the {\it stream function} $\psi({\B x},t)$:
${\B u} ({\B x},t)=-{\bf \nabla} \times \hat {{\B z}} \psi(\x,t)$,
$\omega({\B x},t) = -\nabla^2 \psi({\B x},t)$, where $\hat {{\B z}}$
is a unit vector orthogonal to the $\hat{\B x}$-plane, and $ \nabla^2$
is the Laplacian operator in the plane.  In $\B k$-representation,
$ a(\B k,t)\equiv k\int d{\B R} \exp[-i({ \B R}\cdot{\k} )] {\psi}(\B
R,t)$.  The Fourier transforms of $\B u(\x,t)$ and of
${\bf \omega}(\x,t)$, are denoted as ${\B v}(\k,t)$ and
${ \Omega}(\k,t)$ respectively. These Fourier transforms are expressed
in terms $a(\k,t)$, re-designated for the shortness as $a_{\B k}$:
$\B v(\k,t)=i(\hat {{\bf z}}\times \hat {\k})\, a_\k$,
$\Omega(\k,t)= -k\, a_\k$, where $\hat {\k} =\k/k$.  Now, by \Eq{NS3})
\begin{align}\begin{split}\label{BE2}
 \frac{\partial a_\k }{ \partial t} &= \int
\!\!\frac {d^2 k_1 d^2 k_2}{2\cdot2\pi} \delta (\k+\k_1+\k_2)V_{k12}~
   a^*_{\k_1}a^*_{\k_2}\,, \qquad  
V_{k12} =  \frac{S_{k12}(k_2^2-k_1^2)}{2 kk_1k_2}\,,\ S_{k12} \equiv2 k_1 k_2\,
\sin \varphi_{12} ,~~ \\  
 S_{k12} &= 
S_{2k1}=S_{12k}=- S_{k21}=- S_{1k2}=- S_{21k}\,,
 \quad |S_{k12}| = \sqrt{2(k^2k_1^2+k_1^2k_2^2+k_2^2k^2)
 -k^4-k_1^4-k_2^4} \ .
\end{split}\end{align}
Here the interaction amplitude (or ``vertex'') $V_{k12}$ is expressed
via $S_{k12}$; $|S_{k12}|/4$ is the area of the {{triangular}}
formed by the vectors $\k\,,\k_1$ and
$\k_2$. $\varphi_{12}=\varphi_1-\varphi_2$; $\varphi_k$, $\varphi_1$
and $\varphi_2$ are the angles in the {{triangular}} plane between the
$x_1$-axis and the vectors $\k$, $\k_1$ and $\k_2$ respectively.  The
vertex $V_{k12}$ satisfies two Jacobi identities
	\begin{equation}\label{jac}
          (V_{k12}+V_{2k1}+V_{12k})=0\,,
          \qquad 
	(k^2V_{k12}+k_2^2V_{2k1}+k_1^2V_{12k})=0\ . 
	\end{equation}
These two identities ensure the conservation of energy $\C E$
        in the inviscid forceless limit and the enstrophy $\C H$ given by
\begin{equation}
  \label{ints2} 
  \C E\equiv \int  |a_k|^2 \frac{d ^2 k}{(2\pi)^2}\,, \quad 
\C H\equiv \int  k^2 |a_k|^2 \frac{d ^2 k}{(2\pi)^2}\ .
\end{equation}

Equation (\ref{BE2}) describes the two-dimensional hydrodynamic
turbulence.  One sees that it has the same form as 3D \Eqs{NSE} but
without  additional vector indices.  Therefore the results of this
{{paper}} are applicable for both 2D- and 3D turbulence.
{ The concrete conclusions of our paper depend on the presence
  of the Jacobi identity for the symmetry of the matrix element.  The
  2D turbulence has two quadratic integrals of motion (energy and
  enstrophy) and two Jacobi identities, \eqref{jac} that reflect this
  fact. The 3D turbulence has one integral of motion and just one
  Jacobi identity\,\eqref{1e}. As shown, e.g.  by Krachnan and
  Mongomery\cite{Kraichnan1980} the physical properties of these two
  systems are different, yet they are described by the same technique
  and same triangular resummation.}  To simplify our presentation we
focus in the paper on the 2D turbulence.
 
Following Wyld\,\cite{Wyld1961} we divide the world into the system
under consideration and the thermostat. The action of the thermostat
on the system is modeled by random noise $f(\B k,t)$ and damping
$\gamma_0(k)$. Then we replace in LHS of \Eq{BE2}
{{$\partial a_\k/ \partial t \Rightarrow [\partial/
\partial t+ \gamma_0(\B k) ]a_\k- f(\B k,t)$}} so that we obtain
instead of \Eq{BE2}
        \begin{equation}\label{ME}
	\Big[\frac{\partial }{\partial t}+   \gamma_0(\B k) \Big]a_\k  
        =\int
        \frac {d^2 k_1 d^2 k_2}{2\cdot2\pi} \delta (\k+\k_1+\k_2)V_{k12}~ a^*_{\k_1}a^*_{\k_2}+f(\B k,t)\,, 
\end{equation}
where the average statistics of the noise $f(\B k,t)$ is assumed to
satisfy
$ \< f(\B k,t) f(\B k',t') \>\propto T \gamma_0(\B k,t) \delta(\B k-
\B k') \delta(t-t')$. {{Here $\< \dots \>$ denotes an average
    with respect to the thermodynamic equilibrium ensemble with
    temperature $T$.  }} The presence of the thermostat force and the
damping allows (\ref{ME}) to have nontrivial solutions.  After the
Dyson-Wyld line-resummation, described below, we will disconnect our
system from the thermostat by taking the limit $\gamma_0(\B k)\to
0$. It was shown\,\cite{Wyld1961,Zakharov1975} that the result is
independent of the thermostat parameters.

After the Fourier transformation with respect to time $t$,  
{{\Eq{ME}}} in the ${\bf q}=(\B k, \omega)$-representation
becomes:
  	\begin{eqnarray}\label{BEa}
	 a_\q    =\,^0\C G_\q   \Big [ \frac12\int   \frac {d \q_1   d   \q_2}{ (2\pi)^{d+1}}
  \delta^{d+1}_{q12} V_{k, 1 2}   a^*_1 a^*_2 + f_\q \Big ]\,, \qquad  ^0 \C G_\q= i\big /\big [\omega+i\gamma_0(\B k)\big ]\ . 
	\end{eqnarray} 
	Here $^0\C G_\q$ is the bare Green's function,
        $d \q_j \= d^2 k_j d\omega_j$, and
        $V_{\B k \B 1 \B 2}\=V(\B k ,\B k_1 ,\B k_2)$ {is the
          interaction matrix element describing the strength of
          interactions btween wave numbers $\B k$ ,$\B k_1$ and
          $\B k_2$.}
 
  \subsection{\label{ss:field} Iterative expansion for field variables   $a_\q$  } 
  Introducing the zero-order solution of this equation $^{0\!}a_\q\=
\, ^0 \C G_\q f_\q $ we can get its iterative solution as a formal
infinite series with respect to powers of $^{0\!}a_\q$: $a_\q = \,^0\! a_\q +  \blue{ ^1\!a_\q  }+   \green{^  2\!a_\q  }+
\brown{   ^{3a\!}a_\q  +  ^{3b\!}a_\q   }+\red{ \, ^{4a\!}a_\q+ \, ^{4b\!}a_\q+ \, ^{4c\!}a_\q }+\dots $, where
      
	\begin{subequations}\label{IT2}
          \begin{eqnarray} \label{1A2} \blue{^1\!a_ \q} &=&
            \blue{\frac{ ^0\C G_\q}{2}\int \frac {d {\q_1} d {\B
                  q_2} }{ (2\pi)^{d+1}} \delta^{d+1}_{q56}V_{k56} {
                a_5} { a_6} \ \ \ \Rightarrow \blue{^1\!a_ \q} \propto
              \frac{\C G\, V}2 { a _5\, a _6 }} , \\ \label{2A2} \green{
              ^{2\!}a_\q } & = & \green{ \frac{ ^0\C G_\q}2 \int \frac
              {d {\q_1} d {\q_9} }{ (2\pi)^{d+1}}
              \delta^{d+1}_{q19}V_{k19} G_{\overline 1} \int \frac {d
                {\q_2} d {\q_3}}{ (2\pi)^{d+1}} \delta^{d+1}
              _{156}V_{156} \, a_9 a_5 a_6 } \ \ \ \green{
              \Rightarrow ^{2\!}a_\q\propto\frac{(\C G V)^2}2 \, a_9 a_5
              a_6}\,, \\ \label{3aA2} && \hskip - 1cm \brown{
              \,^{3a\!}a_\q =\, ^0\C G_\q \int \frac {d {\q_1} d {\q_5} }{ (2\pi)^{d+1}} \delta^{d+1}_{q15}V_{k15}
              \,^{2\! }a_1 a _5 \ \ \ \Rightarrow ^{3a\!}a_\q\propto
              \frac{(\C G V)^3}{2 } a_5 a_6 a_7 a_8 }\,, \\ \label{3bA2}
            && \hskip - 1cm \brown{ ^{3b\!}a_\q =\frac{ ^0\C G_\q}2
              \int \frac {d {\q_1} d {\q_2} }{ (2\pi)^{d+1}}
              \delta^{d+1} _{q12}V_{k12} \,^{1 \!}a_1\, ^{1\! } a_2 \
              \ \ \Rightarrow \ ^{3b\!}a_\q \propto \frac{(G
                V)^3}{2^3} a_5 a_6 a_7 a_8 } \,, \\ \label{4aA2} &&
            \hskip - 1cm \red{ ^{4a\!}a_\q= \,^0\C G_\q \int \frac {d
                {\q}_1 d {\q}_2}{ (2\pi)^{d+1}} \delta^{d+1}
              _{q12}V_{k12} \, ^{3a\!}a_1 a _2 \ \ \ \Rightarrow \
              ^{4a\!}a_\q\propto \frac{(\C G V)^3}{2 } { a_5} a_6 a_7 a_8
              a_9} \,, \\ \label{4bAa} &&\hskip - 1cm \red{
              ^{4b\!}a_\q=\,^0\C G_\q \int \frac {d {\q}_1 d {\q}_2}{ (2\pi)^{d+1}} \delta^{d+1} _{q12}V_{k12}
              {^2\!}a _1 {^1\!}a_2 \Rightarrow{\bf{ \
                  ^{4b}a_q\propto}} \frac{(\C G V)^3}{4 } { a_5} a_6 a_7
              a_8 a_9} \,,\\ \label{4cA} && \hskip - 1cm \red{
              ^{4c\!}a_\q= \,^0 \C G_\q \int \frac {d {\q_1} d {\q_2}
              }{ (2\pi)^{d+1}} \delta^{d+1} _{q12}V_{k12} \, ^{3b
                \!}a_1 a _2 \ \ \ \Rightarrow \ ^{4c\!}a_\q\propto
              \frac{(\C G V)^3}{8 } a_5 a_6 a_7 a_8 a_9} \ .
\end{eqnarray}\end{subequations}

Here
$\black{ \ ^0\! a_\q}, \blue{ ^1\!a_\q }, \green{^ 2\!a_\q }, \brown{
  ^{3\!}a_\q}$ and $\red{ \, ^{4\!}a_\q}$ are zeroth, first, second,
third, and fourth-order iterations in the powers of interaction matrix
element $V$. Here the number to the left of $a$ denotes the order of
iterative. For the third and fourth order there are contributions of
different {topologies}, so the letters ``a'',''b'', and ``c'' are used
to differentiate between them.

\begin{figure}
\includegraphics[width=\BiG\textwidth]{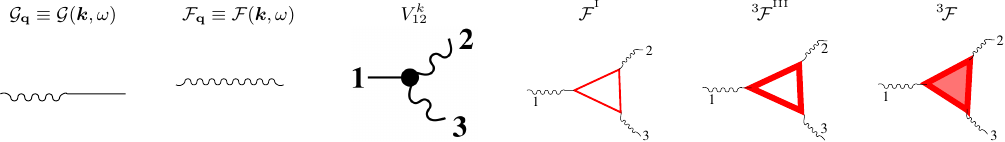} 
\caption{ \label{FF:1basic} Graphical notation for the line-resummed
  Wyld's diagrammatic expansion.  The symbols used are the following:
  $\bullet$ A short wavy lines stand for the canonical variable
  $a_\q=a(\B k,\omega)$.  $\bullet$ A straight line stands for the
  random force field $f(\B r,t)$ that appear in \Eq{ME}. $\bullet$ The
  Green's function $\C G(\B k,\omega)$, which is the response in the
  vorticity to some force is made of a short wavy line and a short
  straight line. $\bullet$ A long wavy line will represent double
  correlation functions $\C F(\B k, \omega)$, of the
  velocities. $\bullet$ The vertex $V_{123}$, \Eq{ME}, is a fat dot
  with three tails. One straight tail belongs to the Green function
  and two wavy tails represent velocities.  $\bullet$ The
  {{triangle}} with three wavy lines represents simultaneous
  three-point correlators of the first order
  $\ ^3\! \C F\Sp {I}_{123}$ (thin {triangle} of the
  third-order $\ ^3\! \C F\Sp {III}_{123} $ (thick triangle) and fully
  dressed three-point correlator (in all orders) $ ^3\! \C F _{123} $
  (red filled triangle).}
 	\end{figure}

\begin{figure}[t]
    \includegraphics[width=\Med\textwidth]{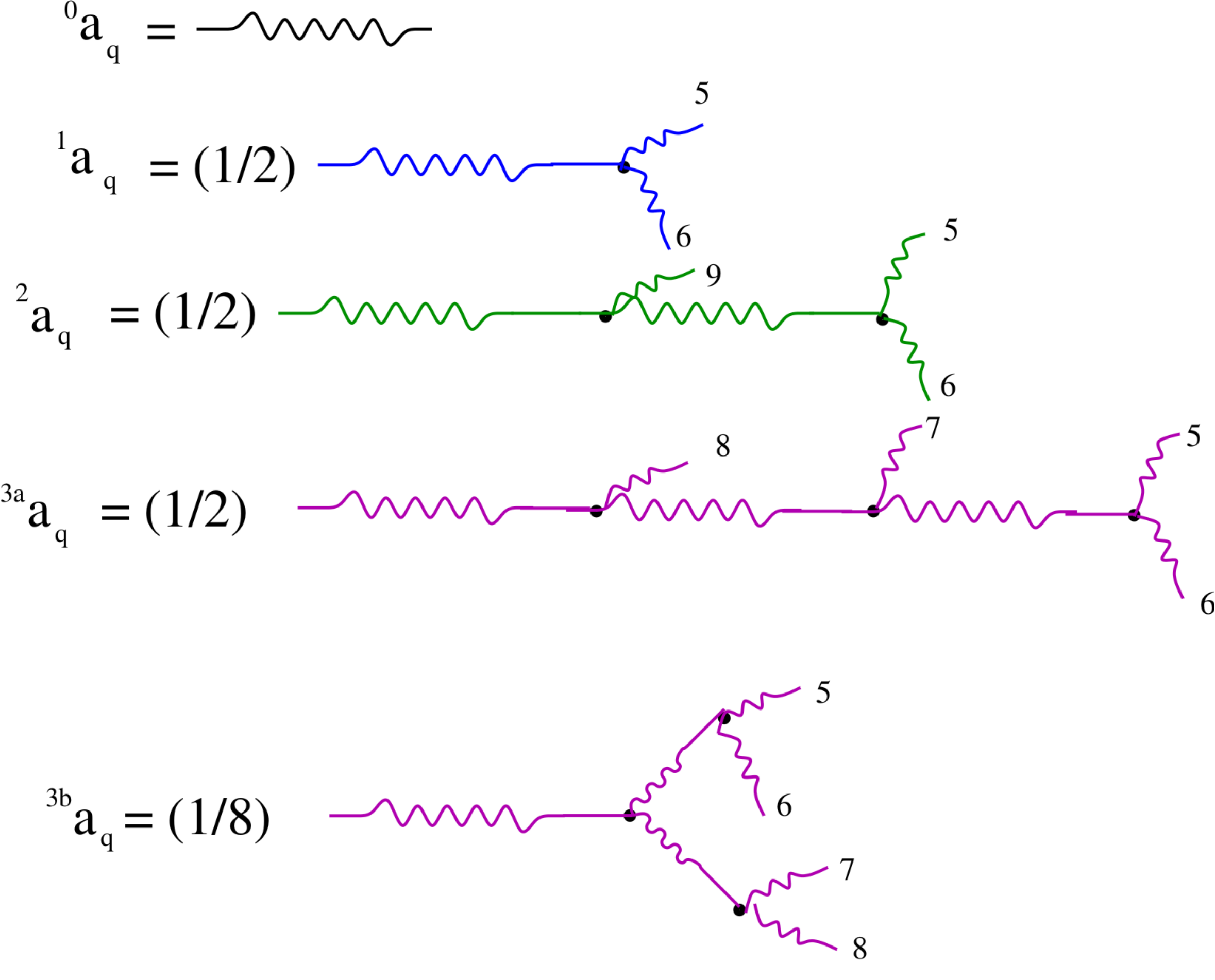}~~~~~~~~~~~
  \includegraphics[width=0.47\textwidth]{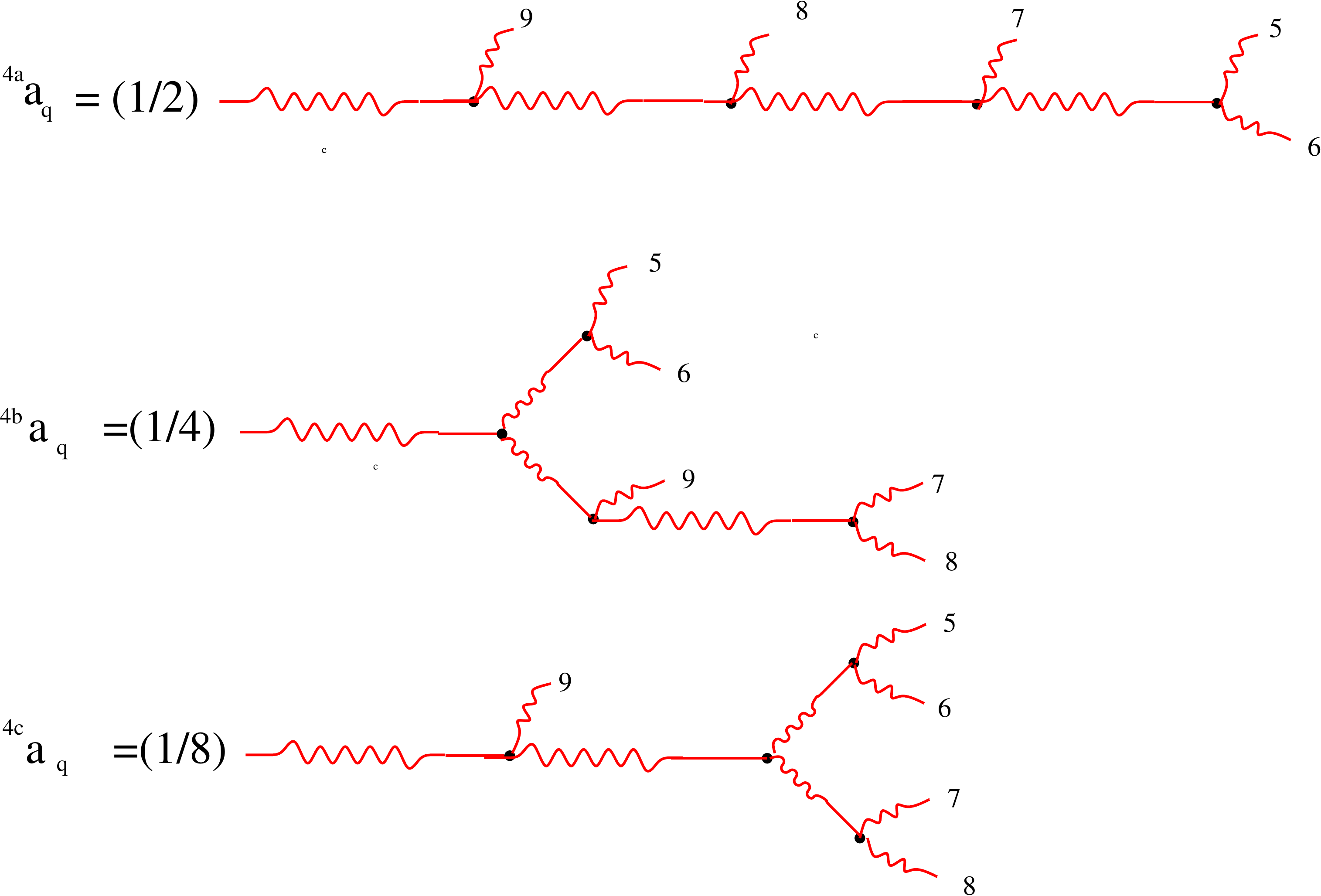}
	\caption{ Color online. Graphical representation of the
          iterative expansion of $a_\q$, given by \Eq{IT2}. We have
          reserved {indices} 1, 2, 3, and 4 ($\q_1,\ \q_2,\q_3$ and
          $\q_4$) for the arguments of the correlation
          functions. Therefore we supplied wavy tails of the trees for
          $^n\! a_j $ with {indices} $j=5,\ 6,$ etc. Here left
          superscript $^n$ denotes the iteration order (the number of
              the {{vertices}} in trees).  }
	\label{FF:1}
      \end{figure}

Using graphical notation shown in \Fig{FF:1basic} we can present each
term in this series in a graphical form, as a ``tree" diagram, as
shown in \Fig{FF:1}. In these diagrams, $^0G_\q$ shown as thin
wavy-straight line, $^0\!a_\q$ as a short thin wavy line which are
connected by vertex $V_{123}$ shown as a fat dot ``$\bullet$" with
the straight tail, belonging to $^0G_{\q_1}$ and two wavy tails,
belonging to $^0\!a_{\q_2}$ and $^0\!a_{\q_3}$.  The key realization
which gives birth to the diagrammatic technique is that instead of
deriving \Eqs{IT2} we could have had drawn {\it all} possible
topologically different trees, {\it without deriving  \Eqs{IT2} analytically first}.

Analyzing {\Eqs{IT2}} and \Figs{FF:1} with the trees, we see that the
trees with the symmetrical elements have a numerical prefactor that is
given by $\dfrac 1N$, where $N$ is the number of elements of the
symmetry group of a diagram. This is a constructive demonstration of
the $\dfrac{1}{N}$ symmetry rule for the trees.  We will see this rule
again when we consider diagrams for the correlation function.  The
symmetry factor appears as a consequence of the
${\B k_1}\leftrightarrow {\B{k_2}}$ symmetry and factor $\dfrac 12$ in
the equation of motion (\ref{ME}). The rigorous proof of the
$\dfrac{1}{N}$ symmetry rule is beyond the scope of the present
{{paper}}.  The $\dfrac{1}{N}$-symmetry rule will play a crucial role
below, as it will lead to the natural grouping of the diagrams into
triads. It would be much harder to see this rule by looking at
analytical expressions alone.

  The next important advantage of a diagrammatic technique is that
  from topological properties of the diagrams one can make conclusions
  about the corresponding analytical expression without detailed
  analysis and even perform a partial resummation of diagrams with
  particular topological properties. This observation leads to the
  Dyson-Wyld line-resummation of \textit{reducible} diagrams. {\it
    Reducible} are the diagrams that contain fragments that can be
  disconnected from the rest of the diagram by cutting two lines. If
  these cut lines are wavy and straight ones, then the infinite sum of
  the corresponding fragments become ``dressed", Green's function
  $\C G_\q$, defined as
  {{$\< \partial a_\q / \partial f_{\q'}\>=(2\pi)^{d+1} \delta^{d+1}
      (\q -\q') \C G_\q$.}}
      {{  This Green's function}}
  can be presented as (see
  e.g. \cite{Wyld1961,Martin1973,Zakharov1972b}) 
	\begin{equation}\label{GnB}
	\C G_\q= i \big / \big [\omega + i \gamma_0(\B k)-\Sigma_\q
          \big ]\,, \quad -\mbox{Im}[\Sigma_\q ] =\Gamma_\q=k^2
        \nu\sb{turb}(k)\,,
	\end{equation}	
	where the ``mass operator" $\Sigma_\q$ is an infinite
        {{sum of}} diagrams {{that begin and end with a
            vertex and determines}} the ``turbulent" dissipation:
        $\nu\sb{turb}$. {{In the case, where}} cut lines in the
        reducible {diagram} are two wavy lines, the
        {{infinite sum corresponds to the}} ``dressed" double
        correlator ${\cal F}_\q$, defined {in the next
          subsection} by \Eq{F2}, shown in diagrams as long thick wavy
        lines.

{{After performing the Dyson-Wyld line resummation, in the
    expansion\,\eqref{IT2} it is possible to replace the bare Green's
    functions $^0\C G_\q$ by their dressed counterpart $\C
    G_\q$. Furthermore, it is possible to replace the bare field
    $^{0\!}a_\q$ by the dressed field $a_\q$. Such modification presents the essence of
    ``dressing'', i.e. moving terms from higher orders of the perturbation theory to lower
    orders and combining them into the ``dressed'' objects. The ``dressed'' version of
    \eqref{IT2} will be used in the rest of the {{paper}}.}}

    \subsection{Diagrammatic expansion of
                correlation functions\label{ss:DT-CF}}
              \subsubsection{\label{ss:Definitions}
                Definitions and procedure}
              We define the two-, three-, four-, and $n$-point
              correlators in $\q=(\B k,\omega) $ space as
     \begin{subequations}\label{Fn}
  \begin{align} \label{F2} 
     &  (2\pi)^{d+1}\delta( {\q_1}+{\q_2}) \ ^{2\!} {\cal F}
    ({\q_1},{\q_2})  
 = \frac{\langle a_{\q_1} a_{\q_2}\rangle}{ 2!} \,, \quad
      (2\pi )^{d+1}\delta( {\q_1}+{\q_2}+{\q_3})  \ ^{3\!} 
    {\cal F}({\q_1},{\q_2},{\q_3})   
     =  \frac{\langle a_{\q_1} a_{\q_2}a_{\q_3}\rangle}{ 3!}\,, 
   \\   \label{F4a}
   &  (2\pi)^{d+1}\delta( {\q_1}+{\q_2}+{\q_3}+{\q_4}) \ ^
       {4\!} {\cal F}({\q_1},{\q_2},{\q_3},{\q_4})
           =  \langle a_{\q_1} a_{\q_2}a_{\q_3}
           a_{\q_4}\rangle/(4!)\,, \\ \label{9c}
  & (2\pi)^{d+1}\delta\Big(\sum\limits_{j=1}^n {\q_j}\Big)\ ^{n\!} {\cal F}
    ({\q_1},\dots, {\q_n})
    \= {\langle \prod\limits_{j=1}^n a_{\q_j}\rangle}/ {( n!)} , ~~~ n = 2,3,
    \dots\,, \quad    ^2{\cal F}(\q)\=\,^2  {\cal F}(\q, -\q)\ .  
         \end{align}\end{subequations}

       Here $d$ is the dimension of space. In the case of 2D
       turbulence, $d=2$.  We have included prefactor $1/ n!$ in the
       {definition (\ref{Fn})} of $n$-point correlation function $^{n\!}\C F$.
       Note that $n!$ is the number of elements of the symmetry group
       of a correlator that is equal to the number of permutations in
       the definition of $\ ^n \C F $ in the definition \eqref{F2}.
       This is precisely the choice that ensures the applicability of
       our   $\dfrac1N$-symmetry   rule  for the correlation functions. 
       As we will see below this
       particular choice simplifies the appearance of final
       expressions for $^{n\!}\C F$. Notice that the notation 
       $^{2\!}\C F ({\q},{\q'})$ involves two arguments, while
       actually, it depends only on one argument, {say} $\q$.
       Therefore in \Eq{9c} we define it in the more traditional way.

       Diagrammatic presentations of $^2\C F$, $^3 \C F$ and $^4\C F$
       can be obtained by gluing together two, three, and four
       trees. The gluing is a graphical representation of the
       averaging over the ensemble of the random force.  On the
       corresponding diagrams of two glued trees the dashed line
       crossing out the double correlator is the point where the
       ``branches'' of two trees were ``glued'' to form a double
       correlator.  The number of possible combinations of the glued
       trees will be of crucial importance in further investigation of
       the diagrammatic series.

       For the Gaussian process, the high-order correlation functions
       can be presented as a product of all possible second
       order correlators. Specifically, it means that
  \begin{equation}
  \begin{split}
  &\langle a_k^* a^*_l a_p a_n \rangle = \C F _k \C
    F_l(\delta^k_p\delta ^l_n + \delta^k_n\delta ^l_p)\,,\ \ 
    \langle
    a_k^* a_l a_p a_n\rangle =0 \,, \quad \langle a_k a_l a_p a_n\rangle =
    0\ .
  \label{eq:wick}
  \end{split}
\end{equation}

  In this {{paper}}, the $n$-point correlators $^n\C F$ also will be
  classified by the number $m$ of interacting vortices in the
  diagrams, shown as superscript from the right: $^n\C F^m$. Thus, the
  lowest and next to the lowest diagrams for $^3\C F$ and $^4\C F$ are
  denoted as $\ ^3{\cal F}\Sp {I}$, $\ ^3 {\cal F}\Sp{III}$ and
  $\ ^4{\cal F}\Sp{II}$, $\ ^4{\cal F}\Sp{IV}$.  We will see that the
  numerical prefactors before the diagrams play a critical role in the
  {{triangular}}-resummation.

 \subsubsection{\label{sss:read}Rules for   reading diagrams}
 Rules for writing down the analytical expressions corresponding to
 specific diagrams are pretty universal across different diagrammatic
 techniques \cite{Lvov1995d}. We focus first on reading the diagram in
 the $({\B r}, t)$ representation. The rules
 are \\
 $\bullet$ {{A diagram is a set of lines connected by three-way
   junctions. Each junction represents an interaction amplitude $V$
   (solid dot in \Fig{FF:1basic})}.}  The wavy lines are the double
 correlators ${^2\cal F}$, while the wavy-straight lines represent the
 Green's
 functions ${\cal G}$.  \\
 $\bullet$  Each propagator is a function of two sets of arguments, say $\B r_1,t_1$ and   $\B r_2,t_2$, associated with its ends. In the stationary and space-homogeneous case, considered in this paper, the propagators depend only on differences of these arguments, e.g. ${\cal G}(\B r_1-\B r_2,t_1-t_2)$. 

 $\bullet$ Double correlator ${^2\cal F}(\B r_1-\B r_2,t_1-t_2)$ is an
 even function of its arguments. {{the Green function measures
     the response of the velocity field (denoted by a wavy line) to the
     forcing (denoted by a straight line). Therefore the Green
     function has the inherent time direction dictated by the
     causality principle.  The direction is from the forcing to the
     velocity field, or from the straight to wavy line. Consequently,
     in the Green function, ${\cal G}(\B r_1-\B r_2,t_1-t_2)=0$ if
     $t_2$ (associated with the forcing) is larger than $t_1$, the
     Green's function value is zero: ${\cal G}(\B r,t)=0$ if
     $t<0$. This a consequence of the causality principle: a response
     of the velocity $\delta v(t_1)$ to the force $\delta f (t_2)$
     must vanish if $t_2>t_1$.}}
 \\
 $\bullet$ Each vertex also has space-time arguments, say $\B
 r_n,t_n$, the same as the legs of three propagators, connected to
 it. In the diagram, one has to integrate over arguments $\B r_n,t_n$
 of all inner vertices.\\ $\bullet$ Since each {{vertex}} has
 its own time we can partition the diagram into time zones.  The
 boundaries of these time zones are denoted by dashed lines on the
 diagrams, as on \Fig{FF:21}. These time zones will play a significant
 role in calculating time integrals corresponding to each diagram, as
 discussed below in Section \ref{s:one-time}.\\ $\bullet$ In $(\B
 k,\omega)$-representation each propagator, say ${\cal G}(\B
 k,\omega)$ [Fourier image of ${\cal G}(\B r,t)$] has only one set of
 arguments, {{and}} each vertex involves delta-functions
  {{of}} the sum of $\B k_n, \omega_n$ arguments
 $(2\pi)^{(d+1)}\delta(\B k_1+\B k_2+\B
 k_3)\delta(\omega_1+\omega_2+\omega_3)$, where $d$ is the
 dimensionality of $\B k$-space.  Finally one has to integrate $\int
 d\omega_n/(2\pi)$ and $\int d\B k_n/(2\pi)^d$ for all intrinsic
 lines.

 We will use these rules to write down analytical expressions
 for all the diagrams we {{consider}} below.

 \subsubsection{Third order correlator  $\ ^3  {\C F}\Sp {I}$ and $\ ^3 \C F\Sp {III}$}
\begin{figure*}[h]
 	\includegraphics[width=\BiG\textwidth]{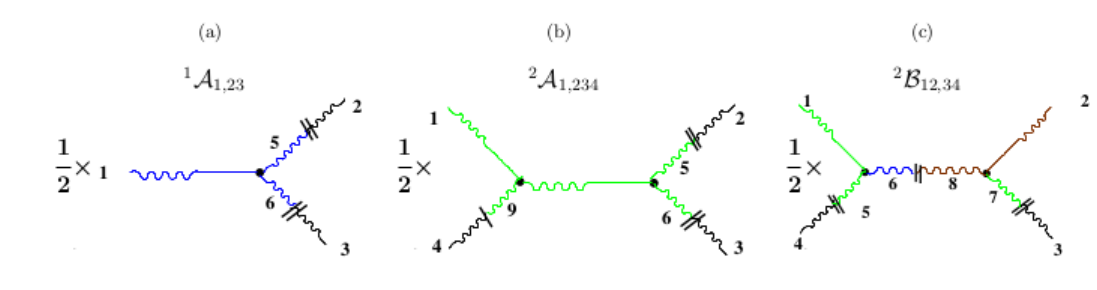} 
         \caption{
           Panel a: The lowest order contributions to the three-point correlator $^{3\!}\C F^{1A}_{123}$ and
          (Panel b) to the four-point correlator $^{3\!}\C
          F^{2A}_{1234}$ and $^{3\!}\C F^{2B}_{1234}$ as {a} result of
          gluing of three and four trees, separated by a double dash.
          All diagrams include prefactors.  }
	\label{FF:3}
      \end{figure*}

\paragraph{First  order diagrams for triple correlator  $^3\C F^1$.
\label{TripleCorrelatorFirstOrder}}
Its first representative  $\ ^3 {\cal F}_{123}^{1A}$ is shown in
Fig.\ref{FF:3}(a) as a diagram $\ ^{1\!}\C A_{1,23}$.  {From definition\,\eqref{F2} one gets}

 \begin{subequations}  \begin{align} \begin{split}  \label{F31}
 	(2\pi)^{d+1}\delta_{123}  \,{^3\! \C F}_{123}^{\rm 1A} &=    \underset{123}{\bf P}{ ^1\!\!\C A_{1,23}}\,,  \quad  
 	(2\pi)^{d+1}\delta_{123}  \,  {^1\!\!\C A_{1,23}}  =  \frac{1}{2}  \< a_1^{ 1 } a_2  a_3 \>\,, \quad \delta_{123}\= \delta(\q_1+\q_2+\q_3)\ .
 	\end{split}  \end{align}
     Here $\underset{123}{\bf P}$ is the permutation operator which,
     when acting on the function, produces a sum of all possible
     permutations of its {indices} divided by the number of all
     possible permutations of the indices. For example
   $ \underset{123}{\bf P}{\cal A}_{1,23} \equiv \dfrac{1}{3!}\left( 
 {\cal A}_{1,23}+{\cal A}_{1,32}+{\cal A}_{2,13}\right. 
 \left. + {\cal A}_{2,31}+{\cal A}_{3,12}+{\cal A}_{3,21}\right). $
  Substituting $^1\!a_1$ from \Eq{1A2} into $ {^1\!{\cal A}_{1,23}}$
  one gets $ {^1\!{\cal A}_{1,23}}=\dfrac1{4}\blue{\C G_1 V_{156}}\<
  \blue{ a_5 a_6 } a_2 a_3\> $.  
Hereafter we colored in \blue{blue} parts, originating from the tree
\blue{$^1\!a_\q$} in \Eq{itA}. We now are to average the resulting
expression using the pairing rule  
\Eq{eq:wick}   which corresponds to gluing together the trees of $^1\!a_\q$ and $
a_\q$. The result is pairs $\overbrace{\blue{\,a_5\,a_6}}$ and $\overbrace{\,
  a_2\,a_3}$ {that give} uncoupled contributions (each of them is
equal to zero). Two equivalent ways to pair
$\overbrace{\blue{\,a_5}\,a_2}$ and $\overbrace{\blue{\,a_5}\,a_3}$
\{denoted for the shortness as $\overbrace{\blue{5}\mbox{-}2}$ and
$\overbrace{\blue{5}\mbox{-}3}$, or even shorter as $
          [\overbrace{\blue{5}\mbox{-}(2},3)]$\} result in
   \begin{equation}\label{F31B}
 ^{1\!}\C A_{1,23}= \frac12\, \blue{ \C G_1 V_{123 }}\, \C {\cal F}_2 \C F_3\ .
 \end{equation} 
 Here $\q_1+\q_2+\q_3=0$,    $\C G_j\= \C G(\q_j)$, and  $\C {\cal F}_j\= \C F(\q_j)$, 
and 
\end{subequations}  
where {{the subscript with an overline denotes the negative of
    the corresponding wave vector,
    i.e. }}$\overline{j}=-\q_j$.
Graphically this result is shown in \Fig{FF:3}(a).  We preserve
notation $^n\!{\cal A}_{1,2\dots}$ for all diagrams of $n\sp {th}$
order in {{vertices}} $V$ with one leg denoting the Green
function $\C G_1$ and any number of wavy tails denoting ${\cal
  F}_j$. {{Here, and in the rest of the paper we separate by
    coma the indices in the correlators corresponding to the Greens
    functions from those corresponding to the double correlators.}}

 \paragraph{Third order diagrams for triple correlator $^3\C F^3$.} 
  In this section, we compute the three-point correlator in the third order in the interaction vertex. As we will show below this object plays a key role in the
 statistical properties of hydrodynamic turbulence. This
 object appears as a result of gluing together three trees and leads to the diagrams which are triangular in shape. 
To calculate  $^{3\!}
\C F^{3}$ we use   \eq{F2} and collect all terms   $\propto V^3$:
  \begin{align}\begin{split}\label{12}
  (2\pi)^{d+1} {\delta^{d+1}_{123}}\, ^3 \! \C F_{123}^{(3)}=&
   \underset{123}{\bf P}\Big [
   {^{3a}\!\!{\cal A}_{1,23}}+ {^{3b}\!\!{\cal A}_{1,23}} + ^3\!\!
   {\cal B}_{12,3} + ^3\!\! {\cal C}_{123} \Big ]\,,\quad {^{3a}\!\!{\cal
       A}_{1,23}}=  \< ^{3a}\!a_1 a_2 a_3 \> / 2\,,   \\  
   \ {^{3b}\!\!{\cal A}_{1,23}}= & \< \ ^{3b}\!a_1 a_2  a_3\> / 2 \,, \quad
    ^3\! {\cal B}_{12,3}=\< a_1^{2} a_2^{1} a_3 \>\,, \quad {^3 {\cal
       C}_{123}}=  {\< a_1^ 1 a_2^ 1 a_3^ 1 \>} / {3!}\ .
 \end{split}\end{align}

\begin{figure*}[t]
 \includegraphics[width=0.8\textwidth]{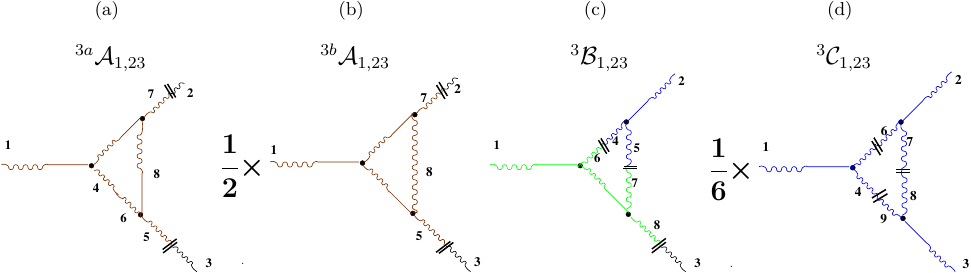}
\caption{\label{FF:4} {{Triangular}} diagrams for the next-lowest,
  third-order triple correlation function as a result of gluing three
  trees, separated by a double dash. Diagrams with one, two, and three
  Green's function in the legs are denoted as $\C A$, $\C B$, and $\C
  C$. All diagrams include prefactor according to $1/N$ symmetry rule.}
 \end{figure*}

  These terms are computed in the Appendix
  (\ref{ThreePointCorrelatorThirdOrder})
  and the  results are given by
  \begin{subequations}\label{16}
  \begin{align} 
      ^{3a}\! {\cal A}_{1,23}  &=   \C G_1   {\cal F}_2 {\cal F}_3 \int  \frac{d \q_4}{(2\pi)^{d+1}}
                                 V_{14{\overline{1+4}}}V_{{\overline{4}}(4-2)2} V_{(2-4)3(1+4)}   \C G_4^* \C G_{2-4} {\cal F}_{1+4}
                                 \, ,   \label{16a}\\  
      ^{3b}\! {\cal A}_{1,23}&=\frac 12  \C G_1  {\cal F}_2 {\cal F}_3
                               \int  \frac{d \q_4}{(2\pi)^{d+1}} \C G^*_4 {\cal F}_{2-4}\C G_{1+4} V_{14\overline{(1+4)}}
                               V_{{\overline{4}}(4-2)2} V_{(4+1)3(2-4)} \,, \label{16b}
      \\
      ^3  {\cal B}_{12,3}&=\C G_1 \C G_2   {\cal F}_3
                           {\displaystyle{\int}}\frac{d\q_4}{(2\pi)^{d+1}}{\cal F}_4 {\cal F}_{4-2} \C G_{1+4} 
                           V_{14\overline{(1+4)}}V_{2{\overline{4}(4-2)}}V_{(4+1)3(2-4)} \,,   \label{16c}\\
      {^3 {\cal C}_{123}}&=  \frac{1}{6}\C G_1 \C G_2 \C G_3
                           {\displaystyle{\int}} \frac{d\q_4}{(2\pi)^{d+1}}{\cal F}_4 {\cal
                           F}_{4-2} {\cal F}_{1+4}  V_{14\overline{(1+4)}} V_{2;{\overline{4}},4-2}V_{3;1+4,2-4} \ .\label{16d} 
                           \ . \end{align}
  \end{subequations}
            {{Recall that  the overline over the indices means that the negative
                of the corresponding wave vector, and the sums of indices imply
                the sum of corresponding wave numbers, i.e.
                $ V_{14\overline{(1+4)}}\equiv V_{ k_1, k_4, -k_1-k_4}$}}.
The corresponding diagrams are  shown in \Fig{FF:4}(a)-(d).

\subsubsection{Four Point Correlation Function}
\paragraph{Second  order diagrams for $ ^{4\!}\C F^{(2)}\propto V^2$} 
These diagrams are proportional to the product of two
{{vertices}} $V$.  They originates from ten terms, which we
divide into two groups $^{2\!}{\cal A}_{1,234}$ with one and two
$G$-tails involving Green's functions.  Hereafter we preserve the
notation $^n\!{\cal B}_{12, \dots}$ for all diagrams of $n\sp {th}$
order in $V$ with two leg $\C G_1$, $\C G_2$ and any number of wavy
tails denoting ${\cal F}_j$. {As before, we separate by coma
  the indices in the correlators corresponding to the Greens functions
  from those from the correlators. Resulting analytical expressions
  for the four point correlator are given by}
  \begin{align} \begin{split} \label{F4b}
 \hskip -.5cm  (2\pi)^{d+1} \delta_{1234} {^4\!\C F}_{1234}^{(2)}=  \underset{1234}{\bf P}\Big [{^2\!\! {\cal A}_{1,234} + ^2\!\!{\C B}}_{12,34} ]\,, \quad
 {^2\!{\cal A}_{1,234}}=  \<\,   ^2 a_1    a_2  a_3 a_4\>/ {3!}\,,   \quad
 {^2\!{\cal B}_{12,34}}= \<\, ^1 a_1\, ^1 a_2    a_3a_4 \> /4\ .
                \end{split} \end{align} 
              The required pairings  are presented  in the Appendix (\ref{FourPointCorrelatorLeading}).
              The results are  diagrams  in \Fig{FF:3}(b) and (c) with 
\begin{equation}
  \label{F4Ba}{^2\!{\cal A}_{1,234}}=
  \frac 12\,    \green{\C G_1 \C G_5 V_{14\overline 5} V_{523 }} {\cal F}_2 {\cal F}_3 {\cal F}_4    \,, \ \                                                          {^2\!{\cal B}_{12,34}} = \frac {1}{2} \, \blue{   \C G_1 \C G_2  {\cal F}_3 {\cal F}_4 {\cal F}_5
    V_{13\overline 5 }
    V_{235}}, \quad \q_5 = (\q_1 + \q_4)
  \ .  
\end{equation}
 These results will be 
used   to obtain a single-time version of the four-point correlator in the second order in the interaction vertex.

\begin{figure*}[t]
             \includegraphics[width=0.8\textwidth]{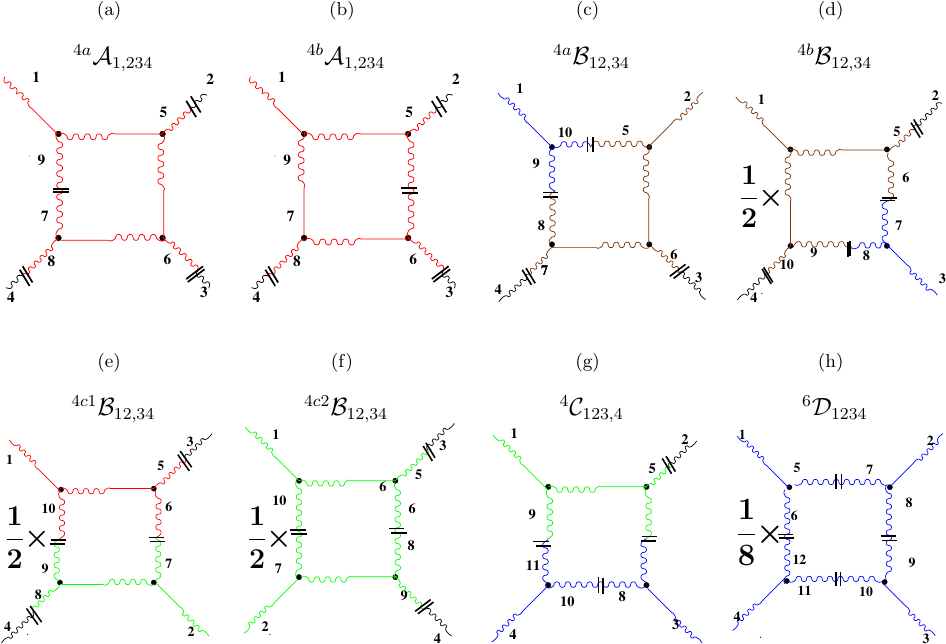} \caption{\label{FF:5}
               The  fourth-order ``square" diagrams for
               the quadruple correlation function
               $^{4\!}\C F^{(4)}_{1234}$ as result of gluing of four
               trees, separated by a double dash.  Corresponding
               analytical expressions are presented in \Eqs{iter}.
                              Diagrams with one,
               two, three, and four Green's functions in the legs are
               denoted as $\C A$, $\C B$, $\C C$ and $\C D$. All
               diagrams include the symmetry prefactor $1/N$.
             }	   \end{figure*}

\paragraph{ Fourth-order diagrams for
  $^{4 \!}\C F^{(4)}_{1234}\propto V^4$ } These diagrams have seven
types of terms:
\begin{align}\begin{split}
                 \label{4Ftot}
	(2\pi)^4 {\delta _{1234}}\, ^4 \! \C {\cal F}_{1234}^{(4)} =
            \underset{1234 }{\bf P}\Big \{ {^{4a}\! \!  {\cal
                A}_{1,234}} + {^{4b}\! \!  {\cal A}_{1,234}}  +
            {^{4a}\!  {\cal B}_{1234}} + {^{4b}\!  {\cal B}_{1234}} +
            {^{4c}\!  {\cal B}_{1234}}  + {^{4 } {\cal C}_{1234}}+
            {^{4} \!  {\cal D}_{1234}} \Big \}\,,   \\
		{^{4a} \! {\cal\red{A}}
		_{1,234}}=  \frac{\< \red{ ^ {4a} a_1  }    a_2   a_3 a_4 \>}{3!},  \    {^{4b} \! {\cal A}_{1, 234}}=  \frac{\<
		\red{ ^ {4b} a_1}  a_2 a_3 a_4 \>}{3!}\,, \ 
	{^{4a} \! {\cal B}_{12,34}}=\frac{\< \brown{ ^ {3a\!} a_1}\blue{^
			1\!a_2}   a_3   a_4 \>}{2},  \    {^{4b} \! {\cal B}_{12,34}}= \frac{\< \brown{^ {3b}\!a_1}\,\ \blue{^ 1\!a_2}  a_3 a_4 \>}{2},  \\  
	{^{4c} \! {\cal B}_{1234}}= \frac{\< \green{^
			{2}\!a_1\, ^ 2\!a_2}   a_3 a_4 \>}{4}, \quad  \ {^{4 } {\cal C}_{123,4}}=  \frac{\< \green{^ {2}\!a_1}\blue{\, ^ 1\!a_2 \, ^
  1\!a_3} a_4 \>}{2}, \ 
	{^{4 } \! {\cal D}_{1234}}=  \frac{\<\green{ ^
			{1}\!a_1\, ^ 1\!a_2 \, ^ 1\!a_3\, ^ 1a_4 } \>}{4!}\ .
	\end{split}\end{align} 
      The resulting diagrams and corresponding analytical expressions
      are computed in the Appendix
      \ref{FourPointCorrelatorFourthOrder}. The diagrams are shown
      in \Fig{FF:5} while the corresponding analytical expressions are
      given by \Eqs{iter}.

  	 \subsection{\label{ss:hight}Diagrammatic rules for plotting high order correlation functions\label{SectionAboutRules}}

         Examining diagrams in \Fig{FF:1} for the {{velocity}}
         field $a_\q$ we see that it is possible to write the $n$-th
         order diagrams for $a_\q$ without going through the
         cumbersome analytic substitutions, presented by \Eq{IT2}: the
         diagrams corresponding to the velocity field $a_\q$ are given
         by all topologically distinct binary trees with $n$ vertices,
         such that all the trunks are made of Green's functions and
         all the end branches are made of $a_\q$s.  Furthermore, every
         portion of the tree that continues in a symmetric fashion
         gets a factor of 1/2 due to the symmetry of the original
         equation of motion.  {{Therefore the}} overall numerical
         prefactor for a tree with $N$ elements in its symmetry group
         is  $1/N$.

         Examining \Figs{FF:3}, \ref{FF:4} and \ref{FF:5} for the
         diagrams for $^{3\!}\C F$ and $^{4\!}\C F$ we formulate the
         rules of the diagrammatic technique, which allows to skip the
         procedure of step-by-step derivation by gluing
         corresponding trees:  \\
         $\bullet$ Diagrams for the $n$-point, $m$-th order correlator
         $^{n\!}\C F^{(m)}$ 
         are all topologically different {graphs} with $m$ vertices and
         $n$ {external} wavy tails.  These wavy tails are either
         the Green functions $\C G$ or double correlations
         $\C F$. \\
         {$\bullet$ Each vertex in the diagram can be reached by the only way via $\C G$ from the outer leg of $\C G$.}\\
         $\bullet$ There are no loops made of the
         $\C G$-functions.\\
         $\bullet$ According to our $\dfrac 1N$-symmetry rule the
         prefactor for a diagram with $N$ elements in his symmetry
         group is $1/N$.

  In particular, diagrams without any symmetry (i.e. with only $N=1$
  identical element of symmetry), including diagrams in \Fig{FF:3}(b),
  \Figs{FF:4}(a) and (c) and \Figs{FF:5} (a,b,c,e, and g) have
  numerical prefactor equal to unity. Furthermore, the diagrams with
  non-trivial symmetry element (i.e. $N=2$) have prefactor $\frac12$,
  as in the diagram $^{1\!}{\cal A}_{1,23}$ with the symmetry $2
  \Leftrightarrow 3$ in \Fig{FF:3}(a), diagram $^{2\!}{\cal
    B}_{12,34}$ with the symmetry $1 \Leftrightarrow 2$ together with
  $3 \Leftrightarrow 4$ in \Fig{FF:3}(c), diagram $^{3b\!}{\cal
    A}_{1,23}$ with the symmetry $2 \Leftrightarrow 3$ in
  \Fig{FF:4}(b), \textit{etc}. The most symmetrical ones are the
  diagrams for $^{3\!}{\cal C}_{123}$ in \Fig{FF:4}(d), (symmetrical
  under permutations of all three arguments) with $P=3!=6$, which
  generates prefactor $\frac16$ and the diagram for $^{4\!}{\cal
    D}_{1234}$ in \Fig{FF:5}(h), which is symmetrical under reflection
  in four lines: horizontal, vertical, and $1-3$ and $2-4$ oblige
  lines and {{rotation by the angles}} $\phi=0, \dfrac\pi 2, \pi,
  \dfrac {3\pi}2$.  Thus for the $^{4\!}{\cal D}_{1234}$ diagram $N=8$
  and the $\frac1N$-prefactor is equal to $\dfrac 18$.

  Analyses of these diagrams and a wide set of additional diagrams not presented here demonstrate that the above-formulated diagrammatic rules work not only for the diagrams as a whole but also for their
fragments. So we expect that this is the general rule for diagrams for {\it
  all} orders and for all of {{diagram's}}  fragments. 
  
We think that this fact follows from the internal structure of the
presented perturbation theory, reflected in the topology of
diagrams. Bearing in mind that the question of numerical prefactor is
of principal importance, allowing
{{triangular}}-resummation of high-order diagrams and that its rigorous
mathematical proof is still absent we decided to check it
constructively for all diagrams, considered in this paper.

\section{\label{s:one-time}
  Simultaneous correlation functions}
In this section, we show how and why the simultaneous correlators can
be further {resummed} up to powers of the simultaneous triple
correlator $^3F$.  As a preliminary step, we introduce all required
simultaneous correlations in the $\B k$-space: $ F(\k)\= F_\k$,
$ ^{3\!}  F ({\k_1},{\k_2},{\k_3})\=\,^{3\!}{ F}_{123} $ and
$ ^{4\!} { F}({\k_1},{\k_2},{\k_3},{\k_4})\={ F}_{1234} $, where
$(2\pi)^{d }\delta( {\k _1}+{\k _2}) { F}(\k_1) \= \< a_{\k_1}
a_{\k_2} \>$,
$(2\pi )^{d }\delta( \k_1 + \k_2 + \k_3 )\ ^{3\!}  F_{123} \= {\langle
  a_{\k_1} a_{\k_2}a_{\k_3}\rangle}/{(3!)}$,
$(2\pi)^{d }\delta( {\k_1}+{\k_2}+{\k_3}+{\k_4})\ ^ {4\!} { F}_{1234}
\= {\langle a_{\k_1} a_{\k_2}a_{\k_3}a_{\k_4}\rangle}/{(4!)}$, {\it
  etc. }  The simultaneous correlation functions relate to
different-time correlators in $\q=(\k,\omega)$-representation as
follows:
                	\begin{subequations}\label{TO}
		\begin{eqnarray}\label{TOa}
                  && F( \k)=\int \frac{d \omega}{2\pi} \C F(\k,\omega)\,,\quad  \label{TOb}
                     ^{3\!}  F( \k_1,\k_2,\k_3)=\int \frac{d \omega_1\, d \omega_2\,d \omega_3}{{(2\pi)^d}}\   ^{3\!}\C  F( \q_1,\q_2,\q_3)
                     \delta  ( \omega_1+ \omega_2+ \omega_3 )  \,, \\  \label{TOc}
                  &&  ^{4\!}  F( \k_1,\k_2,\k_3,\k_4)=\int \frac{d \omega_1\, d \omega_2\,d \omega_3\,d \omega_4}{(2\pi)^{d+1}}  \delta  ( \omega_1+ \omega_2+ \omega_3 + \omega_4)  ^{4\!}\C  F( \q_1,\q_2,\q_3,\q_4)\ .
		\end{eqnarray}
              \end{subequations}
                Therefore to obtain a single-time correlator of any
              order the corresponding multiple-time correlator needs
              to be multiplied by the delta function of the sum of all
              the frequencies and then integrated overall
              frequencies.

	\subsection{\label{ss:1-pole} One-pole approximation}
        \rm For the actual calculation of integrals in \Eqs{TO} one
        needs to know the $\omega$-dependence of $\ G(\k,\omega)$ and
        $\C F(\k,\omega)$.  Therefore to proceed further we adopt the
        so-called {\it{one-pole}} approximation\,\cite{Lvov2000a} in
        which the $\omega$-dependence of the ``mass operator"
        $\Sigma_{\B k, \omega}$ in \Eq{GnB} for the Green function
        $G_\q$ is neglected. Similarly, we further neglect
        $\omega$-dependence of the mass operator
        $\Phi_{\k,\omega}\Rightarrow \Phi_\k $ in the Wyld's equation
        for $\C F_\q= |G_\q|^2 (\Phi_\q+D_\q) $ where su$D_\q$ is a
        correlator of the white noise.  Furthermore in the Dyson
        equation we neglect the double correlator of the white noise,
        since it is much smaller than $\Phi_\q$. As a result we have
        \begin{equation}\label{OnePole}
          \ G_{\k,\omega} = \frac{i}{\omega+ i\gamma_k}\, ,
\ \ \ 
\C F_{\k,\omega}=\frac{\Phi_\k}{\omega^2+\gamma_k^2}= \frac{2\, \gamma_k\,F_\k}{\omega^2+\gamma^2}= 2 \mbox{Re}  \{G_{\k,\omega}\}F_\k\ . 
	\end{equation} 
        Equation\,\eqref{OnePole} replaces $ \C F_{\k,\omega}$ by the
        sum of the Green function and its complex conjugate multiplied
        by $F_\k$. This replacement is a the crucial step that
        {{enables us to group diagrams in the triads that form
            the simultaneous triple correlator $^3F$.}}.  We denote
        $\~{\C G}_\q \=\C G_\q F_\k$ as an ``auxiliary Green's
        functions'', while the original Green function $ G_\q$ is
        called ``true'' Green's function. To distinguish between
        ``true'' and ``auxiliary'' Green's functions in the
        diagrammatic series, the auxiliary Green's functions will be
        distinguished by additional ``dash'' crossing it. {{
            The diagrams with the double correlator will be called ``parent'' diagrams.
            The diagrams which are generated by replacing the double
            correlator with the two auxiliary Green's functions will
            be called ``child diagrams'', or ``children'' for short.}}

        The diagrams with the loop along Green's functions with the
        same orientations give zero contribution to
        $^{n\!}  F_{\dots}$ due to the causality principle.  This is
        true regardless of whether the Green function is ``true'' or
        auxiliary. This property can be seen in $t$- or in
        $\omega$-representation.  It $t$-representation we should
        recognize that the wavy tail of each Green's function has time
        $t\sb w$, while the straight tail has time $t\sb s> t\sb w$.
        Otherwise, Green's function is zero due to the causality
        principle. Therefore wavy tail of Green's function in the next
        loop will have time $t\sb {w,n+1}$ even earlier than
        $t\sb {w,n}$. {{Such Green's functions will
            vanish. Consequently, the value of all loops with the same
            orientation of the Green's functions vanishes and thus the
            diagrams with loops in the Green's functions with the same
            orientation may be omitted from the very beginning.}}
          
        The same conclusion can be obtained in the $\omega$
        representation: similarly oriented number $n$ Green's function
        with frequencies $\omega_1$, $\omega_2=\omega_1+\delta_1$,
        $\omega_3=\omega_1+\delta_1+\delta_2$, etc. (here $\delta_n$
        is the ``incoming" frequency from the connected line in the
        $n\sp{th}$ vertex) are analytical in some
        $\omega_1$-half-plane, again, as a consequence of the
        causality principle. Therefore $\omega_1$-frequency integral
        in the loop indeed vanishes. Similarly, it is possible to show
        that diagrams, involving chain of similarly oriented Green's
        function connected {{to}} any of external tails do not
        contribute to the simultaneous correlators $^{n\!}  F_{\dots}$
        as a manifestation of the causality principle.  This statement
        is, again, true, regardless of whether the Green function is
        ``true'' or ``auxilarly''.

\subsection{\label{ss:Time} Time-zones and  interaction-time integrals in the   diagrams for $^3F_{123} $,    and $^4F_{1234} $} 
 
In this section, we consider the actual procedure of calculations of 
 integrals for interaction times, of the type presented in \Eqs{TO},  in the one-pole approximation. Below we begin with the simplest case of diagrams for $^3F^{(1)}$.

\subsubsection{\label{sss:3}  First-order triad  for  $ ^3F_{123}\Sp {\rm I}$}

 \begin{figure*}
	  \includegraphics[width= \BiG\textwidth]
   {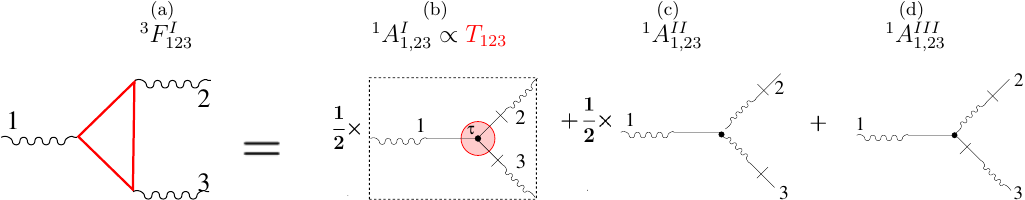} \\
   \includegraphics[width=\BiG\textwidth]{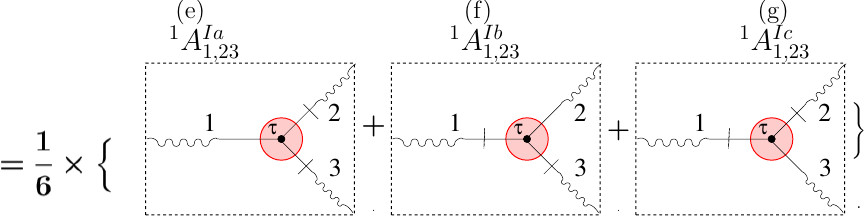}
   \caption{\label{FF:9} The lowest (first) order ``child" diagrams
     for the triple correlator
     $^{1\!}{\cal A}_{1,23}\propto ^3T_{123}$ originated from
     $^{1\!}{\cal A}_{1,23}$ shown \Fig{FF:3}(a).  Only the first of
     them, $^{1\!}  A _{1,23}\Sp {I}$, survives in simultaneous
     correlator $^{1\!}  A_{1,23}$, proportional to the frequency
     integral $I_1$, \Eq{34c}.  Panels (e),(f), and (g) with cycling
     relabeling of the legs 1,\ 2, and 3 present the diagram on {{panel}}
     (a).  
     The dashed line
     connecting the wavy legs of the  Green's functions  stands for the present time
     border: $t_1=t_2=t_3=0$. All times inside this region belong to
     the past, $t<0$. The filled red circle represents the time zone
     $\tau<0$ on this plot.  }
\end{figure*}

\rm After replacement~\eqref{OnePole} in \Eq{F31B} we obtain four
diagrams for $ ^1F_{123}^{(3)}$, shown in \Fig{FF:9}. Three of them,
shown in {{panels}} (b), (c), and (d) vanish after frequency integrations,
as required by \Eq{TOa}. A non-zero diagram in \Fig{FF:9}(a) under the
permutation operator $\underset{123}{\bf P}$ in \Eq{F31} can be
presented as the sum of three diagrams, shown in \Fig{FF:9}(e), (f)
and (g). The set of these three symmetric with respect to the
permutation of their legs diagrams, oriented inside with the straight
line will be referred to below as a ``triad". The simplest triad with
only one vertex inside, as shown in \Fig{FF:9} will be called a
first-order triad. As we see this is nothing but the diagrams for
$ ^1F_{123}^{(3)}$, shown in \Fig{FF:9}({a}) as thin red triangle.

Analitycally, diagrams for $ ^3 F_{123}^{I}$ are as follows:
\begin{equation} ^3F_{123}^{I}=\underset{123}{\bf P} ^{1\!} A_{1,23}, \
^{1\!} A_{1,23} = \frac12 T_{123} V_{123} \, F _2 F_3,\ 
\q_1+\q_2+\q_3=0\, , \label{TripleCorrelatorShort}\end{equation}  where we introduced the triad-interaction time":  \begin{subequations} \begin{eqnarray} \label{Tijk}
\label{34c}  T_{123}&=& \int
  d\omega_1 d\omega_2 d\omega_3 G_1 G_ 2 G_3 \delta( \omega_1 +\omega_2 +
   \omega_3) /{(2\pi)^d} \ . ~~~~~~~
   \end{eqnarray} 
  In the one-pole approximation\,\eqref{OnePole} this integral can be easily taken  to obtain
      \begin{equation}\label{Tijk2}
        T_{ijk}\=1/\Big(\gamma_i+\gamma_j+\gamma_k\Big)\,, \  \gamma_i\= \gamma(k_i)\,, \ \gamma_j\= \gamma(k_j)\,, \dots
\end{equation} \end{subequations}

{{Applying the $\underset{123}{\bf P}$ operation in \Eqs{TripleCorrelatorShort}
    and substituting \Eqs{Tijk} we obtain
    \begin{equation}
      ^3F_{123}^{I}=\frac
    {V_{123} \, F _2 F_3 + V_{231} \, F _1 F_3 + V_{321} \, F _2 F_3 }   {6\Big(\gamma_i+\gamma_j+\gamma_k\Big)} \ .
      \label{TripleCorrelatorFull}
    \end{equation}}}
{{
    It was shown in \cite{Lvov2002b} that the fractional dimension of
    $d=4/3$ the scaling  index of the inverse cascade of energy 
    $F_k\propto 1/k^2$ coincide with the scaling index of the
    thermodynamical equilibrium with the equipartition of enstrophy. 
      In this case the expression 
    in the numerator of RHS of \eqref{TripleCorrelatorFull} vanishes due 
    to the second Jacobi identity (\ref{jac}). Consequently, the value
of the simultaneous triple correlator in the first order vanishes for
the inverse cascade of the energy in the fractional dimension of
$d=4/3$. It was argued in \cite{Lvov2002b} that this is the reason why
the statistics of 2D turbulence is close the Gaussian.}}

We will see below that $\omega$-integral \eqref{34c} and much more
{{complicated}} $\omega$-integrals are much {{easier}}
to calculate in $\B k,t$-representation.  To translate any diagram to
the $({\B k},t)$ the representation we assign different times to the
beginning and to the end of each Green's function: $G_i \equiv G({\bf
k},\tau-t_j), \ \ j=1,2,3.$ In the one-pole approximation Green's
function \eqref{OnePole} has the form
\begin{eqnarray} G({\B k}, \tau) = \exp({\gamma_k \tau})\ \mbox{for} \
\tau<0\ \mbox{and zero otherwise}\ .
\label{GreenFunctionTimeRepresentation}
\end{eqnarray} the Green functions are equal to zero for positive
times as the future can not affect the present (the causality
principle).

Now consider the diagrams in \Fig{FF:9}(e,f,g). Since this is a
one-time correlator, the external ends should have an equal time
assigned to it. Let us assign the time to be zero, as in
$t_1=t_2=t_3=0$ and connect them by the dotted line ``present
time-border", which separates the future (outside of the diagram) and
past time-intervals, inside of the diagram. The time of the vertex,
$\tau$ belongs to the past and goes from $-\infty$ to zero. Now
integral \eqref{34c} in the $\B k,t$-representation can be written as
follows:
 \begin{equation} T_{123} =\int _{-\infty}^0 d \tau G({\B
k_1},\tau)G({\B k_2},\tau) G({\B k_3},\tau) = \int _{-\infty}^0 \exp{[
(\gamma_1+\gamma_2+\gamma_3 ) \tau ] }d\tau=1\big
/\big({\gamma_1+\gamma_2+\gamma_3}\big). \label{I1alternative}\end{equation}
 The answers (\ref{Tijk}) and (\ref{I1alternative}) for the triple
interaction time are equivalent. Naturally, the answer is independent
on whether it is obtained in $t$ or $\omega$ representaton.

 \subsubsection{\label{SSS:3} Third-order triad for $ ^3F_{123}\Sp
{\rm III}$}

\begin{figure*} 
 \includegraphics[width=\BiG\textwidth]{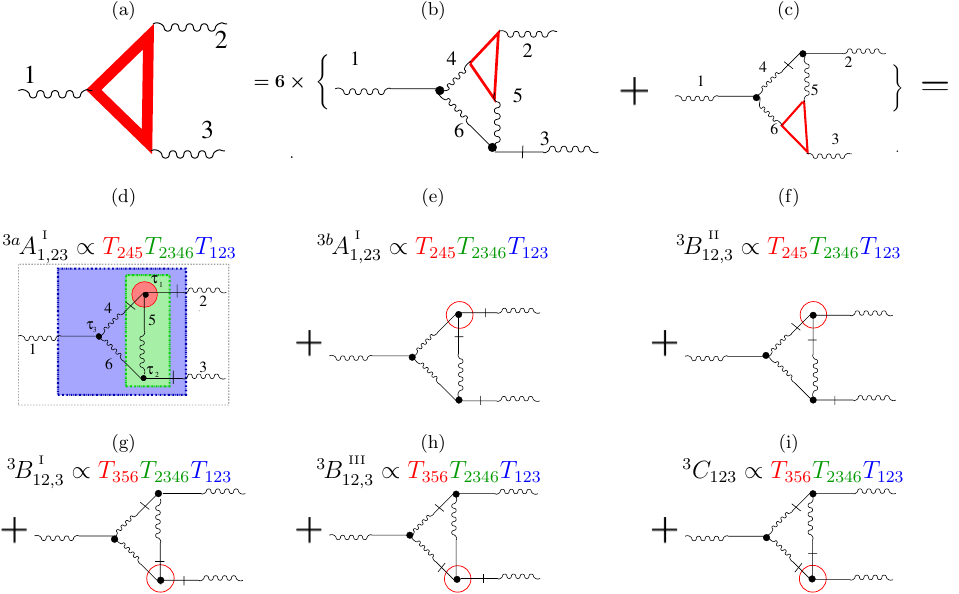} 
 \caption{\label{FF:11} The next-lowest ($3\sp{rd}$) order ``child"
   diagrams for the simultaneous triple correlator $^{3\!}
   F_{123}\Sp{III}$, denoted in panel (a) as a thick red
   triangle. Panels (b) and (c) show its representation via $^{3\!}
   F_{123}\Sp{ I}$, denoted as thin red triangles, while the last two
   lines show original (not summed {{yet}} ) diagrams for $^{3\!}
   F_{123}\Sp{III}$.  Panel (b) is the sum of panels (d,e,f) while
   panel (c) is the sum of panels (g,h,i).  Three chronologically
   nested ({{red}}-green-blue) {{time zones}} are shown in
   panel (d) which produce the product of three interaction times:
   factor \green{$T_{2346}$} appears from the integration of $G_2 G_3
   G_4 G_6$ over $\tau_2$, etc.  These notation are used on all
   subsequent Figures.  } \end{figure*}

 To calculate the third-order triple correlator $^3F\Sp {III}$ we take
(\ref{OnePole}) and substitute it into \Eqs{16}. Graphically, this
corresponds to replacing all three double correlators of $ ^{3a}\!
{\cal A}_{1,23} $ by the pair of auxiliary Green's functions, run in
either direction (i.e. eight possibilities). The
 six of {{the total of }} $4\cdot 8=32$ possibilities give
 nonzero contributions for simultaneous correlation functions.  The
 resulting six diagrams are shown in \Figs{FF:11} and denoted as
 $ ^{3a}\! { A}_{1,23}\Sp {I} $, $^{3b}\!  { A}_{1,23}\Sp {I} $,
 ${ B}_{12,3}\Sp {I} $, ${ B}_{12,3}\Sp{II} $,
 ${ B}_{12,3}\Sp {III} $, and ${ C}_{1,23}\Sp I $.

\begin{figure*}
 	\begin{center}
 		\includegraphics[width=\BiG\textwidth]{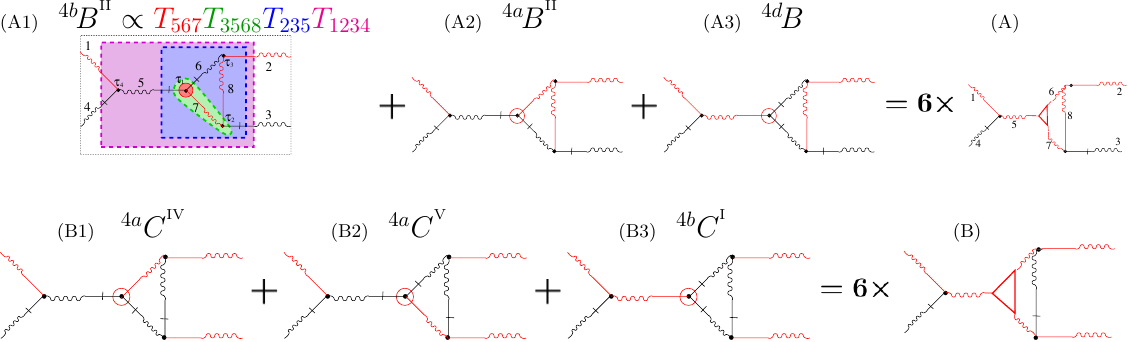}\\
   \includegraphics[width=\BiG\textwidth]{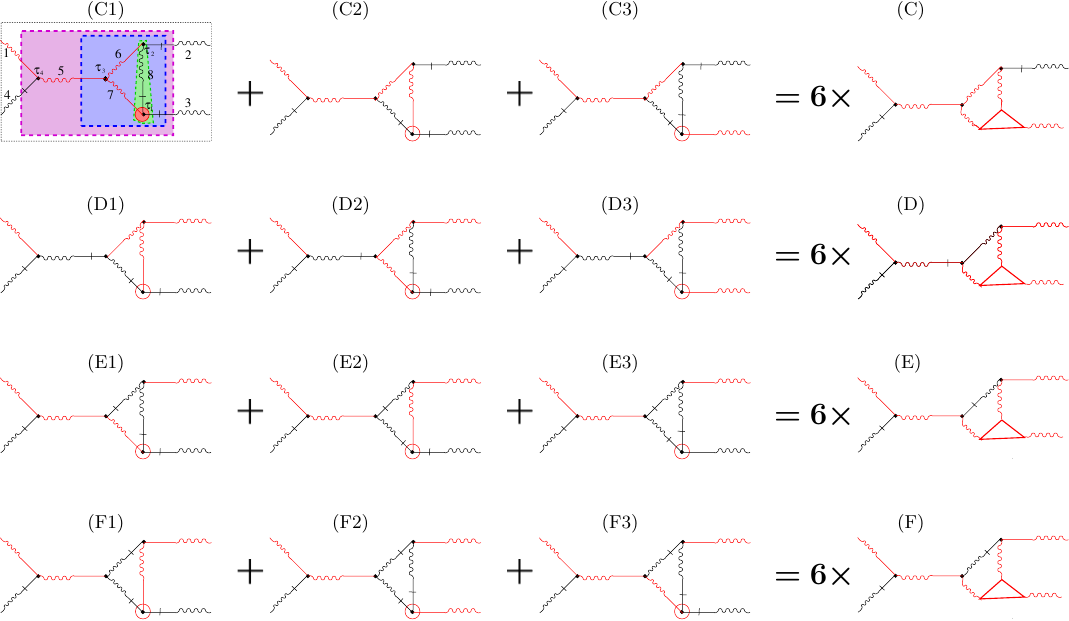}
 			\end{center}
                        \caption{\label{FF:21} {First group of the
                            skeleton diagrams with the earliest time zones
                            in the (678)-triangle, denoted by red
                            circle. Similar to Figure (\ref{FF:11})
                            the time zones are colored as
                            red-green-blue-magenta from the earliest
                            to the latest time zones. The time factors
                            are colored accordingly to the time zones
                            they originate from.  As in the previous
                            \Figs{FF:9}, \ref{FF:11}, and \ref{f:9}
                            interaction times $T$ appear from the
                            integrations of the product of the Green
                            functions, entering the appropriate time
                            zone by straight lines, over the time
                            $\tau_j$ of their vertex.  Therefore there
                            are as many time zones as there are
                            vertices.  These time factors are colored
                            accordingly to the time zones they
                            originate from.  }
 }
 \end{figure*}

 As explained above, Green's functions have inherent time direction
 {{in corresponding diagrams}}: time flows in the direction from
 a wavy to a straight line.
 Therefore the beginning of the Green
 function has an earlier time than the end of the Green function
 that enters the vertex. To calculate one-time correlators, we replace
 the double correlator with the sum of the two auxiliary Green functions
 oriented in the opposite directions. These Green's functions,
 therefore, partition the diagram for a multi-point correlator in the
 distinct time zones. Making an arbitrary choice that the external
 legs of the diagram correspond to time $t=0$ we have earlier times
 inside the diagram. In fact, we have telescopically nested time zones
 that flow from the earliest time zone to the present time zone. In
 our diagrams, we color the earliest time zone as red, the {{later}} as
 green, and even the {{later}}  as blue. We color the latest time zone, if
 present, as magenta.  The number of nested time zones is equal to the
 number of interaction {{vertices}}.  In some diagrams, the ordering of
 the zones is not uniquely defined by Green's functions. For such
 diagrams, as explained below and in figure captions, all possible
 ordering of time zones must be taken into account in calculating
 interaction-time integrals for the simultaneous correlators.
 
 As before we connect all three external wavy {{legs}} of the
 Green's functions by the black dotted line, denoting the present
 time-border $t_1=t_2=t_3=0$.  The times of three {{vertices are }}
 denoted as $\tau_1$, $\tau_2$ and $\tau_3$ in
 \Fig{FF:11}(a).{{Each of these times belong to }} a particular
 time-zone, colored in red, green, and blue.

 {According to the causality principle,} all of these time zones belong
 to the past: $\tau_1<0$, $\tau_2<0$, and $\tau_3<0$. The present time
 depends only on the past time, and does not depend on the future.
 Green's functions  $G(\B k_4,\tau_{21})$,
 $G(\B k_5,\tau_{23})$, and $G(\B k_6,\tau_{31})$ (hereafter
 $\tau_{ij}\= \tau_i-\tau_j$) {{in this diagram}} prescribe
 chronological order of the time zones: $\tau_2<\tau_3<\tau_1<0$.

Armed with this arrangement we can easily compute time integral in
\Fig{FF:11}(A1)
{{
    $$I\Sb{A1}= \int\limits _{-\infty}^0 \!\! \!d\tau_1 G(\B k_1,\tau_1) 
    \times\int\limits _{-\infty}^{\tau_1} \!\!\! d\tau_3 G(\B
    k_3,\tau_3) \int \limits _{-\infty }^{\tau_3} d\tau_2 G(\B
    k_2,\tau_2)G(\B k_4,\tau_2-\tau_1)G(\B k_5,\tau_2-\tau_3)G(\B
    k_6,\tau_3-\tau_1)$$ }} written in the $t$-representation for the
  diagram \Fig{FF:21}(A1).  Most Green's functions, except for
  $G(\B k_1,\tau_1)$, cross borders between time zones such that their
  fragments belong to different zones.  Using the decomposition rule
  $G(\tau)=G(\tau-\tau')G(\tau')$ for the Green function in the
  one-pole approximation, \Eq{GreenFunctionTimeRepresentation}, we can
  present {{these Green's functions}} as the product of the
  Green's functions such that each of them belongs to the one-time
  zone {{only}}. Namely:
  $G(\B k_2,\tau_2)= G(\B k_2,\tau_{23})G(\B k_2,\tau_{31})G(\B
  k_2,\tau_1)$,
  $G(\B k_3,\tau_3)= G(\B k_3,\tau_{31})G(\B k_3,\tau_1)$, and
  $G(\B k_4,\tau_{21})= G(\B k_4,\tau_{23})G(\B k_2,\tau_{31}) $.  Now
  interaction-time integral $I$ can be factorized as follows:
  $I =\, ^3T_{123}\, ^3T_{245}\, ^4T_{2346}$, where
  \begin{subequations}\label{time-int}
  \begin{align}\begin{split} \label{22a}
                 T_{123} =\int\limits _{-\infty}^0 \!\! \!d\tau_1  G(\B k_1,\tau_1)  G(\B k_2,\tau_1) G(\B k_3,\tau_1) \,, \quad   \, ^3T_{123} =\int\limits _{-\infty}^0 \!\! \!d\tau_{23}  G(\B k_2,\tau_{23})  G(\B k_4,\tau_{23}) G(\B k_5,\tau_{23})\end{split}\end{align}
             are the triad interaction times, defined by \Eq{I1alternative}.
             We   introduce the quadric interaction time $   T_{ijkl}$ 
\begin{equation}\label{4T} 
    T_{ijkl}= \int \limits _{-\infty}^0 d\tau\  G(\B k_i,\tau)  G(\B k_j,\tau) G(\B k_k,\tau) G(\B k_l,\tau)     = 1\big / \big [\gamma_i +\gamma_j +\gamma_k +\gamma_l \big]\ .
  \end{equation}\end{subequations} 
In our case, $^4T_{2346}$ originates from the $\tau_{31}$-integration
over the earlier time-border of the intermediate time-interval (filled
in \Fig{FF:11}a with green) with four Green's functions directed inside
of it. The oldest time interval with three incoming Green's function
$G(\B k_2)$, $G(\B k_4)$, and $G(\B k_5)$ {{produces}} $^3T_{245}$, while
the earliest time interval gives $ T_{123}$ with wave-vectors of the
external legs.
Clearly, time integrals depend only on the diagram
topology and are independent of the particular type of the Green
function: true or auxiliary. Therefore time integrals are the same for
the diagram in \Fig{FF:11}(b) and many others in \Fig{FF:11}.
Corresponding full analytical expressions for these diagrams can be
found in Appendix \ref{SquareDiagramsFourPointCOrrelator}, see
{\Eqs{36}, \eq{BTriads},\eq{Ctriads}.

\subsubsection{\label{ssss:3}  Second-order diagrams  for  $ ^4F_{123}\Sp {\rm II}$}

   \begin{figure*}
 	\begin{center}
 		\includegraphics[width=\BiG\textwidth]{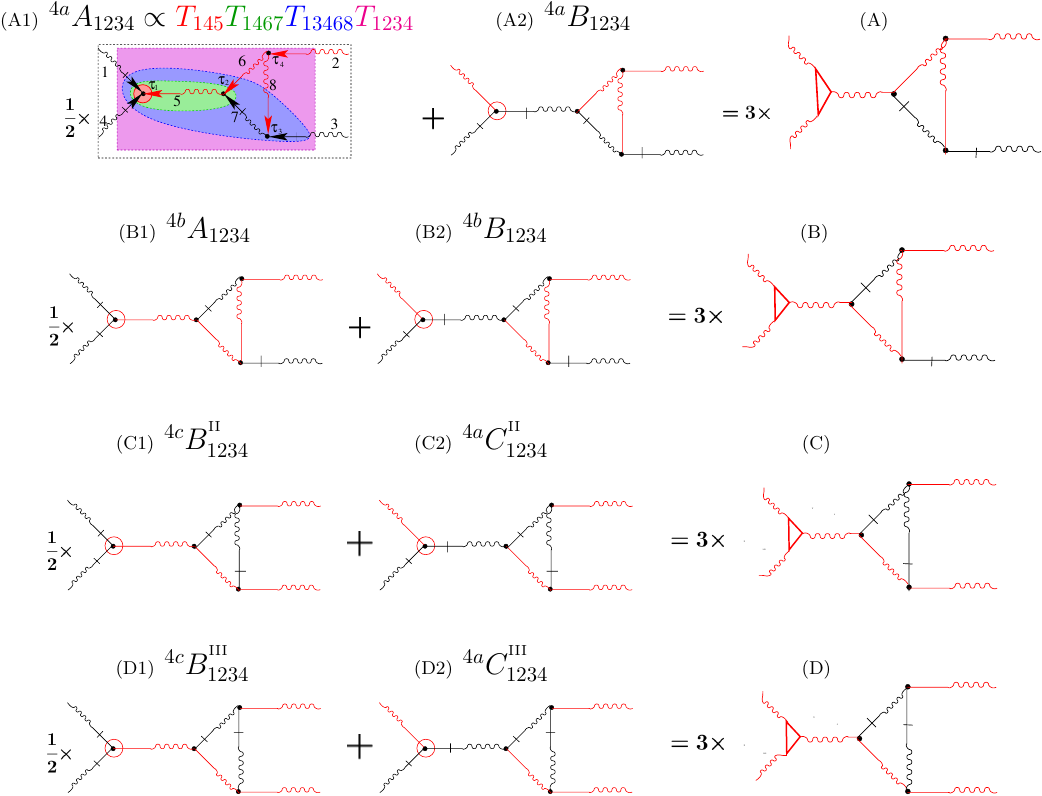}
 		 	\end{center}
                        \caption{\label{FF:20}Second group of the
                          {{triangular}} diagrams with earliest
                          (145)-time zone, denoted as a red circle.
                          Notation, color codes of the time zones,
                          and of interaction times are the same as in
                          previous \Figs{FF:11} 
                          and           \ref{FF:21}.  }
\end{figure*}

\begin{figure*} \begin{center}  
	 \includegraphics[width=0.8\textwidth]{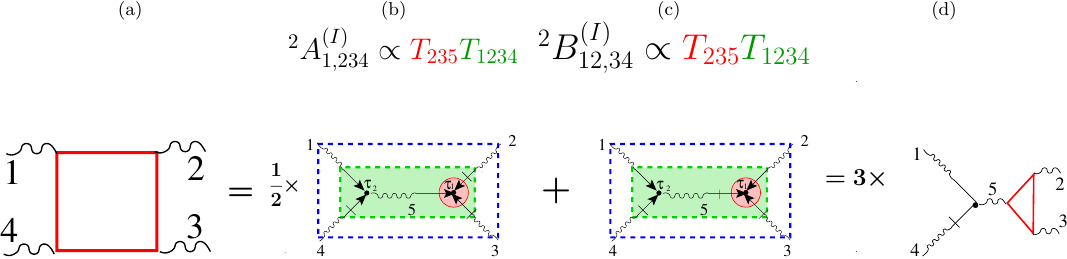}
 \end{center}
 \caption{ \label{f:9} Diagrams for the quadruple correlator
   $^{4\!}\C F_{1234}\Sp{II}$, denoted as a thin red square in panel
   (a) and expressed via ``child" diagrams in panels (b) and (c),
   {{contributing to the simultaneous correlators}}. As in
   \Figs{FF:9} and \ref{FF:11} interaction times $T$ {{are}}
   the same in panels (b) and (c). Diagrams in panels    {{(b) and (c)}} can be summarized in the diagram in the
   panel (d), which involves triple correlator
   $^{3\!}\C F_{235}\Sp{I}$, shown as a thin red triangle.  }
          \end{figure*}

\begin{figure*}
	\begin{center}
		\includegraphics[width=\BiG\textwidth]{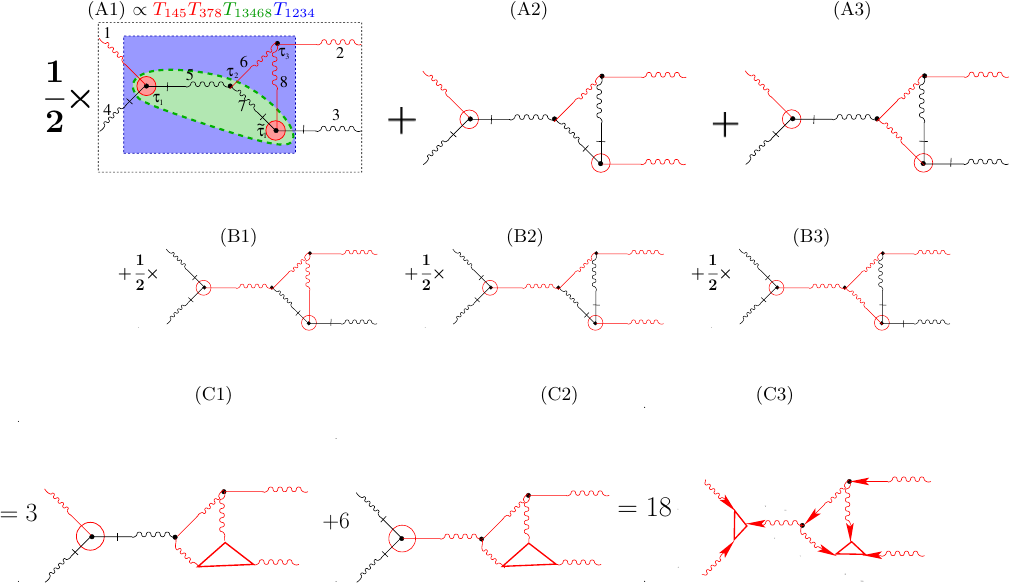}
			\end{center}
                        \caption{\label{FF:23} Third group of the
                          diagrams with two earliest time
                          zones. Notation, color codes of the time
                          zones, and of interaction times are the same
                          as in previous figures. New element here is
                          that there are two earliest time zones,
                          plotted as red circles. Consequently, there
                          are two red zones, one (intermediate) green
                          zone, and a later blue zone.  The times of
                          the earliest zones are $\tau_1$ and
                          $\tilde{\tau_1}$. The relationship between
                          $\tau_1$ and $\tilde{\tau_1}$ is not fixed
                          and we have two independent integrations
                          over $\tau_1$ and $\tilde{\tau_1}$ producing
                          the product \red{$T_{145}T_{378}$.}  }
\end{figure*}

  \begin{figure*} 
	\includegraphics[width=\BiG \textwidth]{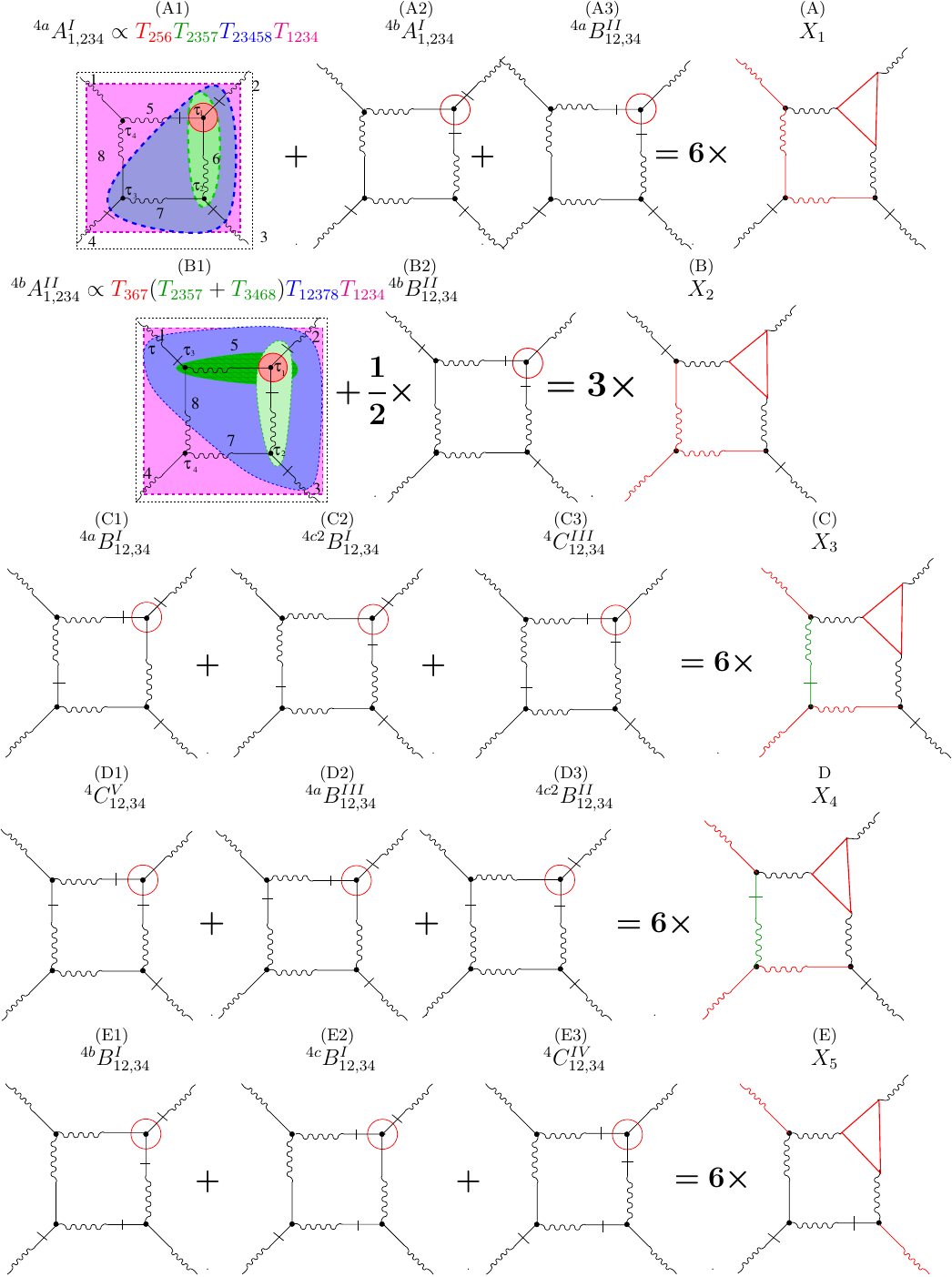}  
	\caption{\label{FF:17} Five  subgroups of the square diagrams for $^{4\! }  F_{1234}^{(4)}$ with cancellation in each line. 
	 Notation, color codes of the time zones, and  of interaction times
	 are the same as in previous figures. The new element appears in panel (B1):
	 the relationship between $\tau_2$ and $\tau_3$ is not
	 dictated by the orientation of the Green functions. Consequently,
	 integrations over {{$\tau_2$}} and $\tau_3$ are performed differently in
	 these two cases. If $\tau_2<\tau_3$ then the $\tau_2$ integration 
	 occurs over the region colored in light green, while
	 $\tau_2>\tau_3$ region is colored in darker green.
	 Such a partition of integration regions leads  to the sum of two
	 contributions \green{$(T_{2357}+T_{3468})$} originated from these
	 two green zones. 
	}
	\end{figure*}
\begin{figure*}
  \includegraphics[width=0.8\textwidth]{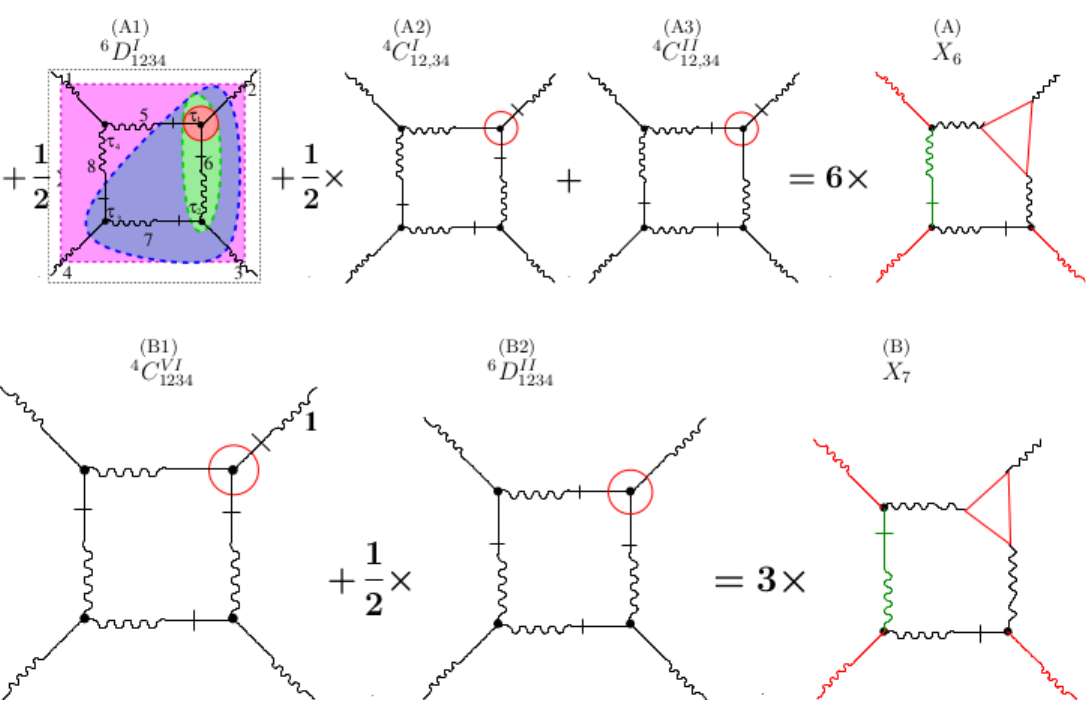}
	\caption{Next two subgroups of the square diagrams for $^{4\! }  F_{1234}^{(4)}$ which sums up to the triple correlator shown on the right of each line  as a red triangle.} 
	\label{FF:18} 
\end{figure*}

 \begin{figure*}
	\begin{center}
		\includegraphics[width= 0.7\textwidth]{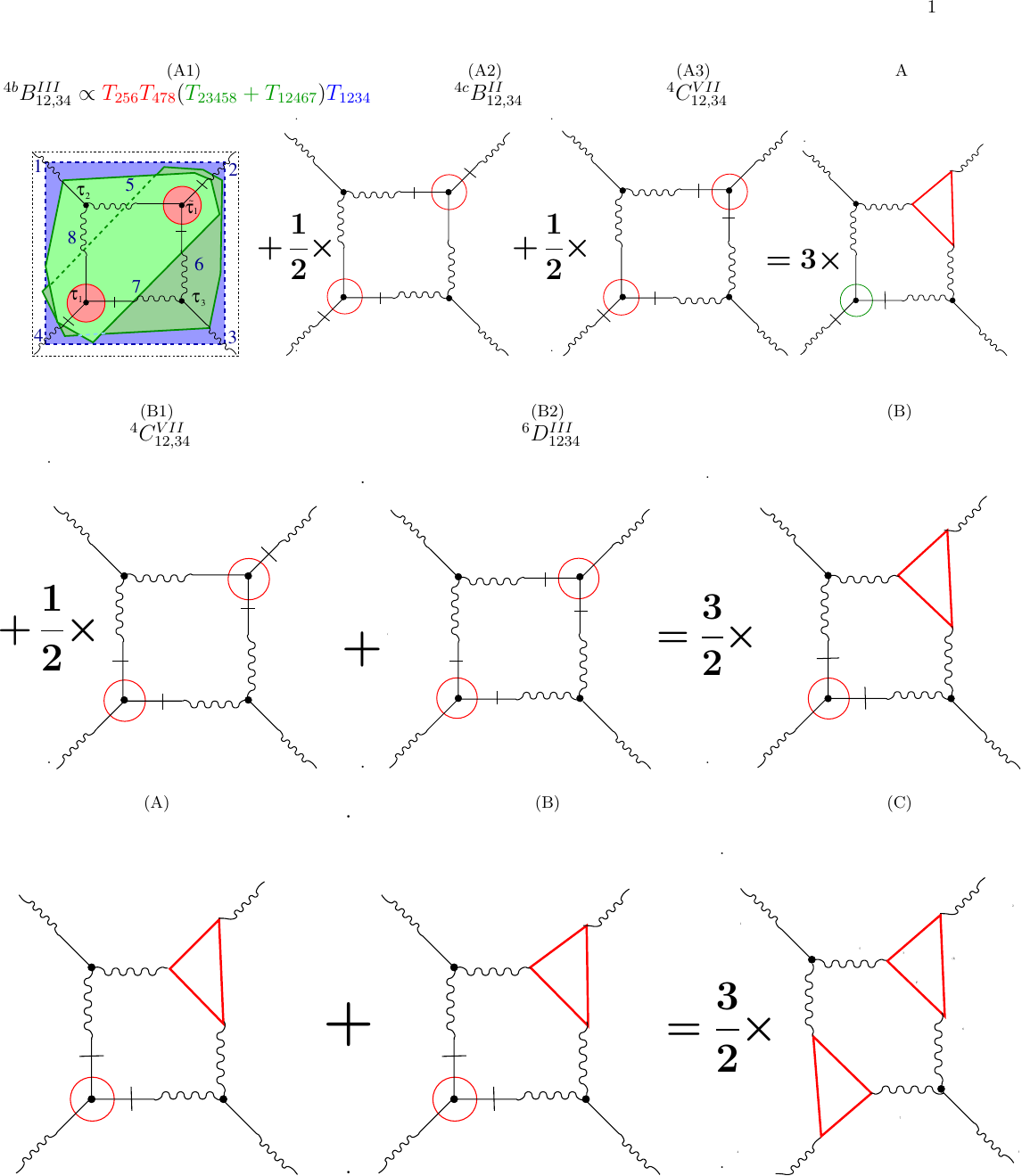}
			\end{center}
                        \caption{\label{FF:19}Last group of the square
                          diagrams for $^{4\! }  F_{1234}^{(4)}$ {with
                            two earliest time zones producing the
                            product {$T_{256}T_{478}$}.  As in
                            \Fig{FF:17}(B1) the relationship between
                            $\tau_2$ and $\tau_3$ resulting in the sum
                            of two contributions
                            \green{$(T_{23458}+T_{12467})$} originated
                            from two overlapping green zones.  }}
	 \end{figure*}
We now consider diagrams for $ ^4F_{1234}\Sp{II}$  shown in
  \Figs{f:9} panels (a) amnd (b). There are two vertices with times $\tau_1$, belonging
  to the earliest red-filled time zone, and $\tau_2<\tau_1$ in the green
  zone. Time integration over the three Green's functions $ G_2$,
  $ G_3$, and $ G_5$, entering red zone leads to a factor $T_{235}$,
  while  four Green's functions $ G_1$, $ G_2$, $ G_3$, and $ G_4$,
  entering green zone {{produces}} $T_{1234}$.

  Diagrams for the fourth-order contributions to $ ^4F_{1234}\Sp{IV}$
  are more complicated. They include four vertices and require four
  integrations over $\tau_1$, $ \tau_2$, $ \tau_3$, and $ \tau_4$
  with, generally speaking, more complicated topology of the time
  zones, not necessarily chronologically nested.  Therefore before
  presenting analytical expressions for their interaction times we
  present in the forthcoming \Sec{ss:3} the diagrammatic rules on how
  to reconstruct these expressions from the topology of the diagrams
  without their explicit calculations.

 \subsection{\label{ss:3} {Diagrammatic rules for the reconstruction of interaction times from the topology of the time-zones }}

 Finding {{appropriate}}  ime zones that allow factorizing time
 integrals for the interaction times in diagrams for $^3F_{123}$ and
 $^4F_{123}\Sp{II}$ described above together with more complicated
 situations in numerous diagrams for $^4F_{123}\Sp{II}$, we came up
 with a set of rules of how to avoid explicit integrations over
 $\tau_1$, $ \tau_2$, $ \tau_3$, and $ \tau_4$ in diagrams for
 $^4F_{1234}\Sp{IV}$.  {Diagrams for $^4F_{1234}$ of any {{order}}, as
   well as the diagrams for higher-order correlation function, can be
   divided into two major groups: weakly connected diagrams like those
   shown in 
{Figs. (\ref{f:9}-\ref{FF:23})}, and
compact diagrams in 
{Figs. (\ref{FF:17}-\ref{FF:19})}.
   Unlike
   compact diagrams, weakly connected diagrams can be divided into two
   parts by cutting just one line.  In the diagrammatic rules,
   formulated below, we will show how one can find all-time integrals
   in presented here two groups of four order diagrams} in particular
 and in even higher order diagrams, in general, just by simple
 analysis of their topological structure.  The rules are as follows:
\begin{enumerate}
   \item Partition the diagram to telescopically nested time zones as dictated by the 
     Green's functions.
     
    {Each time zone has its ``own" vertex inside it characterized by
      the vertex time $\tau_j$. Therefore} the number of time zones is
    equal to the number of interactive {{vertices}}. Resulting time
    integrals will be a product of distinct time factors corresponding
    to each zone. Each time factor is an interaction time of the wave
    numbers of Green's functions entering the zone.

  \item In most cases, like in some diagrams on Figs.
    \ref{FF:21}- \ref{FF:18}, times
    of the vertices are fully ordered:
    $\tau_1<\tau_2<\tau_3 < \tau_4 <0$. In these cases, time zones are
    uniquely chronologically nested, having the earliest (red filled
    in our diagrams) time zone with $\tau_1$, early (green colored)
    zone with $\tau_2>\tau_1$, and recent (blue colored) zone with
    $\tau_3$ and, finally the very recent (magenta-colored) zone with
    $\tau_4$.  According to rule 1) integration over these times gives
    the product of four interaction times, in a particular case of
    \Fig{FF:20}(A1) this is
    $\red{T_{145}}\green{T_{1467}}\blue{T_{13468}}\magenta{T_{1234}}$. Here
    for concreteness, we colored interaction time according to the
    color of the corresponding time zone.
   
\item In all the cases considered above and in general, the earliest
  (red) zone {{always}} has  three incoming Green's functions, producing
  triple interaction time, e.g. $\red{T_{145}}$, in
  \Fig{FF:20}(A1). The very recent zone in the triple correlator
  $^3 F_{123}$ produces $T_{123}$ {{and}} in the quadruple correlator
  $^4 F_{1234}$ produces $T_{1234}$. Therefore the most recent
  interaction-time is of the same order as the correlator generating
  it with the same wave-vector arguments. This statement is true for
  any-order correlations.
   
\item it may happen that a number of time zones have the relationship
  between times that are not uniquely determined by the Green
  functions. Then there are two possibilities
  \begin{enumerate}
  \item Some diagrams may have two and more earliest time zone, like
    {{those}} in \Figs{FF:23}(A1) and in \Fig{FF:19}(A1) with two earliest time
    zones with $\tau_1$ and $\widetilde \tau_1$. In these cases,
    time-integral over all vertex times factorizes with the product of
    integrals over $\tau_1$ and over $\widetilde \tau_1$ from minus
    infinity to zero, producing the product of two (or as many as the
    number of the earliest time-zones) triple interaction times.

    \item If Green's functions do not uniquely define the ordering of
      times corresponding to the {{vertices}}, then the time zones are to
      be drawn separately for each possible ordering of interaction
      times.  For such a case the resulting analytical expression
      contains the {\it sum} of corresponding interaction times, see
      e.g. diagrams in \Fig{FF:17}(B1) and \Fig{FF:19}(A1).  Note that
      such branching of regions of integrations may happen at any
      level except the earliest and the very recent time zones.
    \end{enumerate} 
\end{enumerate}

To summarize, the } time integral for the diagram with $n$ vertices
{{equals}}  to the product of $n$ interaction times, corresponding to all
uniquely defined time zones Green's functions entering them. If some
zones are only partially overlapping the frequency integral includes
the sum of their interaction times.

\section{ \label{s:LO} Triangular-resummation of the   triple-line reducible triads }

\subsection{ \label{ss:triple}
  {{Triangular }}resummation of diagrams for triple correlators
  $^3F_{123}$}
In this subsection, we will bring all the things we have considered
together and introduce triangular  resummation of the
triple correlator, {{the latter being}} the main focus of this
{{work}}.  It has three appearances, shown on \Fig{FF:1basic}:
empty thin red triangle for first order contribution $^3F\Sp{ I}$,
empty thick red triangle for third order contribution $^3F\Sp {III}$
and filled thick red triangle for the full correlator $^3F$.
 
   One of the main points of this paper may be recognized by comparing
   diagrams for $^3F\Sp {I}$ on \Figs{FF:9} and for $^3F\Sp {III}$ on
   \Figs{FF:11}.  Note that our definition of the correlator involves
   the permutation operator $\bf P_{123}$, defined by \Eqs{F31}
   {{and}} \Eq{12}.  First notice that the first and second lines
   of the \Figs{FF:11} contain vertex marked by a red circle, and the
   Green's functions connected to this vertex form precisely the
   diagrams of the $^3F\Sp {I}$ shown at \Figs{FF:9}. {{As we will
       see below, see \Eq{28}, this fact}} allows us to write the compact expression
   for $^3F\Sp {III}$.

   \begin{align}\label{3F3}\begin{split}
                             \ ^{3\!}  F_{123}\Sp{III}=&  {\bf P}\int \frac{d \B k_4}{{(2\pi)^d}}\, ^4T_{2346} V_{146} \times             \big[ V_{425}F_2+V_{245} F_4\big ] \, ^{3\!}F_{356}\Sp{ I}\,, \quad 
                                                         \B k_5=\B k_4+ \B k_2 
                                                         {\rm \ and \ } \B k_6=\B k_4- \B k_1\,,
                         \end{split}	\end{align} 
                       expressed in the terms of the first-order
                       correlator as follows from first two diagrams
                       in the last line in \Fig{FF:11}. This fact has
                       deep consequences as we will see below.

                       The permutation operator $\B P_{123}$ acting on
                       the two diagrams of \Fig{FF:11} produces six
                       diagrams. {{These}} diagrams can be
                       grouped into a triad of diagrams that involves
                       three vertices, have rotational $C_3$ symmetry,
                       and three tails of Green's functions
                        chronologically ordered
                       inside  the triangle from the
                       present time $t=0$ in the simultaneous
                       correlator back to all past times $t_j< 0$, as
                       required by casualty principle.  We will refer
                       to this object as a ``third-order triad" and
                       depict it by the thick red triangle.

                       Notice that the bare Green's and bare double
                       correlation in the Dyson-Wyld line resummation
                       are called reducible fragments, which can be
                       separated from the body of a diagram by cutting
                       two lines. {{Bearing}} this in mind we
                       can clarify them as ``double-line" reducible
                       diagrams.  Such a name immediately suggests the
                       existence of ``triple line reducible diagrams".
                       Indeed, our triad diagrams can be separated
                       from the body of a diagram by cutting three
                       Green's functions entering this time zone.  We
                       call this object ``triple-line reducible
                       triads" (of $C_3$ symmetrical groups of
                       diagrams). Up to now, we met in \Fig{FF:9}
                       first-order triple-line reducible triads (with
                       one vertex) and in \Figs{FF:11} third-order
                       triple-line reducible triads with three
                       {{vertices}}, shown, e.g. in
                       \Figs{FF:11}(a) by a blue square.

Let us again examine {{panel}} (a) on \Fig{FF:11}. The red oldest time
zone has three incoming Green's functions entering it with straight
lines. The red time zone is in turn inside the earlier, blue time zone
which has three straight line entering it. This is an example of the
 triangular telescopically nested time-ordered reducible
triads.
Higher-order diagrams for the simultaneous triple
{{correlators}}  will have multiple zones
nested in {{similar}} manner. These zones will
sum up the fully dressed triple correlator.

Indeed, analysis of the higher-order diagram for $^3 F$ shows that
besides two first-order triads in the last line of \Fig{FF:11} (thin
red triangles) one finds diagrams in which instead of the first-order
triads one meets third-order triads (thick red triangles). These
diagrams represent five-order triads, which in  turn can
be found in  even higher diagrams, etc. This possibility
originates first from the fact that perturbation diagrammatic series
involves {\it all} topologically possible diagrams, and second because
all diagrammatic rules, including $\dfrac 1N$-symmetry rules and
time-integration rules, are applicable not only to the whole diagrams
but also to any of {{its}} fragments.
Therefore there is a mechanism for the infinite resuming of
telescopically nested, chronologically-ordered three-line reducible
triads, {appearing { {instead}} of the
  earliest time zones resulting in the fully} dressed triple
correlator $^3 F_{123} $.

 The Dyson line resummation leads to a fully dressed Green's function. The Wyld
resummation leads to the fully dressed double correlator. The triangular resummation of the three-line reducible triads suggested here leads to a fully dressed simultaneous triple correlator,  as we will show in \Sec{ss:full}.

\subsection{\label{sss:triple}{{{Triangular}}
  resummation of diagrams for quadruple correlators $ ^4F_{1234} $}}

  \subsubsection{\label{sss:4}Identifying   $^3F \Sp {\rm I}$ triad in diagrams for $^4F\Sp{\rm II}$ } 
  Analyzing the second-order diagrams $^{2\!}{\cal A}_{1,234}$ and
  $^{2\!}{\cal B}_{12,34}$ for $^{4\!}\C F_{1234}$, shown in
  \Figs{FF:3}  {(b) and (c)} we, as before, replace, the (three) double
  correlators by a sum of two auxiliary Green's functions according to
  \Eq{OnePole}. {{In such a case}}  each of the diagrams produces $2^3$
  diagrams. Out of those $2\times 2^3=16$ diagrams only three of them,
  shown in \Figs{f:9}(a,b) survive for the same-time case after
  frequency integration required by \Eq{TOc}.  Note that the diagram
  \Fig{f:9}(b) appears {\it twice} in different orientations. This
  removes the factor $\frac{1}{2}$ in front of it. The disappearance
  of the $\frac12$ factor occurs as a manifestation of the $\dfrac
  1N$-symmetry rule because the diagram in \Fig{f:9}(b) lost
  reflecting symmetry with respect to the vertical line, present
  (together with prefactor $\dfrac 12$) in the diagram depicted in
  \Fig{FF:3}(c).

Analytically diagrams in \Fig{f:9} can be written as $ {^4\!  F}_{1234} \Sp{II}=\underset{1234}{\bf P}\Big [~{ ^2\!\!}  A
_{1,234} + ^{2\!\!}B_{12,34} \Big]$,  where
 \begin{subequations}\label{32}
\begin{eqnarray}\label{32a} 
    \label{32b} ^{2\!} A _{1,234}
  = \frac 12\,  V_{14\overline 5} V_{523 } { F}_2 { F}_3 { F}_4 T\,, 
  \ \ { ^2\!}  B_{12,34} =  F _3 F _4 F _ {1+4}
  V_{13\overline 5 } V_{235}T \,, \quad T \=\, T_{1234} \,   T_{235}\,, 
\end{eqnarray}  
with   $\q_5=  \q_1+\q_4 =-(\q_2+\q_3)$.  The time
integral $T$ here was found with the help of
diagrammatic rules, formulated in \Sec{ss:3}.  It
reduces to a product of the triad  and quartic
interaction times.  Together with diagrams in 
\Figs{FF:9}(e,f,g) this allows us to recognize 
that the sum diagrams \Figs{f:9}  {(b) and (c)} include the
correlator $^3F_{235}\Sp {I}$, shown in
\Figs{f:9}(d) as red empty thin triangle.
Analytically this reads:
\begin{equation}\label{26}
 {^4\!  F}_{1234} \Sp{II}= 3 \, T_{1234}(V_{145} F_4 )   \, ^{\!\!} F\Sp {I}_{235}  \ .
\end{equation}
\end{subequations} 
We see that the first contribution of $^4\!  F_{1234} \Sp{II}$ to the
four-point correlator contains the first contribution
$ ^3 F\Sp{ I}_{235}$ to the three-point correlator. We will show below
that this statement generalizes to higher orders as follows: the
$(n+1)$ order of $^4\!  F_{1234} ^{n+1}$ involve $n$-order
contribution of $ ^3 F^n$.  Consequently, the fully dressed
fourth-order correlator depends on the fully dressed third-order
correlator. This is the essence of the
{{triangular}} resummation and it underlines the
key role played by the third-order correlator.
 \subsubsection{
     Identifying  $^3F\Sp {\rm I}$ and  $^3F\Sp{\rm III}$   triads  in the weakly connected spine diagrams  for $^4F\Sp{\rm IV}$.} 
   Recall that diagrams in \Figs{f:9}(b) and (c) are weakly connected
   in the sense that they can be divided into two parts by cutting
   only one line, sometimes referred to as ``spine".  Using
   diagrammatic rules for $^{4\!}F^{(4)}_{1234}$ formulated in
   \Sec{ss:hight} we found all the weakly connected spine diagrams
   shown in Figures \ref{FF:21}, \ref{FF:20}, and \ref{FF:23}. We
   divide the diagrams into these three figures by the position of the
   earliest (red) time zone relative to the spine $G_5$.  Figure
   \ref{FF:21} shows 18 diagrams with the earliest time-zone to the
   right of the Green function $G_5$ in the (678)-triangle,
   \Fig{FF:20} includes 10 diagrams with the (145)-time zone to the
   left of $G_5$, while \Fig{FF:23} involves six diagrams with two
   earliest time on either side of $G_5$.

      The diagrams of Figures \ref{FF:21},\ref{FF:20} and \ref{FF:23}
      are grouped in such a way that the triple correlator
      $^3 F\Sp {I}$ is identifiable in each line of the
      Figure. Namely, each line contains equivalent diagrams except
      {for} the position of the true of Green's function
      entering the earliest time zone. Consequently, each line sums up
      to the diagram in the right column containing the third order
      correlator in the third order $^3F\Sp{III}$ shown as a thin red
      triangle.
     
     Consider the  diagrams in \Fig{FF:21}.  The six
     resulting diagrams on the right of the Figure have the same
     structure {{connecting two parts by leg $5$. One part is
         the block of $G_1G_4 V_{145}$ with three legs. The second
         part consist of of the structures in which we recognize one
         of the diagrams for $^ 3F\Sp{III}$, shown \Fig{FF:11}.
       }}
          
       Therefore
       {{similarly to to \Eq{26}}} the sum of all diagrams in
       \Fig{FF:21} for the $^4F\Sp{IV}$ (denoted as
       $^{4,\alpha}F\Sp{IV} _{1234}$) can be presented via
       $^3{F\Sp{III}} $.
     
 \begin{subequations}\label{4F}
 \begin{equation}\label{27}
   ^{4,\alpha}\!   F_{1234}\Sp{IV}   = 3 \, T_{1234}(V_{145} F_4 )   \, ^{3\! } F\Sp {III}_{235}  \ .
\end{equation}
Comparing \Eqs{26} and \eqref{27} we see that i) The fourth-order
correlator $^4\!  F_{1234} ^{n+1}$ of any order always includes
quadruple interaction time, $T_{1234}$, originated from integration in
the latest time-zone with four external legs of the Green function
$G_1G_2G_3G_4$; ii) the earliest time zone, (235) in this case,
denotes the place where the triple correlator appears after the
{{triangular}} resummation.

Considering diagrams with the earliest (145) time zone in \Fig{FF:20}
we see {{that}} five lines of diagrams (A), (B), (C), (D), and
(E) have {{the }} same structure,
{{summed}} to the triple correlator
$^3F\Sp {I}_{145}$ times three point objects, denoted as $X\Sp {A}$,
$X\Sp {B}$, $X\Sp {C}$, $X\Sp {D}$, and $X\Sp {E}$.
{{The sum of these diagrams is given by}}:
\begin{align}\label{4Fb}
    \begin{split}
  ^{4,\beta}\!   F_{1234}\Sp{IV}   &=  3 \, T_{1234}  \, ^{3\!\!} F\Sp {I}_{145}  \big [  X\Sp {A}_{5,23}+X\Sp {B}_{5,23}+X\Sp {C}_{5,23}+X\Sp {D}_{5,23}+X\Sp{ E}_{5,23}  \big ]\,, \quad \B k_5=\B k_1+\B k_4\,, \\
      X\Sp {A}_{5,23} &=  V_{567} V_{6\overline 2 8 } V_{8\overline 3 7}F_3 F_4 \,,  \quad X\Sp {B} _{5,23} =
                      V_{756} V_{268 } V_{8\overline 3 7}F_3 F_6 \,, \ \dots \ . 
    \end{split}\end{align}
    {Equations for the rest of the terms in the RHS of \Eq{4Fb} can be easily reconstructed from their diagrammatic representation in \Fig{FF:20}.}

The last group of the  weakly-connected spine  diagrams with two earliest time zones is shown in
\Fig{FF:23}.  These diagrams sum
up  into one diagram with the product of two triple correlators  shown in  on the
\Figs{FF:23}(C3).   The corresponding analytical expression is 
\begin{equation}\label{4Fc}
  ^{4,\gamma}\!   F_{1234}\Sp{IV}  = 9 \,T_{1234} \,^3 F\Sp {I} _{145}\int \frac{d \B k_6}{(2\pi)^d}T_{13468}\,^3 F\Sp {I} _{378}
  V_{268}V_{657}\ .
\end{equation}\end{subequations}
Remarkably, this contribution is proportional to the {\it square}
of the triple correlator $^3 F\Sp {I}$.

 \subsubsection{ Identifying  $^3F\Sp {I}$   in the compact square diagrams for $^4F\Sp{\rm IV}$}  
As seen in \Fig{FF:5} there are eight  compact  ``square" diagrams for the
 quadruple correlator $^4F\Sp{IV}$.
 Consequently, they produce $8\cdot 2^4=128$ child diagrams but only
  22  of them, shown in \Figs{FF:12} and \ref{FF:13}, contribute
 to the simultaneous correlator.   Similar to the previous subsection, 
 here  we show how 
 all of them can be grouped in triads, each
 of which represents the  triple correlator $^3F\Sp {I}$.  We
   identify the  triads of  
 diagrams such  that  all elements 
   in the diagrams in each triad are identical, except one vertex, where 
 the ``true'' Green's function
 occupies each of the three positions in turns, and the other two
 positions are occupied by auxiliary Green's functions. 

 First of all, we separate all diagrams into two groups with one
 earliest time zone, shown in \Figs{FF:17} and \ref{FF:18}, and two
  earliest time zones, shown in \Fig{FF:19}.

The biggest group with one such zone will be  further  divided 
into several topologically different sub-groups as follows.  
 Since the diagrams are under the permutation operator, we 
redraw the diagrams  (by rotations or by mirroring)  in such a way 
that the earliest time zone will be placed in the upper right corner of
the diagram.  We then label  all lines as shown
e.g. in \Figs{FF:17}(A1).

We classify the diagrams by the number of external true Green's
functions, labeled by 1,2,3 and 4. We call the diagrams to be "Green's
function identical" if their external Green's functions are
identical. This set of lines does not include 2, 5 which connect to
the earliest time zone.  {We remind that all {{vertices}} must be
  connected to one of the external true Green's functions by the true
  Green's functions with the same orientations.
 
In a one-Green's function subgroup of the diagrams, there are two options: with $G_1$
and with $G_2$.  and $G_3$. $G_3$ subgroup is not new: it coincide
with $G_1$-diagram by mirroring in 2-4 line which connect $G_2$ and
$G_4$.

Consider the first $G_1$ subgroup. From general requirements
 {{its}} topology must include $G_7$ and $G_8$
true Green's functions which connect 3- and 4-vertices to 1-vertex
with true Green's function $G_1$. This subgroup has only three
diagrams, shown in \Fig{FF:17} panels (A1), (A2) and (A3).  These
three diagrams form the first triad which sum up to the diagram
\Fig{FF:17}(A) which involves triple correlator $^{3\!}F^{(1)}_{256}$.
 
Next $G_2$ subgroup must include $G_7$ and $G_8$ true Green's
functions which connect 3- and 4-vertices to 1-vertex with true
Green's function $G_2$. This subgroup also has only two diagrams,
shown in \Fig{FF:17} panels (B1) and (B2), forming the triad that sums
up to the diagram (\ref{FF:17})(B) which involve triple correlator.
 
There are nine diagrams with two true Green's functions labeled 1, 3,
and 4. The first six diagrams involve $G_1$ and $G_2$. Three of them,
collected in (C1), (C2) and (C3) panels, have auxiliary Green's
function $G_8$, oriented down by a straight line. They are summarized
in a diagram \Fig{FF:17}(C) which involves the same triple correlator
$^{3\!}F^{(1)}_{256}$.  {{The remaining }}
three diagrams with auxiliary Green's functions $G_8$, oriented up,
are summarized in the diagram \Fig{FF:17})(D).  Last two-Green's
function triad with $G_1$ and $G_3$ create diagram in
(\ref{FF:17})(E) with the same triple correlator
$^{3\!}F^{(1)}_{256}$.
 
Three Green's function subgroup with $G_1$, $G_2$ and $G_3$ are shown
in \Fig{FF:18}. They create two subgroups with $G_8$ oriented up and
down. The first triad creates a diagram in \Fig{FF:18}(A).
Considering the diagram in \Fig{FF:18}(B1) as two ones (with prefactor
$\frac12$) and rotating one of them around 2--4 lines we have a second
{{triad}} that creates diagram \Fig{FF:18}(B).

\begin{figure*}
\includegraphics[width=0.5\textwidth]{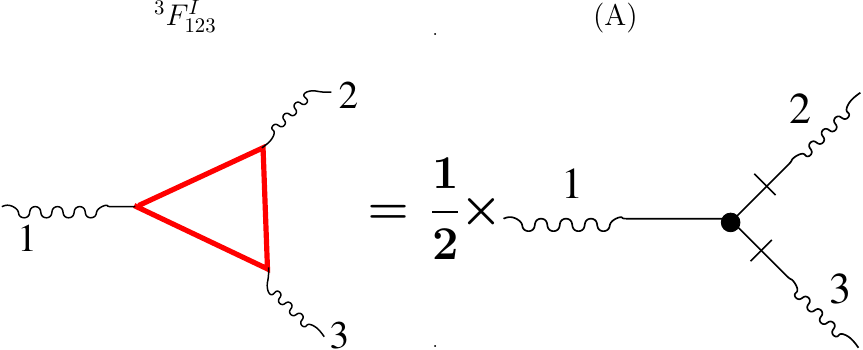} 
\includegraphics[width=\BiG\textwidth]{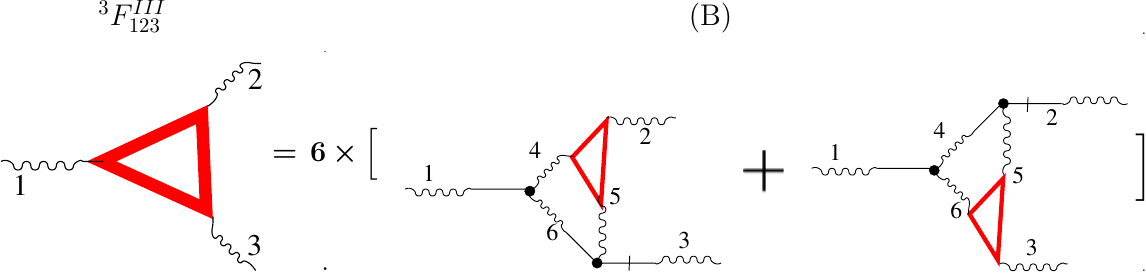} 
\includegraphics[width=\BiG\textwidth]{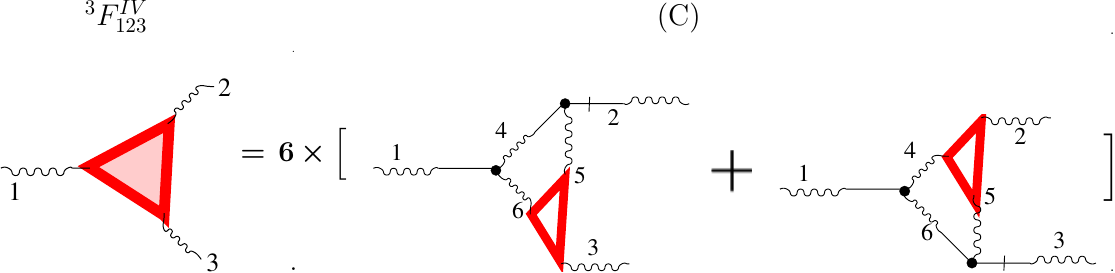} 
\includegraphics[width=\BiG\textwidth]{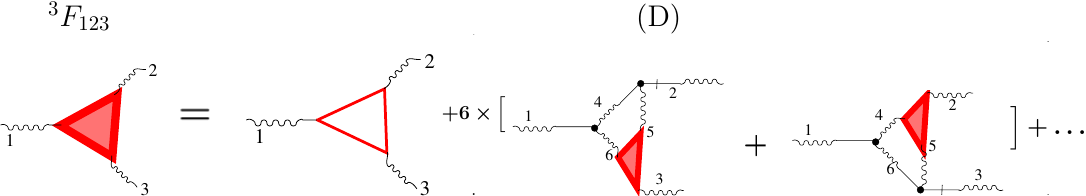} 
\caption{\label{f:8}{First triple-irreducible diagrams in the 
    triangular resummation for $^{3\!}F^{(\infty)}$: panel (A1) panel
    (A2) $^{3\!}F_{123}\Sp {II}$, panel (A2) $^{3\!}F_{123}\Sp {II}$
    and panel (B) $^{3\!}F_{123}\Sp {IV}$. Panels (C)
    {{yields}} integral equation \eqref{28} as
    a result of full {{triangular}}resummtion for $^{3\!}F_{123} $.}  }
 \end{figure*}

\begin{figure*}
 \includegraphics[width=\BiG\textwidth]{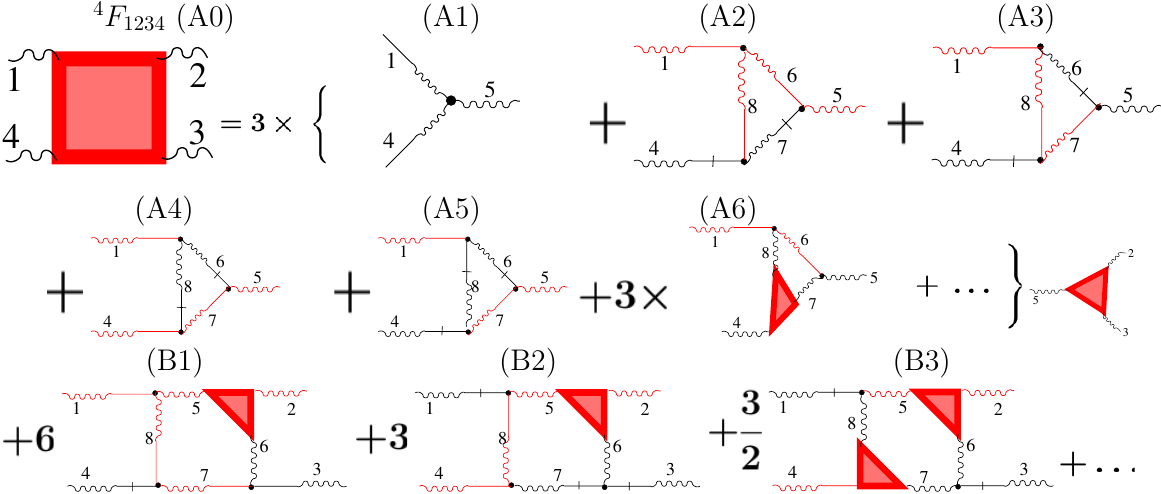} 
 \caption{\label{F:30}Full {{ triangular}} resummation for the four-point correlator $^{4\!}F_{1234}$. 
 }
 \end{figure*}
 
Analytical expressions for $^{4\delta \!}F\Sp{IV}_{1234}$, originated from
diagrams $X_1,X_2,\dots X_7$ depicted in \Figs{FF:17} and \ref{FF:18}
are as follows:
{{
\begin{align}\begin{split}\label{562}
               ^{4\delta\!}F\Sp{IV} _{1234}&=  T_{1234} \underset{1234}{\bf P}\Big \{  \int \frac{d \B k_5}{{(2\pi)^d}}\, \ \ 
                                             ^{3\!}F^{(1)}_{256} \,  J _{1234}^{\square}\Big \}\,, 
                                             \quad  {\bf k_6}  ={\bf k_2} + {\bf k_5},\,, \quad 
                                             J_{1234}^\square  = \sum\limits_{i=1}^7 X_i\,,  \\  
               X_1 &=  6 \, T_{2357}T_{23458}V_{1 \overline{5} 8}V_{\overline 7 3 6}V_{\overline 8 4 7}F_3 F_4\,, 
                     X_2= 3 \, (T_{2357}+T_{1268})T_{12378}V_{8 1 \overline 5}V_{\overline 7 6 3 }V_{4 7 \overline 8}F_1 F_3\,,  \\ 
               X_3&=  6 \, T_{2357}T_{23458}       V_{1 \overline {5} 8}V_{\overline {7} 3 6}V_{4 7 \overline {8}}F_3 F_8\,, \quad
                    X_4= 6 \, T_{2357}T_{23458}V_{1 \overline 5 8}V_{\overline 7 3 6}V_{4 7\overline 8}F_3 F_8\,,  \\ 
               X_5&=  6 \, T_{2357}T_{23458}V_{1 \overline 5 8}V_{\overline 7 3 6}V_{\overline 8 4 7}F_4 F_7\,, \ \ 
                    X_6= 6 \, T_{2357}T_{23458}V_{1 \overline 5 8}V_{3 6 \overline 7}V_{4 7\overline 8}F_7 F_8\,,  \\ 
               X_7&= 6 \, T_{2357}T_{23458}V_{1 \overline 5 8}V_{3 6 \overline 7}V_{4 7\overline 8}F_7 F_8\ . \end{split}\end{align}
         }}
       The second group of diagrams with two earlier time zones consists
       only {{ of}} four diagrams.  As shown in \Fig{FF:19} we   used the   diagram
       $^{4\!}C\Sp{VII}$   in  two panels (A3) and (B1) putting
       prefactor $\frac12$ in front of them.  After that diagrams (A1), (A2)
       and (A3)   are  summed to diagram (A) and diagrams (B1) and (B2) -- to
       diagram (B). In its turn, diagrams (A) and (B) can be summed to
       diagram (C) which involves two triple correlators
       $^{3\!}F\Sp{I}_{256}$ and $^{3\!}F\Sp{I}_{478}$. An analytical
       expression for the sum of all diagrams with two earliest time zones
       is as follows:
 	\begin{align}\begin{split}
                       ^{4\varepsilon\!}F\Sp{IV} _{1234} =&  T_{1234} \underset{1234}{\bf P}\Big \{  \int \frac{d \B k_5}{{(2\pi)^d}}\,  ^{3\!}F\Sp{ I} _{256} \,   ^{3\!}F\Sp {I} _{478} 
                                                            \times \big (T_{23458}+T_{12467} \big ) \,
                                                            V_{15\overline 8}V_{3\overline 6 7}\Big \}\ . \end{split}
 	\end{align}
  
 {Similar to \Eq{4Fc} here  the
contribution to the fourth-order-four-point correlator comes from the product of {\it{two}} three-point correlator of the first order $^3F_{123}\Sp {I}$.  Analysing the structure of the diagrammatic technique we expect that in the higher-order diagrams  terms with a product of three and more correlators  $^3F_{123}\Sp {I}$ will appear.  }

\subsection{\label{ss:full}Full triangular
  resummations for $^3F_{123}$ and $^4F_{1234}$}

In the previous sections, we demonstrate how the sum of six initial
diagrams for $^3F_{123}\Sp {III}$   presented \Fig{FF:11} fuses into
two diagrams in \Fig{FF:11}  {(b) and (c)} involving $^3F\Sp {I}$. Similarly, the
sum of two diagrams for $^4F_{1234}\Sp {II}$ in \Fig{f:9} combines to
one diagram \Fig{f:9}(d) with $^3F \Sp {I}$. Moreover, the sum of
eighteen diagrams for $^4F_{1234}\Sp {IV}$ in \Fig{FF:21} combines
into six diagrams, involving $^3F\Sp{I}$ which, in its turn fuse into
just  one diagram with $^3F \Sp {III}$, analytically presented by
\Eq{27}. In exactly the same way,  the rest of the diagrams for
$^4F_{1234}\Sp {IV}$, shown in {Figs. \ref{FF:20}-\ref{FF:19}} was summarised in these figures to
diagrams, involving the triple correlator $^3F\Sp {I}$.
These findings
are not a miracle, but the deep consequence of fundamental features of
the perturbation approach reflected in the diagrammatic technique and
the crucial role that is played by the three point correlator.

Namely, the diagrammatic series involves all topologically possible
diagrams, satisfying general restrictions, described in
\Sec{ss:hight}.  These restrictions together with the $\frac1N$-rule,
prescribing numerical prefactor have local character, i.e. they are
applicable to the entire diagram, or to any of 
{{its}} fragments. For example, any diagram for $^3F_{123}$ has
three external Green's functions $G_1$, $G_2$, and $G_3$ entering the
diagram by straight lines.  Similarly, any earliest time zone also has
three ``boundary" Green's functions, say $G_k$, $G_l$, and $G_m$
entering the zone in the same way. Therefore the sum of all diagrams
inside  the earliest time zone of $(2n+1)$-order [with
$(2n+1)$ vertices] gives exactly the triple correlator of
$(2n+1)$-order, $^3F^{n+1}_{klm}$. This is  exactly
{{what}} happened in all diagrams with the first-order earliest
time zones (colored in red), summarised to $^3F\Sp {I}$, while the
diagrams \Fig{FF:21} with the third-order earliest time zones,
(colored in blue) were summarised to $^3F\Sp {III}$.

Consider first full { {
    triangular}} resummation for the triple correlator $^3F_{123}$.
Panel (A) in \Fig{f:8} just {{resembels the}}
diagram in \Fig{FF:9}(b) for $\ ^3F\Sp {I}_{123}$, while \Fig{f:8}(A2)
shows resummations diagrams for $^3F\Sp {III}_{123}$ with the result
$^3 F\Sp {III}_{123}\propto ^{3\!} F\Sp {I}_{123}$, as indicated by
\Eq{3F3}. The next step is shown in \Fig{f:8}(B). 
{{Comparing panels (B) and (C) we see the pattern which
    illustrates the essence of the triangular resummation. Namely the
    $^{3\!} F\Sp {III}_{123}$ depends on the ${^1\!} F\Sp {I}_{123}$
    in the same way as $^{3\!}F\Sp {V}_{123}$ depends on
    $^{3\!}F\Sp {III}_{123}$.  }} We can continue this iteration {\it
  ab {{infinitum}}}, see the result in \Fig{f:8}(C). 
Clearly, this procedure does not create high-order diagrams for
$^3 F\Sp {III}_{123}$ of a more complicated topological structure.
These diagrams are replaced by ``\dots" in line (C) of \Fig{f:8},
which presents the entire series for
$^3F = \,^{3\!}F\Sp {I}_{123}+\, ^{3\!}F\Sp {III}_{123}+\,^{3!}F\Sp{V}_{123}+\dots$.  Analytically we have

\begin{align}\label{28}\begin{split} ^{3\!}F_{123} = \,^{3\!}
    F_{123}^{I} + 6 \underset{123 }{\bf P}\int \frac{d \B
      k_4}{(2\pi)^d}T_{2346} V_{146} \times \big[
    V_{425}F_2+V_{245} F_4\big ] \, ^{3\!}F_{356} + \dots\
    . \end{split} \end{align}
This equation represents the essence of
{{triangular}} resummation for the
fully-dressed triple correlator $^3F_{123}$, as it represents
$^3F_{123}$ through the infinite series that {{itself}} involves
 $^3F_{123}$.  If one neglects higher-order
contributions replaced in \Eq{28} by dots, then this equation becomes
{\it closed} equation for $^3F_{123}$. Clearly, this procedure is an
uncontrolled approximation. Currently, this equation looks linear
since the higher-order terms are represented as dots
``$\dots$". Analyzing higher-order contributions we found diagrams
with two, three, and more earliest time zones, giving birth to
contributions to \Eq{28} proportional to $(^{3\!}F) ^2$,
$(^{3\!}F)^3$, etc.

{{ Based on \Eqs{28}  consider  inverse energy  
cascade  in the fractional dimension
$d=\frac43+ x$ close to the critical dimension $d_0=\frac 43$.   Here there are two limiting cases:  i)  constant energy flux $\varepsilon$ or ii)   constant energy $^{2\!}F$.   When $\varepsilon=$const in the limit $x\to 0$ the energy $^{2\!}F\to \infty$. We focus on the latter case, when  the   energy $^{2\!}F(\B k)$  is kept  finite.  Then in the limit $x\to 0$    the energy 
flux $\varepsilon$ (together with $\ ^{3\!} F$)  vanishes.  In this case \Eq{28} becomes {\it{nonlinear homogeneous}} equation with powers of full triple correlators on its right-hand side. This equation has a trivial  solution
      $\ ^{3\!}  F_{123}=0$.  This is a demonstration that the triple
      simultaneous correlator is zero under these assumptions ($d=\frac43$ and finiteness of the energy) not only in the leading order (as shown in \cite{Lvov2002b}) but in all orders, i.e fully dressed correlator $\ ^{3\!} F=0$.  }} 
 
  The diagrams for the four-point simultaneous correlator
  $^{4!}F_{1234}$ can also be {{triangular}}
  {{resummed}}. This is shown in \Fig{F:30}. We denote
  $^{4\!}F_{1234}$ as a red thick-filled square. The simplest diagram,
  in {{panel}} (A1), originates from the infinite resummation
  of the diagram for $^4F_{1234}\Sp{II}$ in \Fig{f:9}(d), where the
  $^{3\!}F\Sp {I}$ is summed up to the full correlator. Graphically
  this part of the {{triangular }} resummation
   is depicted by replacing the thin
  {{triangle}} by thick filled {{triangle}}. The
  {{triangular }}resummation includes all diagrams
  in \Fig{FF:21}. Four lines of diagrams in \Fig{FF:20} (after
  inversion over spine Green's function $G_5$) serve as the first
  terms of the triangular  resummation that leads to
  diagrams in \Figs{F:30}(A2,A3,A4,A5). In the same way diagrams in
  \Fig{FF:23} produce diagram \Fig{F:30}(A6), proportional to
  $(^3F)^2$, compact square diagrams in \Figs{FF:17} and \ref{FF:18}
  gives after resummation diagrams in \Fig{F:30} panels (B1) and
  (B2). Remarkably, compact square diagrams in \Fig{FF:19} with two
  earliest time-zones create diagram in \Fig{F:30}(B3). This diagram
  is proportional to the {\it square} of {{the full triple
      correlator $^3F$}}. Higher order terms, represented by dots, can
  also be {{triangle}} resummed.

Analytical expressions for various contributions to $^4F_{1234}$,
denoted as $^4F_{1234}\Sp{A1}$, \dots $^4F_{1234}\Sp{A6}$,
$^4F_{1234}\Sp{B1}$, $^4F_{1234}\Sp{B2}$, and $^4F_{1234}\Sp{B3}$ can
be straightforward reconstructed from the corresponding
diagrams. below we will present a few examples:
 \begin{align}\label{29}\begin{split}
^{4\!}F_{1234}\Sp{A1}=& 3\, ^{3\!}F_{235} F_4 V_{145} \, \,, \   ^{4\!}F_{1234}\Sp{A2}=
 3\, ^{3\!}F_{235} F_4 \!\! \int \!\! \frac{d \B k_6}{(2\pi)^2}V _{168}V_{657}V_{847}F_7\,, \dots \  ^4F_{1234}\Sp{A6}=
 9 \, ^{3\!}F_{235}   \!\! \int \!\! \frac{d \B k_6}{(2\pi)^2}V _{168}V_{657}  \, ^{3\!}F_{235}\,, \\ 
 ^{4\!}F_{1234}\Sp{B1}=& 6\, F_3 F_4   \!\! \int \!\! \frac{d \B k_5}{(2\pi)^2}V _{156}V_{847}V_{673} \, ^{3\!}F_{256}\,, \dots \quad  ^4F_{1234}\Sp{B3}=
 \frac 32   \!\! \int \!\! \frac{d \B k_5}{(2\pi)^2}V _{158}V_{367}  \,^{3\!}F_{256}\,^{3\!}F_{478}\ .
 \end{split}\end{align}
These equations present fully dressed quadruple correlator
$^{4\!}F_{1234}$ as a series in powers of fully dressed triple
correlator $^{3\!}F$, explicitly involving linear and quadratic
contributions. Some diagrams with five and more vertices will give
contributions of third, fourth, and higher powers of $^{3\!}F$.
  
{{In particular, this means that in the
    inverse energy cascade in a dimension near the critical,
    $d =\frac43+x$, the dressed fourth order correlator
    $^ {4\!}F_{1234}$ vanishes in the limit $x \to 0$ with finite
    energy. Moreover, there is every reason to believe that this
    statement is valid for all high-order correlators $^{n\!}F $ with
    $n=4\, , 5\, , \dots$. If so, the statistics of turbulence in
    $d =\frac43$ becomes Gaussian when all comulants vanish and is
    very close to the Gaussian statistics for $d =\frac43+x$. We think
    that this explains the experimental observation that statistics of
    inverse energy cascade is close to Gaussian even for $d=2$.  }}

 \section{\label{s:sum}Summary}
 {{In this paper we reconsider the perturbation theory for the
     hydrodynamic turbulence via the Dyson-Wyld diagrammatic technique
   }}  presenting {{a}} detailed
 analyses of the three-point and four-point velocity correlation
 functions in leading and {{the}} next
 order. {{This}} corresponds to the first and
 {the} third orders in the interaction amplitude for the triple
 correlator and to {{the}} second and {the} fourth order
 for the four-point correlator.  {{This allowed us to recognize
     the crucial role played by the triple correlator and the energy
     flux over scales and clarify their role in determining the entire
     statistics of turbulence. }} In the framework of the
 Dyson-Wyld diagrammatic technique  we performed  the following steps:\\
%
 $\bullet$~We showed how to build diagrammatic series for the complex
 amplitude of strongly interacting fields. \\
 $\bullet$~We  showed how to build  diagrammatic series for the
 three-point, four-point, and higher-order correlation function in the
 velocity. This is achieved by averaging over an ensemble of random
 force, or {\it gluing} together diagrams for the velocity field (trees).  In doing so we
 have demonstrated constructively the natural emergence of the
 $\frac{1}{N}$ symmetry rule,  which prescribes the numerical
 prefactor of a diagram {{via number $N$ of elements of their symmetry group. }}
\\
$\bullet$~We then considered {\it simultaneous} velocity
correlators. In doing so, we {{needed}} to perform the
integration over all possible frequencies of multiple time
correlators. To achieve this goal, it is imperative to know the
frequency dependence of the double correlator and Green's function.
We have assumed {\it {{the}} one-pole} approximation for the
frequency dependence of  double correlator and Green's
function. This assumption led to the decoupling of all the double
correlators into the sum of two auxiliary Green's functions pointed in
different directions.  This, in turn, allowed us to formulate
time-integration
rules by introducing time zones and time boundaries. These rules allow one to reconstruct the time integrals from the topology of the diagrams without explicit calculations of the high-order time integrals. \\
 $\bullet$~We then considered in detail the simultaneous triple
 correlator of the velocity in the first and third order in the
 interaction vertex. We have shown how the diagrams can be grouped
 into triads, thus leading to the triple correlator.  This grouping
 allowed us to formulate the
 {{triangular}}-resummation. The
 {{triangle}}-resummation does for triple-reducible
 diagram what Dyson-Wyld ressumation does for double-reducible
 diagrams. Namely, the {{triangular }}resummation
 replaces the bare triple correlator by its
 ``dressed'' counterpart.\\
 $\bullet$~We considered a four-point correlator in {{the
   }}second and {{the}} fourth order. We showed that the
 {{triangular}}-resummation is equally applicable to
 four-point simultaneous correlators. We did it by identifying a
 third-order simultaneous velocity correlator in the diagrammatic
 series for four-point correlators. We did it constructively for
 {{the}} second-order and {{the}} fourth-order diagrams. 
  Analysing the structure of the diagrammatic technique we
 {{showed}} that this pattern continues for all correlators for
 any order in the perturbation theory. Namely, all the diagrams can be
 grouped together in triple-reducible diagrams, that is the groups
 where the diagrams are equivalent except for one {{vertex}}
 where {{the}} Green's function is rotated. These groups can be
 {{resummed}} to
 become third-order simultaneous correlators.

 {{
 $\bullet$~Considering diagrammatic series of the perturbation theory
 for any-order correlation functions $^{n\!}F$ we demonstrated that
 they can be reordered (resummed) such that $^{n\!}F$ explicitly
 includes one or more series for $^{3\!}F$. From physical viewpoint
 this means that $^{n\!}F$ is a polynomial in powers of $^{3\!}F$,
 without zero-order term.  In particular, this means that if $^{3\!}F$
 vanishes (like it happens in thermodynamic equilibrium) then all
 irreducible diagrams $^{n\!}F$ (i.e. cumulants of the high-order
 correlators) vanishes as well and statistics of turbulence become
 Gaussian.  Baring in mind that the energy flux over scales
 ${\varepsilon}(\B k)$ is proportional to $^{3\!}F$ we conclude that
 this flux governs not only the energy distribution over scales
 (i.e. $^{2\!}F$), as Kolmogorov assumed in 1941, but the entire
 statistics of hydrodynamic turbulence. We stress that this conclusion
 is not based on any truncation of the diagrammatic series but on
 their analysis as a whole.
}}

 $\bullet$ As explained in \cite{Lvov2002b}, the turbulence inverse
 energy cascade {{coincides}} with the thermodynamical
 equilibrium for the fractional dimensions of $d=\frac{4}{3}$.  Therefore
 it is exactly Gaussian. The difference between $\frac43$ and $d=2$ is
 a small parameter $\epsilon=d-\frac{4}{3}$.  The last
 observation explains therefore why the 2D turbulence is close to
 Gaussian state.  {{As $^3 F^{I}_{123}\propto\epsilon$, the
     triangular resummation determined by \Eqs{28} shows}} that the
 full correlator $^3F_{123}$ is also small.  If $^3F_{123}$ is small,
 then the Figure \ref{F:30} explains why the fourth order cumulant is
 also small.  In this way, in 2D turbulence, cumulants $^n F$ become
 presented as a series in powers of the small parameter
 $\epsilon=d-\frac43 = \frac23$, {{thus}} demonstrating the
 closeness of the 2D statistics to the Gaussian case.
              
 Note also that the {{triangular}} resummation of
 the triads is possible only for the one-time correlators, otherwise,
 for example, diagrams, shown in \Figs{FF:9}(b), (c), and (d) have to
 be accounted for. From a theoretical viewpoint, this is the
 consequence of the fact that in the thermodynamic equilibrium, the
 only simultaneous statistics are Gaussian (with vanishing all
 cumulants of the correlation functions).

 We hope that this paper provides solid theoretical foundation for the
 further analytical study of the statistics of highly developed 2D and
 3D hydrodynamic turbulence {{and other systems of hydrodynamic
     type.}}

\paragraph*{Acknowledgment}

We would like to express our gratitude to the anonymous referees for
their valuable feedback, which significantly helped us to improve this
manuscript. YL acknowledges support from the NSF DMS award 2009418.

   \bibliography{bib.bib}

\section{Appendix}
\renewcommand{\theequation}{A\arabic{equation}} \setcounter{equation}{0} 
\subsection{Averaging products of trees}
Here we perform in detail gluing  the rest of the
trees into diagrammatic series for correlation function. The gluing steps may be avoided alltogether. We present them here for references and
to show how the {{$1/N$}}   symmetry  rule appears.

   \subsubsection{Three Point Correlators {{of the}}
   Third Order
   \label{ThreePointCorrelatorThirdOrder}} In this subsection we
 compute the three-point correlation function in the third order in
 the interaction vertex. This object is obtained by gluing together
 three trees and averaging over the ensemble of random force. The
 advantage of the diagrammatic technique is that the gluing {{of
     the trees representing perturbation expansions}}, again, can be
 ommitted alltogether. {{Consequently}} the
 diagrams for these three-point {{correlators of the }}third
 order in the vertices can be drawn from scratch. {{Construction of these diagrams
     can be achieved }}by exhausting all
 possible topologies consistent with the diagrammatic rules. We
 present the details of the calculations here to illustrate
 constructively the mechanism of appearance of the symmetry
 $\frac{1}{N}$ rule. 
   
   The terms
      \begin{subequations}
 \begin{eqnarray} \label{ThreePointCorrelatorThirdOrderTerms}
  && \hskip -.4cm (2\pi)^{d+1} {\delta^{d+1}_{123}}\, ^3 \! \C F_{123}^{(3)}=
   ^3 \!\!  {\cal F}_{123}^{(3)}\,, \ \  {\cal F}_{123}^{(3)}=
   {^{3a}\!\!{\cal A}_{1,23}}+ {^{3b}\!\!{\cal A}_{1,23}} + ^3\!\!
   {\cal B}_{12,3} + ^3\!\! {\cal C}_{123} \,,  \\ &&{^{3a}\!\!{\cal
       A}_{1,23}}= \frac{ \< ^{3a}\!a_1 a_2 a_3 \>}2\,,
   \ {^{3b}\!\!{\cal A}_{1,23}}=\frac {\< \ ^{3b}\!a_1 a_2 \, a_3\>}2 \nn
   \ \   ^3\! {\cal B}_{12,3}=\< a_1^{2} a_2^{1} a_3 \>\,, \ {^3 {\cal
       C}_{123}}= \frac 1 {3!} {\< a_1^ 1 a_2^ 1 a_3^ 1 \>} \,,
 \end{eqnarray}
 \end{subequations}
with one ($^{3a\!\!}{\cal A}_{1,23}$ and $^{3b\!\!}{\cal A}_{1,23}$),
two ($^{3\!}{\cal B}_{12,3}$), and three ($^{3}{\cal C}_{123}$)  external $\C
G$-legs respectively.  Resulting diagrams are shown in \Fig{FF:4}. For
consistency with previous notation, we denote by $ ^{3\!\!} {\cal A}$
third-order three-point correlators with one external Green's function,
by $\ ^3 {\cal B}$ third order correlators with two external Green
functions.  Consistently with previous definitions of $\C A$- and $\C
B$-terms we use notation $^n {\cal C}_{123, \dots}$ for all diagrams
of $n\sp {th}$ order in {{vertices}} $V$ with three $\C G$-legs $\C G_1$,
$\C G_2$ and $\C G_3$ and any number of wavy tails denoting ${\cal
  F}_j$.

We now consider each of these terms separately, one by one.
 \paragraph*{\textbf{1.} $^{3a}\!\C A$-terms.}  The   $^{3a}\!{\cal A}_{1,23}$ term after substitution of \Eq{3aA2} for $^{3a}\!a_1$ can be presented as
\begin{equation}
^{3a}\!{\cal A}_{1,23}  =\frac{ \brown{\C G_1 \C G_{\dots} ^2V^3_{\dots} }} {4  } \<   \brown{(a _4  a _5a _6  a _7)_1}   (a _2 a _3)_2  \> \ .\nonumber
\end{equation}
 We now construct double correlators by pairing the fields (wavy
 lines).  To get an irreducible contribution we pair $[(
   \overbrace{7\mbox{-}(2},3)_2]$ (two options), then $[(
   \overbrace{3\mbox{-}(5},6)_2]$ (two options) and finally $
 \overbrace{4\mbox{-}6} $ (one option).  The resulting diagram is
 shown in \Fig{FF:4}(a) and corresponding analytical expression is
 given by \Eqs{16a}.
 \paragraph* {\textbf{2.} $^{3b}\!\C A$-terms.} The   $^{3b}\!{\cal A}_{1,23}$ term after substitution of \Eq{3bA2} for $^{3b}\!a_1$ can be presented as
	\begin{equation}
	 ^{3b}\!{\cal A}_{1,23} =\frac{ \brown{\C G_1 \C G_{\dots}
              ^2V^3_{\dots} }} {16 } \< \brown{(a _4 a _5a _6 a _7)_1}
          (a _2 a _3)_2 \> \ . \nonumber
\end{equation}
However, due to the different topology of the tree for $^{3b}a_q$ we should
pair fields differently. For example, as follows:
$[(4,5,6,\overbrace{7)\mbox{-}2}]$ (four options), next
$[(5,\overbrace{6)\mbox{-}3}]$ (two options) and finally
$\overbrace{4\mbox{-}5}]$ (one option). The resulting analytical
  expression is given by \Eqs{16b}
as shown diagrammatically in  \Fig{FF:4}(b).
\paragraph* {\textbf{3}. $^3\!\C B$-terms.} The $ {^3 {\cal B}_{12,3}}$ term after
substitution of \Eq{1A2} for $^{1}\!a_1$ and \Eq{2A2} for $^{2}\!a_1$
can be presented as
\begin{eqnarray} 
^3 {\cal B}_{12,3} =\frac{ \green{\C G_1 V_{\dots}}\blue{\C
      G_2}G_{\dots}V^2_{\dots}} {2^2 } \< \blue{(a _4 a _5)_1}\green{(
    (a _6 a _7a _8)_2 )} a _3 \>.~~~~~~~\nonumber
\end{eqnarray}
To get an irreducible contribution we  should pair  for example  $[
  \overbrace{6\mbox{-}(4},5)]$ (two options), next $[
  \overbrace{3\mbox{-}(6},7)]$ (two options) and finally
$\overbrace{7\mbox{-}5}]$ (one option).  As the result, we have four
  equivalents contribution to $^3 {\cal B}_{12,3}$ as shown in \Eqs{16c}
  as shown in \Fig{FF:4}(c).

\paragraph* {\textbf{4}.$^3 \C C$-terms.} The  $ {^3  {\cal C}_{123}} $ term  can be presented as 
	\begin{equation}
	{^3 {\cal C}_{123}} =\frac{(\blue{\C G_1 \C G_2 \C G_3 V^3_{\dots}})}{2^3
          \cdot 3!} \< \blue{(a _4 a _5)_1 (a _6 a _7)_2 (a _8 a _9)_3
        } \> \ .
        \nonumber
\end{equation}

Pairing, for example as follows: $[ \overbrace{4\mbox{-}(9},6,7,8)]$
(four options), $[(5, \overbrace{6\mbox{-}(7},8)]$ (two options) and
$[ \overbrace{5\mbox{-}8}]$ (one options) we have 8 equal terms. The
resulting diagram is \Fig{FF:4}b and the corresponding analytical
expression given by \Eqs{16d}.

 \subsubsection{Four Point Correlator leading terms
  \label{FourPointCorrelatorLeading}}

Here we show how to glue four trees together to form the diagrammatic
series for the four-point correlator in the second order in the
interaction vertex. These steps are performed here in
detail to demonstrate explicitly how the diagrammatic
series for correlation function appear and why they have the factor
 {{corresponding}} to the
$\dfrac{1}{N}$-symmetry rule.  These steps are equivalent to those
described in the section (\ref{TripleCorrelatorFirstOrder}) and may be
omitted altogether by drawing the diagrams from scratch for the
correlation functions as explained in the section
(\ref{SectionAboutRules}.

Consider first the expression for ${^2\!{\cal
    B}_{12,34}}$. Substituting expressions for $\ ^1 a_1$ and
$\ ^1 a_2$ from \Eq{1A2} and \Eq{2A2} we obtain

$$
{^2\!{\cal B}_{12,34}}=\frac 1{16} \blue{\C G_1 \C G_2 V_{156} V_{278}} \<\blue{ ( a_5 a_6 )_1  (a_7 a_8 )_2} (  a_3a_4 )_3\> 
  \ .
  $$
Here and below subscripts {{$1$ and $2$}} identify the tree from
which the analytical structure  appeared. Pairing
fields in each $(\dots)_j$ group leads to uncoupled
contribution. There are four equivalent ways to pair $[\overbrace{
    4\mbox{-}( 5},6,7,8)]$, two ways to pair
$[\overbrace{3\mbox{-}(7},8)]$ and one way to pair $ \overbrace{
  6\mbox{-}8}$.  The result, diagrammatically shown in \Fig{FF:3}(c),
has corresponding analytical expression {is given in  (\ref{F4Ba})}. 
 
 Next, substituting $\green{^2\!a_\q}$ from \Eq{2A2} in 
 \Eq{F4b} we get
\begin{eqnarray}\label{F4Baa}{^2\!{\cal A}_{1,234}}&=&\frac{ (\green{\C G V}) }{2\cdot 3!}\<  (\green{a _5  a _6  a _9})_1 (a _2  a _3  a _4 )_2 \>   \ .
 \end{eqnarray} 
Required irreducible diagrams can be obtained from all possible (six)
pairing of all three fields $(\dots)_1$ with all three fields
$(\dots)_2$: $[( (5 ,6,\overbrace{9)_1\mbox{-}(2},3,4)_2]$. The
resulting analytical expression is given in
(\ref{F4Ba})
and is diagrammatically shown in \Fig{FF:3} (b).

\subsubsection{Fourth-order Four Point Correlator
   \label{FourPointCorrelatorFourthOrder}}
In this section, we glue four trees to {{obtain}} the 
expressions for the four-point correlators in the fourth order in the vertices. 
   
      Notice that we have met here the new type of diagrams 
${^{4}   {\cal D}_{1234}}$, 
  	  where we preserved {{this}} notation
  	  for all diagrams of $n\sp {th}$
          order in {{vertices}} $V$ with four $\C G$-legs, $\C G_1$, $\C G_2$,
          $\C G_3$ and $\C G_4$ and any number (including zero, as in this
          case) of wavy tails denoting ${\cal F}_j$.
   
          \paragraph* {\textbf{1.} $^{4a}\! {\cal A}_{1,234}$-term.}
          After substitution \Eq{4aA2} for \red{$^{4a}\!a_1$} the
          $ ^{4a} \! {\cal A}_{1,234} $ term can be presented as
  	\begin{subequations}\label{4aAa}
  	\begin{eqnarray}\label{4aAa2} ^{4a} \! {\cal A}_{1,234} =   \frac{\red{G_1 G_{\dots}^3 V^4_{\dots}}}{2\cdot 3!} 
  	 \<  a_2 a_3 a_4 \red{(
             a_5  a_6 a_7  a_6 a_8  a_9)_1 } \> .~~~
  	\end{eqnarray}
        Here we red-colored terms originated from
        {\red{$^{4a}\!a_1$}}, and bracketed {$\red{(\dots)_1}$}
        corresponding field. A particular topological position of
        these {{terms}} and the rest {{of the }}fields
        {{are}} shown in \Fig{FF:5}(a), where
        (at this moment) we have to separate all wavy lines into two
        parts.  Pairing, for example as follows:
        $[(4,3 \overbrace{2)\mbox{-}5}]$ (three options),
        $[(4, \overbrace{3)\mbox{-}6}]$ (two options), and
        $[ \overbrace{4\mbox{-}(7},8)]$ (two options) and
        $[ \overbrace{{ 8}\mbox{-}9}]$ (one option) we have 12 equal
        terms.  {{The}} factor $12$ fully
        compensate {{the}} denominator in \Eq{4aAa} giving
        prefactor unity in the diagram for
        $^{4a} \! {\cal A}_{1,234}$.  The results can be schematically
        presented as

\begin{equation} \label{4aAb} ^{4a} \! {\cal A}_{1,234} =    \red{G_1} {\cal F}_2 {\cal F}_3 {\cal F}_4  \red{G_{\dots}^3 V^3_{\dots}{\cal F}_{\dots}} \ . 
  	\end{equation}
       \end{subequations}
        
Explicit analytical expression for $ ^{4a} \! {\cal A}_{1,234}$ can be
reconstructed from \Fig{FF:5}(a) and is given by \Eq{itA} below.  The
diagram for $ ^{4a} \! {\cal A}_{1,234}$ has only one element of
symmetry, the identity, therefore the factor $1$ in front of it is
consistent with the $\dfrac 1N$-rule.
	
\paragraph*{\textbf{2.} $^{4b}\! {\cal A}_{1,234}$-term.}  After substitution  \Eq{4bAa} for \red{$^{4b}\!a_1$} the  $ ^{4b} \! {\cal A}_{1,234} $ term  can be presented similarly to \Eq{4aAa}
as follows
\begin{subequations}\label{4bAb}
  \begin{eqnarray}\label{4bAb2} ^{4b} \! {\cal A}_{1,234}=
    \frac{\red{G_1 G_{\dots}^3 V^4_{\dots}}}{4\cdot 3!}  \< a_2 a_3
    a_4
                  \red{(  a_5  a_6 a_7  a_6 a_8  a_9)_1} \>  .~~~
  		\end{eqnarray} 

                However, {{as shown in \Fig{FF:5}(b)}} the
                topology of {{the corresponding}}
                tree is different. This
                different topology dictates a different way of
                pairing, for example
                $[(4,3, \overbrace{2)\mbox{-}(5},6]$ (six options),
                $[ \overbrace{6\mbox{-}(7},8))]$ (two options),
                $[ \overbrace{8\mbox{-}(3},4)]$ (two options) and
                $[ \overbrace{4\mbox{-}9}]$ (one option) we have 24
                equal terms.  Again, this fully compensates the
                denominator in \Eq{4bAa} giving prefactor unity in the
                diagram for $^{4b} \! {\cal A}_{1,234}$.
 
\end{subequations}

           \paragraph*{\textbf{3.}  $^{4a}  {\cal B}_{12,34}$-term.}  
           
After substitution of 
 \Eq{1A2} for  {\blue{$^{1}\!a_1$}} and   \Eq{2A2} for  {\green{$^{2}\!a_2$}} the  $ ^{4a} \! {\cal B}_{12,34} $ term  can be presented similarly to \Eqs{4aAa} and \eqref{4bAa} as follows
  	 
  		\begin{eqnarray}\nn {^{4a} \! {\cal B}_{12,34}}=   
  		\frac{\blue{G_1} \green{G_2}G_{\dots}^2 V^4_{\dots}} {2^3} \< a_3 a_4 \green{( a_5  a_6 a_7  a_6 a_8 )_2 }
                  \blue{( a_9  a_{10})_1} \> .  
  		\end{eqnarray} 
                Here Green's functions and free fields originated from
                \blue{$^{1}\!a_1$} and {\green{$^{2}\!a_2$}} are
                colored in blue and green and are taken in parentheses
                \blue{$(\dots)_1$} and \green{$(\dots)_2$}.  Their
                particular positions on the diagram for
                $^{4a} \! {\cal B}_{12,34}$, that dictate
                their pairing 
                {{configuration}},  {{are}}
                shown \Fig{FF:5}(c).  The result, that leads to this
                diagram is independent of the particular choice of
                strategy.  For concreteness we pair free fields
                $ ^ 0\!a_j$ in the {{following}} way:
                $[ \overbrace{6\mbox{-}(3},4)]$ (two options),
                $[ \overbrace{{\cyan 7}{\red 4}\mbox{-}(7},8)]$ (two
                options), $[ \overbrace{8\mbox{-}(9},10)]$ (two
                options) and finally $[ \overbrace{5\mbox{-}10}]$ (one
                option).
The resulting diagram is presented in \Fig{FF:5}(c) with the
corresponding analytical expression given by \Eq{itC} below. Note that again
the numerical prefactor is equal to unity.  \\
  
\paragraph*{\textbf{4.}  $^{4b}  {\cal B}_{12,34}$-term.} After substitution  
 of \Eq{3bA2} for \brown{$^{3a}\!a_1$} and  \Eq{1A2} for   \blue{$^{1}\!a_3$} the  $ ^{4b}   {\cal B}_{12,34} $ term  can be presented similarly to \Eq{4aAa}  as follows
  	   		\begin{eqnarray}  {^{4b}  {\cal B}_{12,34}}=   \frac  {\brown{G_1}\blue{G_3} G_{\dots}^2 V^4_{\dots} }{32}\<   \brown{ (    a_5  a_6 a_9  a_{10} )_1 }    \blue{( a_7 a_{8})_3 }   a_2 a_4 \>   .  ~~
  		\label{4bBaI}
  		\end{eqnarray} 
Pairing in the way: $[ \overbrace{2\mbox{-}(5},6,9,10]$ (four
options), $[ \overbrace{4\mbox{-}(10},9)]$ (two options), and $[
  \overbrace{9\mbox{-}(8},7)]$ (two options) and finally $[
  \overbrace{4\mbox{-}10}]$ (one option) we have 16 equal terms, while
denominator in \Eq{4bBa} is equal to 32. Therefore the results for
${^{4b} \! {\cal B}_{12,34}}$ has prefactor 1/2 as graphically shown
in \Fig{FF:5}(c) with the corresponding analytical expression given by
\Eq{itD} below.

 Since the diagram has mirror symmetry with respect to 1-3 diagonal,
 the factor $\dfrac 12$ in front of a diagram is consistent with our
 $\dfrac 1N$-rule.
  	
 \paragraph*{\textbf{5.}  $^{4c} {\cal B}_{12,34}$-term.}  Using twice
 \Eq{2A2} for \green{$^{2}\!a_j$} the $ ^{4c} {\cal B}_{12,34} $ term
 is shown in \Fig{FF:5}(e) (if one brakes the wavy
 lines). Analytically it can be presented as follows:
   \begin{eqnarray} ^{4c} \! {\cal B}_{12,34}=  \frac{\green{G_1 G_2  G_{\dots}^2 V^4_{\dots}}}{2^4}  \< \green{ ( a_5  a_6  a_{10})_1
                    ( a_7 a_8 a_{9})_2 } a_3 a_4 \>.~~
       \label{4cBa}
  		\end{eqnarray}

 The additional prefactor of $1/2$ is consistent with our $\frac
 1N$-rule since the diagram has two elements of symmetry:
 identity and rotation by $\pi$ radians which maps diagram onto
 itself.  Explicit analytical expression for $^{4c1} \! {\cal
   B}_{12,34}$ is given by \Eq{itE} below and can be reconstructed
 from \Fig{FF:5}(e).

 {{In \Eq{itE}}} we denote {{the result}} as
 $^{4c1} {\cal B}_{12,34}$ because there is another contribution to
 $^{4c} \! {\cal B}_{12,34}$, {{which}} originate from a
 different way of pairing $[ \overbrace{4\mbox{-}(9},5,6,8)]$ (four
 options), then $[ \overbrace{3\mbox{-}(5},6))]$ (two options),
 $[ \overbrace{6\mbox{-}8}]$ (one options) and finally
 $[ \overbrace{7\mbox{-}10}]$ (one option). Now we have again $8$
 equal contributions to Eq.(\ref{itF}) with has denominator 1/16. This
 gives again prefactor 1/2 reflecting mirror symmetry with respect to
 the horizontal line in the diagram $\ ^{4c2\!}{\cal B}_{12,34}$ shown
 in Fig.\ref{FF:5}(f) with the corresponding analytical expression
 given by \Eq{itF} below.  In total
\begin{equation}\label{B}^{4c} \! {\cal B}_{12,34}=\,^{4c1} \! {\cal B}_{12,34}+\,
^{4c2} \! B _{12,34}\ .
\end{equation}
\paragraph*{\textbf{6.}  $^{4 }   {\cal C}_{123,4}$-term.}  After substitution 
 of \Eqs{2A2} and \eqref{1A2} 
for \green{$^{2}\!a_1$}  \blue{$^{1}\!a_3$} and \blue{$^{1}\!a_4$} the  $ ^{4} \! {\cal C}_{123,4} $ term  can be presented similarly to \Eq{4aAa}  as 
 \begin{eqnarray}\nn {^{4} \! {\cal C}_{123,4}}=  \frac{\green{G_1}\blue{G_3\, G_4} G_{\dots}  V^4_{\dots}}
  		{16} \< \green{( a_5 a_6 a_{9})_1} a_2 \blue{( a_7
                  a_8)_3 ( a_9 a_{11})_4 }\>   .
  		\end{eqnarray} 
Pairing, for example as follows: $[ \overbrace{2\mbox{-}(5},6]$ (two
options), $[ \overbrace{6\mbox{-}(7},8, 10,11)]$ (four options), $[
  \overbrace{8\mbox{-}10},11]$ (two options) and $[
  \overbrace{9\mbox{-}11}]$ (one option) we have 16 equal
terms. Therefore the results for ${^{4} \! {\cal C}_{1,234}}$ has
prefactor unity, with analytical expression shown in \Eq{itG} below
and depicted graphically in \Fig{FF:5}(g). \\

\paragraph*{\textbf{7.}  $^{4} {\cal D}_{1234}$-term. } Substitution  \Eq{1A2} for \blue{$^{1}\!a_j$} into \Eq{4Ftot} for  $ ^{4} \! {\cal D}_{1234} $ gives 
	\begin{subequations}\label{4Db} 
  		\begin{eqnarray}\nn
  	   ^{4} \! {\cal D}_{1234}&=& \frac{\blue{ G_1 G_2 G_3 G_4
                      V^4_{\dots}}} {2^4 \cdot 4!}       \times \<
                  \blue{ ( a_5 a_6)_1 ( a_7 a_8)_2 ( a_9 a_{10})_3 (
                    a_{11} a_{12})_4} \> .
  		\end{eqnarray}    
                Pairing, for example as follows:
                $[ \overbrace{5\mbox{-}(7},8,9,10,11,12)]$ (six
                options) followed by the pairing
                $[ \overbrace{8\mbox{-}(9},10,11,12)]$ (four options)
                and $[ \overbrace{10\mbox{-}(11},12)]$ (two
                options). Pairing finally $[ \overbrace{6\mbox{-}12}]$
                (one option) we have 48 equal terms.  Therefore the
                results for $^{4} \! {\cal D}_{1234}$, shown in
                \Fig{FF:5}(h), has prefactor $\frac{48}{2^4 4!}= 1/8$,
                consistent with the $\dfrac 1N$-rule.
 \end{subequations}
The resulting diagrams are shown in
  \Fig{FF:5}(h). Corresponding analytical expression are as follow:  
 \begin{subequations}\label{iter}
  			\begin{eqnarray}\label{itA}
  			\ ^{4a\!}{\cal A}_{1,234}&=&G_1{\cal F}_2{\cal
                          F}_3{\cal F}_4{\displaystyle\int}\frac{d
                          \q_5} 
                          {(2\pi)^{d+1}}V_{18{\overline{5}}}V_{52{\overline 6}}V_{63{\overline 7}}V_{74\overline{8}}G_5 G_6 G_7 {\cal
                            F}_8, \\
  			\label{itB}
  			\ ^{4b\!}{\cal A}_{1,234}&=&G_1{\cal F}_2{\cal
                          F}_3{\cal F}_4{\displaystyle\int}\frac{d {
                            \q_5}}{(2\pi)^{d+1}}V_{18{\overline{5}}}V_{52{\overline
                            6}}V_{{\overline
                            7}36}V_{\overline{8}74}G_5{\cal F}_6
                        G_7^* G_8^*,  \\
  			\label{itC} 
                          \ ^{4a\!}{\cal B}_{12,34}&=&G_1G_2{\cal F}_3{\cal F}_4
                    {\displaystyle\int}\frac{d {  \q_5}}{(2\pi)^{d+1}}V_{18{\red{\overline{5}}}}V_{25{\overline 6}}V_{63\overline{7}}V_{7;4,{\overline 8}}{\cal F}_5{\red{G_6}}G_7{\cal F}_8, \\ 
  			\label{itD}
  			\ ^{4b\!}{\cal
                          B}_{12,34}&=&\frac{1}{2}G_1{\cal F}_2
                        G_3{\cal F}_4{\displaystyle\int}\frac{d \q_5}
                        {(2\pi)^{d+1}}
                        V_{18{\red{\overline{5}}}}V_{\red{5}2{\overline 6}}
                        V_{36\overline{7}}V_{\overline{8}74}G_5
                        {\cal F}_6{\cal F}_7G_8^*, \\
  			\label{itE}
  			\ ^{4c1}{\cal B}_{12,34}&=&\frac{1}{2}G_1{\cal
                          F}_2 G_3{\cal F}_4{\displaystyle\int}
                        \frac{d \q_5} {(2\pi)^{d+1}}
                        V_{18{{\overline{5}}}}V_{\red{5}2{{\overline{6}}}}
                        V_{36{{\overline{7}}}} V_{74\overline{8}} G_5 {\cal
                          F}_6G_7{\cal F}_8, \\
  			\label{itF}
  			\ ^{4c2}{\cal B}_{12,34}&=&
                        \frac{1}{2}G_1{\cal F}_2 {\cal F}_3G_4
                             {\displaystyle\int}
                             \frac{d   \q_5}{(2\pi)^{d+1}}
                             V_{18{\red{\overline{5}}}}
                             V_{{5}2{\red{\overline{6}}}}      V_{\overline{7}36}   V_{47\overline{8}}      G_5 {\cal F}_6G_7^*{\cal F}_8,   \\
  			\label{itG}
  			\ ^{4c}{\cal C}_{123,4}&=&G_1{\cal F}_2 G_3 G_4 {\displaystyle\int}      \frac{d { \q_5}}{(2\pi)^{d+1}} V_{1;8,{\red{\overline{5}}}}V_{52\overline{6}}V_{36\overline{7}}   V_{47\overline{8}} G_5 {\cal F}_6{\cal F}_7{\cal F}_8,   \\
  			\label{itH}
  			{\cal D}_{1234}&=&\frac{1}{8}G_1G_2G_3G_4{\displaystyle\int}      \frac{d {  \q_5}}{(2\pi)^{d+1}} V_{18\red{\overline{5}}}V_{25\overline{6}}V_{36\overline{7}}   V_{47\overline{8}} {\cal F}_5{\cal F}_6{\cal F}_7{\cal F}_8\ .  \label{SquareD} 
  			\end{eqnarray}
  		\end{subequations}
  	 
  	 	Here $\q_6=\q_2+\q_5$, $\q_7 = \q_3+\q_6=\q_2+\q_3+\q_5$ and $\q_8=\q_5-\q_1$.

\subsection{Calculations of the simultaneous triple correlator in the third order in the vertex}

To write down the corresponding analytical expression we will choose
the  {{notation}} and direction of wave
vectors according to {\ $\q_5=\q_1+\q_4$ and $\q_6=\q_4-\q_2$.
  Diagrams $^{3a \!}  A _{1,23} $ and $^{3b \!}  A _{1,23} $ (both
  with prefactor $\frac12$) produce two identical (under the
  permutation operator) twins.

Sums of these diagrams are shown in \Figs{FF:11}(a) and (b), now with
prefactor unity. These diagrams 
{{have the}}  Green's function $G_5$ {{oriented in different ways}}. 
{{The corresponding analytical expressions are given by}}
\begin{subequations}         \label{36}
              \begin{align}\begin{split} 
     ^{3a \!}  A _{1,23}\Sp {I} = F _2   F _3    \!\!    \int  \!\!  \frac{d \k_4 T_{123}T_{356}T_{2346}}{(2\pi)^{d }} 
                V_{14{\overline{5}}}V_{{{6}};\overline{ 4}2} V_{53\overline{6}}
                             F _{\red{4}}     \,, \ \nonumber\\                                    ^{3b \! } A _{1,23}\Sp {I}  =  F _2   F _3       \int  \frac{d \k_4 T_{123} T_{246}T_{2345}      }{(2\pi)^d}  V_{14{\overline{5}}}V_{\overline 426} V_{53\overline 6}   F _{6}\ .  
        \end{split}\end{align}
      \end{subequations}
      Diagram $^3 \C B _{12,3}$, presented in \Fig{FF:4}(c) has three
      {{child diagrams}} shown {{in}} \Figs{FF:11} ,
      {{panels}} ({\bf c}\,,a), ({\bf c}\,,b), and ({\bf c}\,,c)
      with the following analytical expressions
 
        \begin{eqnarray} \label{BTriads}
     ^3 B _{12,3} \Sp I  =& F _2 \int \frac{d\k_4 T_{123}T_{356}T_{2346} }{(2\pi)^ d }
                  V_{14\overline{5}} V_{{{6}\overline 4 2}}
                  V_{35\overline 6} F_{\red 4} F _{6} \,, \ \ \  ^3 B
                  _{12,3} \Sp {I{{I}}} = F _{\red{ 3}} \int
                  \frac{d\k_4 T_{123} T_{246}T_{2345} }{(2\pi)^ d } V_{14\overline{5}}
                  V_{{26\overline 4 }} V_{35\overline 6} F
                            _{5} F _{6} \,, \\
                                ^3 B _{12,3} \Sp {III}  =& F
                  _{\red{3}} \int \frac{d\k_4 {{  T_{123} T_{356}T_{1346}}}
                  }{(2\pi)^ d } V_{14\overline{5}}
                  V_{{{2}}\overline 4 6} V_{3 5 \overline 6} F
                  _{{\red{4}}} F _{\red{5}} \ .
        \end{eqnarray}
        
{The procedure of ``multiplication'' ensures that all the
  ``children''} inherit the same combinations of the {{vertices}} and
double correlators from its common parent, but differ in the frequency
integrals.

Last diagram in \Fig{FF:4}, {{panel}} (d) with prefactor $\frac16$
produces six identical twins and results in diagram \Fig{FF:11} ({\bf
  d}) now with prefactor unity. {{The analytical expression corresponding to these
    diagrams is given by}}
\begin{eqnarray}\label{Ctriads}
     ^3  C_{123}&= &    \int  
     \frac{d\k_4T_{123} T_{246}T_{2345}}{(2\pi)^{d }}V_{145}V_{2{\overline{4}}6}V_{35\overline 6}
        F   _4   F_{5} F _{6} \ .
            \end{eqnarray}  
    
            Note that diagrams for $^{3b}\!  A _{1,23}\Sp {I}$,
            $^{3}\!  B _{1,23}\Sp{II}$,and $^3 C _{123} $, shown in
            \Fig{FF:11}) ({\bf b}), ({\bf c}\,b), and ({\bf d}), have
            identical directions {{for}} all
            Green's functions and therefore have the same frequency
            integral while diagrams for $^3 B _{12,3}\Sp{I }$ and
            $^3 B _{12,3}\Sp{III }$, shown in \Fig{FF:11} ({\bf c}\,a)
            and ({\bf c}\,c) have the same (but different from the
            previous set of diagrams) orientation of the Green
            functions.

     \subsection{Square Diagrams: Details of Calculations\label{SquareOneTIme}}

         \subsubsection{\label{sss:L3}Next-lowest ($4\sp{rd}$) order diagrams for quadruple   correlator 
         \label{SquareDiagramsFourPointCOrrelator}}

       Each of the eighth diagrams for the quadruple correlator,
       depicted in \Fig{FF:5} with analytical
       expressions\,\eqref{iter} involve four double
       correlator. Consequently, they produce $8\cdot 2^4=128$ child
       diagrams but only {{22}} of them, shown in \Figs{FF:12} and
       \ref{FF:13}, contribute to simultaneous
       correlator.
       
               The detailed are presented in the
           Appendix (\ref{FourPointCorrelatorFourthOrder}).

 First diagram $ ^{4a\!} A_{1,234}$ has only one surviving child shown
 in \Fig{FF:12}({\bf a}):
                  \label{F4c}
    			\begin{eqnarray}
    		^{4a\!}  A_{1,234}\Sp {I}= F_2 F _3 F _4 \!\! \int \!\!
                          \frac{d \k_5 T_{123} T_{356}T_{1346}}
                               {(2\pi)^d}V_{18{\overline{5}}}V_{52{\overline
                                   6}}V_{63{\overline
                                   7}}V_{74\overline{8}} F_8 .
                               ~~~~~ \label{4aAb2}
    		\end{eqnarray}

    	Diagram $ ^{4b\!}  A_{1,234}$,  depicted on \Fig{FF:5}b
        has two surviving children,  \Fig{FF:12}({\bf
              b}\,a) and \Fig{FF:12}({\bf b}\,b).  Corresponding
        analytical expressions  are given by 
        			 \label{4bAc}\begin{align}\begin{split}
    					^{4b\!}  A _{12;34}\Sp{I} = F_2  F   _3F_4  \!\! \int \!\! \frac{d {  \k_5   T_{1234}T_{256}T_{2357}T_{23458}  }}{(2\pi)^d} V_{18{\overline{5}}}V_{52{\overline 6}}V_{{\overline 7}36}V_{\overline{8}74}  F  _6       ,   \ \  \nonumber\\
                                                            ^{4b\!}  A _{12;34}\Sp {II}=F_2  F   _3F_4  \!\! \int \!\! \frac{d {  \k_5}
 T_{1234}T_{367}T_{2357}T_{12378}}{(2\pi)^d} V_{18{\overline{5}}}V_{52{\overline 6}}V_{{\overline 7}36}V_{\overline{8}74}  F  _6    .  
    					\end{split}\end{align}  
These analytical expressions follow the same pattern: they
 have the same combination of vertices and double correlators.
 This appears to be the  general rule for all children of the
 same parent diagram.  These analytical expressions for the
     ``child'' of the same parent nevertheless do  differ in
 frequency integrals.  Diagrams $^{4a\!}B_{12,34}$ [\Fig{FF:5}c and
   \Eq{itC}], have already three surviving children, shown in 
     \Fig{FF:12}, {{panels}} ({\bf c}\,a), ({\bf c}\,b), ({\bf c}\,c).

     Note that the total of 22 square diagrams for the four-point
     correlator of their fourth-order presented on \Fig{FF:12} and
     \Fig{FF:13} depend only on very few (8 to be exact) frequency
     integrals. This is due to the fact that the frequency integral
     depends only on the position and orientation of the Green
     functions, and is independent on the vertices and on whether the
     Green's function is "true" or "auxiliary". Consequently, there
     are a lot of repeated frequency integrals as seen 
     {{in}} \Figs{FF:12} and \Figs{FF:13} and analytical
     expressions {for these diagrams.}

   \begin{figure*}  
    	  \includegraphics[width=0.7\textwidth]{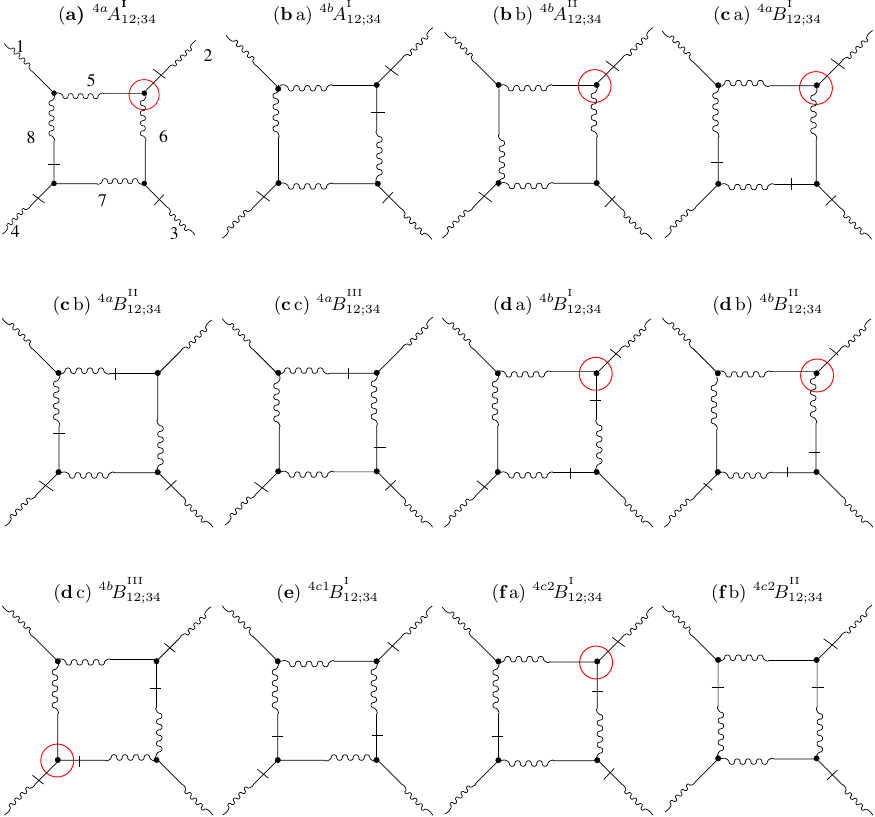}  
          \caption{
            \label{FF:12} First group ($ A$ and $ 
          B$) of next-lowest ($4\sp{th}$) order ``child" diagrams for
          the simultaneous quadruple correlator $^{4\!} F_{1234}$.
                  }
   \end{figure*}
  
 \begin{figure*}  \includegraphics[width=0.7\textwidth]{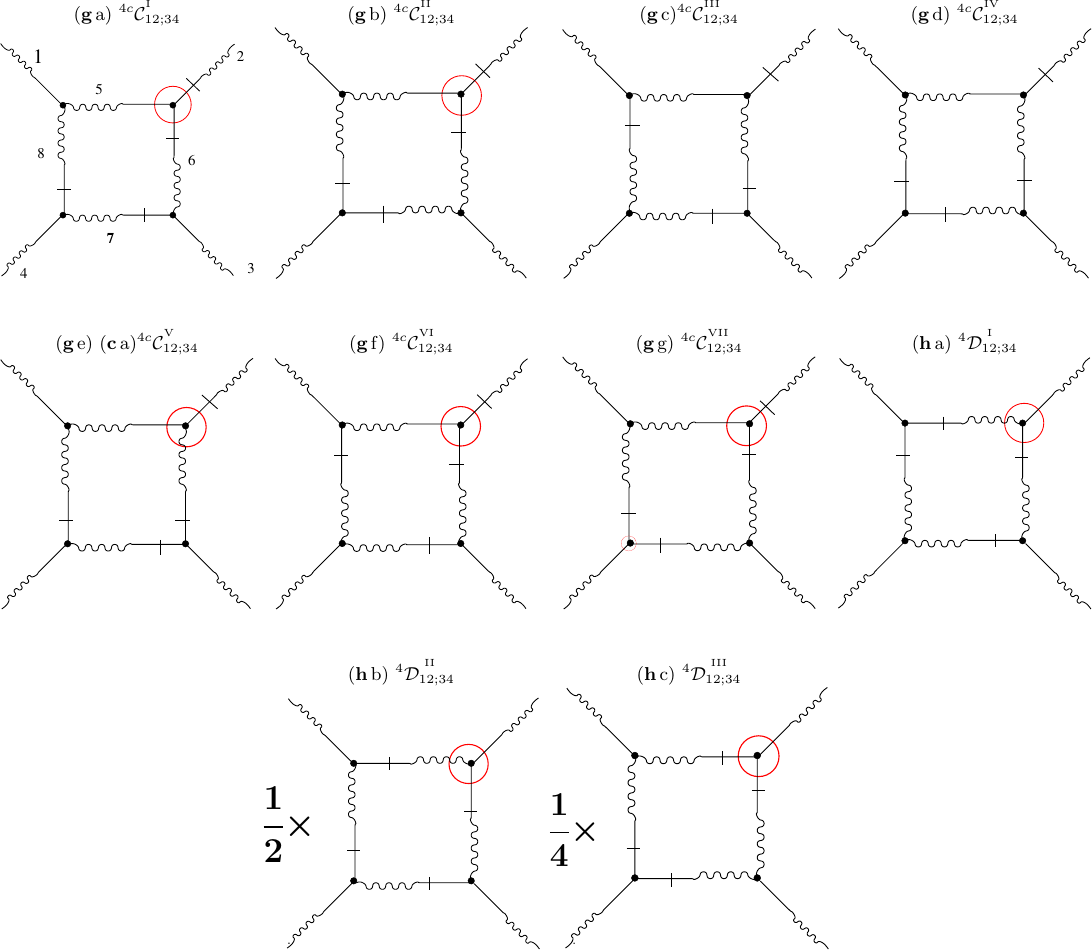}
 	 
   \caption{\label{FF:13}
     Second group ($\C C$ and $\C D$) of the
     next-lowest ($4\sp{th}$) order ``child" diagrams for the
     simultaneous quadruple correlator $^{4\!} F_{1234}$. }
         	\end{figure*}

        

{{The analytical expressions corresponding to the panels (ca), (cb) and (cc) of the \Figs{FF:12} are given by}}

        \label{4aB}\begin{align}\begin{split}
\white{.}\hskip -.2 cm       	^{4a\!}B_{12,34}\Sp {I  }= 
                                  F _3  F   _4 \!\! \int  \!\! \frac {d\k_5 {{ T_{1234}T_{367}T_{23458}\left(T_{3468}+T_{2357}\right) 
                                  }}}{(2\pi)^d} V_{18{{\overline{5}}}}V_{25{\overline 6}}V_{63\overline{7}}V_{74{\overline 8}}F_5 F_8,  \\ 
  \white{.}\hskip -.2 cm       	^{4a\!}B_{12,34}\Sp {II}= F_3 F_4  \!\! \int  \!\!\frac{d    \k_5  T_{1234}\  T_{478}\,T_{3468}\,T_{23458}  }{(2\pi)^d}  V_{18{{\overline 5}}}V_{25{\overline 6}}   V_{63\overline{7}}V_{74{\overline 8}} F_5 F_8 ,  \\ 
                                  \white{.}\hskip -.2 cm        	^{4a\!} B_{12,34}\Sp {III}=F_3F_4 \!\! \int  \!\! \frac{d {  \k_5 
T_{1234}T_{367}T_{23458}\left(T_{3468}+T_{2357}\right)
                                  }}{(2\pi)^d} V_{18{{\overline{5}}}}V_{25{\overline 6}}V_{63\overline{7}}V_{74{\overline 8}}F_5F_8.  \label{Square4aB} 
       	\end{split}\end{align} 
      
      Parent diagram $^{4b\,}\C B_{12,34}$, \Fig{FF:5}({\bf d}), is
      symmetric with respect to rotation around {{the}} first
      and {{the}} third leg. Therefore in accordance with the
      $\dfrac 1N$ rule, it has prefactor $\frac12$. It has four
      children. The first two of them are rotationally symmetric and
      became identical (i.e. twins) under the permutation
      operator. They both contribute to diagram
      $ ^{4b\,} B_{12,34}\Sp {I}$, \Fig{FF:12} ({\bf d\,}a), now without
      symmetry and with prefactor unity. Last two children are shown
      in \Fig{FF:12}, {{panels}} ({\bf d\,}b) and ({\bf d\,}c),
      both symmetric and with prefactors $\dfrac 12$ as required by
      general diagrammatic rules. Their analytical expressions are:

\label{XX}\begin{align}\begin{split}
       &    \white{.}\hskip -.4 cm    	\ ^{4b\!}B_{12,34}\Sp {I }= F_2 F_4\!\!
                    \bigintsss \!\!
                \frac{d \k_5 T_{1234}\  T_{478}\,T_{3468}\,T_{23458} } {(2\pi)^d}
                V_{18{{\overline{5}}}}V_{{ 5}2{\overline 6}}V_{36\overline{7}}V_{\overline{8}74} F_6F_7
                    , \label{4bBa}
               \\ 
 &               \white{.}\hskip -.4  cm  \ ^{4b\!}B_{12,34}\Sp {II
   }=\frac{ F_2 F_4}{2}\!\! \bigintsss \!\!\frac{d \k_5
    T_{1234}T_{367}T_{2357}T_{12378}}
                {(2\pi)^d}V_{18{{\overline{5}}}}V_{{5}2{\overline
                    6}}V_{36\overline{7}}V_{\overline{8}74} F_6F_7
   , 
                         \\ 
   &                 \white{.}\hskip -0.4  cm  \ ^{4b\!}B_{12,34}\Sp {III
                }=\frac{ F_2 F_4}{2}\!\! \bigintsss \!\! \frac{d \k_5T_{1234}T_{367}T_{23458}\left(T_{3468}+T_{2357}\right)}
                {(2\pi)^d}
                V_{18{\overline{{5}}}}V_{{5}2{\overline 6}}V_{36\overline{7}}V_{\overline{8}74}
                F_6F_7.
                      \end{split}
          \end{align}

     Due to the nature of the ``multiplication'' procedure the
         ``children'' repeat the parent's genome (combination of
         {{vertices}} and double correlators) but differ in
         frequency integrals.
                  \\
     
Next diagrams $^{4c1}{\cal B}_{12;34}$ [\Fig{FF:5}(e) and \Eq{itE}]
(with prefactor $\frac12$) has two twins both contributing to 
   \Fig{FF:12}({\bf e}),  with corresponding analytical
      expression given by    \
\begin{eqnarray}\label{4c1-e}
    	 ^{4c1}B_{12,34}\Sp {I  }=F_2 F_4 \!\! \int  \!\!    \frac{d   \k_5 T_{123} T_{356}T_{1346} } {(2\pi)^d}    V_{18{{\overline{5}}}}V_{{5}2{{\overline{6}}}}      V_{36{\overline{{7}}}}      V_{74\overline{8}} F_6 F_8. 
    		        \end{eqnarray}
where {{the factor}}  $\frac12$   was
      replaced  {{by unity}}.

Diagrams $^{4c2}{\cal B}_{12;34}$ [\Fig{FF:5}(f) and \Eq{itF}] has two
sets of two twins, shown in  \Fig{FF:12}({\bf}\,fa) and ({\bf
  f}\,b)  respectively. 
{{Following the }}    $\dfrac 1N$ rule, {{we now replace }}
    $\frac12 \Rightarrow 1$ {{in both diagrams}}.
We therefore have
 \begin{eqnarray}
   \hskip - .3 cm   	  ^{4c2}B_{12,34}\Sp {I  }&=&F_2  F_3  \!\!\! \bigintssss  \!\!    \frac{d   \k_5  T_{1234}\  T_{478}\,T_{3468}\,T_{23458}  }{(2\pi)^d}    V_{18{{\overline{5}}}}V_{{5}2{{\overline{6}}}}      V_{\overline{7}36}   V_{47\overline{8}}      F_6F_8,  \nonumber\\
     	 ^{4c2}B_{12,34}\Sp {II  }&=&F_2 F_3 \!\!\! \bigintssss  \!\!    \frac{d   \k_5T_{1234}T_{256} T_{478}\left(T_{12467}+T_{23458}\right)}{(2\pi)^d}      V_{18{{\overline{5}}}}V_{{5}2{{\overline{6}}}}      V_{\overline{7}36}   V_{47\overline{8}}      F_6F_8.\nonumber \\ \label{Square4c2Bb}  
    		\end{eqnarray}

                The most ``prolific" diagram $^{c}\C C_{123,4}$
                [\Fig{FF:5}(g) and \Eq{itG}] have as much as seven
                children, shown in \Fig{FF:13}({\bf g}\,a--g).
                Consistent with the logic of our multiplication
                the procedure they differ only in frequency integrals.
                      {The corresponding analytical expressions are given by}}
     \begin{subequations}\label{4C}
     	\begin{eqnarray} 
          ^{4c}C_{123,4}\Sp {I  }&=&
                                     F_2\!\!\! \int  \!\!    \frac{d   \k_5
 \, T_{1234}\  T_{478}\,T_{3468}\,T_{23458} }   {(2\pi)^d}  V_{18{{\overline{5}}}}V_{52\overline{6}}V_{36\overline{7}}   V_{47\overline{8}} F_6F_7F_8, \nonumber\\
          ^{4c}C_{123,4}\Sp {II  }&=&F_2\!\!\! \int  \!\!    \frac{d { \k_5
 {{{ T_{1234}T_{256}  T_{12378}  \left(T_{2357}+T_{1268}\right)}}}
 }}{(2\pi)^d} V_{18{{\overline{5}}}}V_{52\overline{6}}V_{36\overline{7}}   V_{47\overline{8}} F_6F_7F_8, ~~~~~~   \nonumber\\
          ^{4c}C_{123,4}\Sp {III  }&=&F_2\!\!\! \int  \!\!     \frac{d { \k_5
 \,^T_{1234}\  T_{256}\,T_{2357}\,T_{23458}}}{(2\pi)^d}  V_{18{{\overline{5}}}}V_{52\overline{6}}V_{36\overline{7}}   V_{47\overline{8}} F_6F_7F_8,   \nonumber\\ 
                                       ^{4c}C_{123,4}\Sp {IV  }&=&F_2\!\!\! \int  \!\!     \frac{d { \k_5
 T_{123} T_{356}T_{1346}
    T_{123} T_{356}T_{1346}     }}{(2\pi)^d}  V_{18{{\overline{5}}}}V_{52\overline{6}}V_{36\overline{7}}   V_{47\overline{8}} F_6F_7F_8,  ~~~~~~\nonumber\\
                     ^{4c}C_{123,4}\Sp {V }&=&F_2\!\!\! \int  \!\!     \frac{d { \k_5 T_{1234}\ T_{367}\,T_{2357}\,T_{12378}     }}{(2\pi)^d} V_{18{{\overline{5}}}}V_{52\overline{6}}V_{36\overline{7}}   V_{47\overline{8}} F_6F_7F_8,  \nonumber\\
          ^{4c}C_{123,4}\Sp {VI  }&=&F_2\!\!\! \int  \!\!    \frac{d { \k_5
T_{1234}T_{256} T_{478}\left(T_{12467}+T_{23458}\right)
      }}{(2\pi)^d} V_{18{{\overline{5}}}}V_{52\overline{6}}V_{36\overline{7}}   V_{47\overline{8}} F_6F_7F_8,   ~~~~~~ \nonumber\\
          ^{4c}C_{123,4}\Sp {VII  }&=&F_2\!\!\! \int  \!\!    \frac{d {
                                       \k_5  T_{1234}T_{367}T_{23458}\left(T_{3468}+T_{2357}\right)      }}{(2\pi)^d}  V_{18{{\overline{5}}}}V_{52\overline{6}}V_{36\overline{7}}   V_{47\overline{8}} F_6F_7F_8  .  ~~~~~~~  
 	\end{eqnarray}\end{subequations} 
      Finally, we give analytical expressions for $2^4-2=14$ children
      of diagram $^{4\!}\C D_{1234}$ with prefactor $\frac18$
      [\Fig{FF:5}(h) and \Eq{itH}], shown \Fig{FF:13}({\bf h}\,a),
      ({\bf h}\,b), and ({\bf h}\,c). First, eight identical (under
      the permutation operator) twins together contribute to
      $\ ^6 D^{I}_{12,34}$ shown on \Fig{FF:13}({\bf h}\,a).  This
      diagram has no symmetries and therefore has prefactor unity.
      Diagram \Fig{FF:13}({\bf h}\,b) includes four twins and has
      prefactor $\dfrac 12$ instead of the parent prefactor $\dfrac 18$,
      while last diagram {{ \Fig{FF:13}({\bf h}\,c)}} includes
      only two twins and has prefactor $\dfrac 14$.  

  We now write down all analytical expressions for the $D$ square
  diagrams:
      \begin{eqnarray}
            ^{4c}D_{1234}\Sp{I }&=&\!\!\int \!\!\!  \frac{d { \k_5
  T_{1234}\  T_{256}\,T_{1268}\,T_{12467} 
      }}{(2\pi)^d}
      V_{18{{\overline{5}}}}V_{25\overline{6}}V_{36\overline{7}}
      V_{47\overline{8}} F_5F_6F_7F_8, \nonumber\\
        \ \ \ \  ^{4c}D_{1234}\Sp{II }&=&\!\!\int \!\!\!  \frac{d { \k_5
\, T_{1234}\ T_{158}\,T_{1268}\,T_{12378} }}{2(2\pi)^d}
      V_{18{{\overline{5}}}}V_{25\overline{6}}V_{36\overline{7}}
      V_{47\overline{8}} F_5F_6F_7F_8 \,,\nonumber      \\ 
        ^{4c}D_{1234}\Sp{III }&=&\!\!\int \!\!\!  \frac{d { \k_5
T_{1234}T_{367}T_{23458}\left(T_{3468}+T_{2357}\right) }}{4( 2\pi)^d}
      V_{18{{\overline{5}}}}V_{25\overline{6}}V_{36\overline{7}}
      V_{47\overline{8}} F_5F_6F_7F_8.\nonumber\\          \label{4Da}  
	\end{eqnarray}
        Here as before  $\q_6=\q_2+\q_5$,  $\q_7 = \q_3+\q_6=\q_2+\q_3+\q_5$     and $\q_8=\q_5-\q_1$.
\end{document}